\documentclass[12pt,psfig,twoside]{report}

\usepackage{url}
\usepackage{romannum}

\usepackage{amssymb}
\usepackage[cmex10]{amsmath} 
\usepackage{mathtools}
\usepackage{amsthm}  
\usepackage{amsfonts}
\usepackage{bbm}
\usepackage{pdfpages}
\usepackage{pgf}
\usepackage{pgffor}
\usepackage{tkz-graph}
\usepackage{tikz}
\usepackage{fp}
\usepackage{textcomp}
\usetikzlibrary{patterns,positioning,shapes,arrows,arrows.meta,graphs,svg.path,calc,fit}
\usepackage{pgfplots}
\usepackage{mathtools} 
\usepackage{pst-sigsys}
\usepackage{stfloats}
\usepackage{graphicx}
\usepackage{subcaption}
\usepackage{epsfig}
\usepackage{epstopdf}
\usepackage{booktabs}

\usepackage{cite}
\usepackage{setspace}
\usepackage{comment}
\usepackage[draft]{todonotes}
\usepackage{IEEEtrantools}
\usepackage{tabu}
\usepackage{xcolor}
\usepackage{enumerate }

\usepackage{comment}

\usepackage{cancel}
\usepackage{caption}

\newtheorem{theorem}{Theorem}[chapter]

\newtheorem{lemma}{Lemma}[chapter]

\newcommand{\bF}{{{\boldsymbol{F}}}}
\newcommand{\bX}{{{\boldsymbol{X}}}}
\newcommand{\bP}{{{\boldsymbol{P}}}}
\newcommand{\bfv}{{{\boldsymbol{f}}}}
\newcommand{\bfA}{{{\boldsymbol{A}}}}
\newcommand{\bfU}{{{\boldsymbol{U}}}}
\newcommand{\bfD}{{{\boldsymbol{D}}}}
\newcommand{\bfV}{{{\boldsymbol{V}}}}
\newcommand{\bfG}{{{\boldsymbol{G}}}}
\newcommand{\bfB}{{{\boldsymbol{B}}}}
\newcommand{\bfP}{{{\boldsymbol{P}}}}
\newcommand{\bfC}{{{\boldsymbol{C}}}}

\newcommand{\mycomment}[1]{}  

\edef\cohtheory#1{%
	\noexpand\newcommand\expandafter\noexpand\csname #1func\endcsname[1][*]{%
		\noexpand\MakeUppercase{#1}^{##1}}%
}

\DeclareMathOperator{\tr}{Tr}
\DeclareMathOperator{\Ev}{E}
\newcommand{\R}{\mathbb{R}}
\newcommand{\C}{\mathbb{C}}
\newcommand\x{\times}
\newcommand{\V}[1]{\ensuremath{\mathbf{#1}}}

\newcommand{\E}{\ensuremath{\mathbb{E}}}
\newcommand{\Fc}{\ensuremath{\mathcal{F}}}

\newcommand{\m}{m}

\newcommand{\norm}[1]{\left|\left| #1 \right|\right|}

\newcommand{\specstat}{\ensuremath{\Psi}}
\newcommand{\vx}{\ensuremath{\mathbf{x}}}
\newcommand{\Xk}{\ensuremath{X_K}}
\newcommand{\Xkn}{\ensuremath{X_{K_n}}}
\newcommand{\Gk}{\ensuremath{G_K}}
\newcommand{\Gkn}{\ensuremath{G_{K_n}}}

\makeatletter
\newcommand\mathcircled[1]{%
	\mathpalette\@mathcircled{#1}%
}
\newcommand\@mathcircled[2]{%
	\tikz[baseline=(math.base)] \node[draw,circle,inner sep=1pt] (math) {$\m@th#1#2$};%
}
\makeatother

\makeatletter
\def\calcLength(#1,#2)#3{%
	\pgfpointdiff{\pgfpointanchor{#1}{center}}%
	{\pgfpointanchor{#2}{center}}%
	\pgf@xa=\pgf@x%
	\pgf@ya=\pgf@y%
	\FPeval\@temp@a{\pgfmath@tonumber{\pgf@xa}}%
	\FPeval\@temp@b{\pgfmath@tonumber{\pgf@ya}}%
	\FPeval\@temp@sum{(\@temp@a*\@temp@a+\@temp@b*\@temp@b)}%
	\FProot{\FPMathLen}{\@temp@sum}{2}%
	\FPround\FPMathLen\FPMathLen5\relax
	\global\expandafter\edef\csname #3\endcsname{\FPMathLen}
}
\makeatother

\makeatletter
\newcommand{\gettikzxy}[3]{%
	\tikz@scan@one@point\pgfutil@firstofone#1\relax
	\edef#2{\the\pgf@x}%
	\edef#3{\the\pgf@y}%
}
\makeatother

\tikzset{%
	block/.style    = {draw, thick, rectangle, minimum 	height = 1em, minimum width = 2em},
	fact/.style		= {draw, thick, isosceles triangle, minimum height = 0.5em, minimum width = 1em, inner sep=2pt} ,
	sum/.style      = {draw, circle, inner sep = 1pt, minimum size = 1em}, 
	junc/.style     = 
	{coordinate}, 
	input/.style    = {coordinate}, 
	output/.style   = {coordinate}, 
	txt/.style   	= {rectangle, inner sep=0pt, minimum size=2pt}, 
	point/.style  	= {circle, inner sep=1.5pt, fill=black},
	edgeto/.style	= {->,>=triangle 45} ,
	edge/.style	= {-} ,
	bitsline/.style = {-{Latex[angle=60:10pt]}, double,double distance=2pt, line width=1pt},
	switchline/.style	= {->,>=triangle 45,dashed} ,
	nodesblock/.style 2 args = {
		draw, rounded corners, dashed, 
		inner sep=0pt, outer sep=0pt,
		fit=(#1) (#2)},
	fitblock/.style={draw=black, rounded corners, dashed,
		inner xsep=1em,
		inner ysep=1.25em}
}

\newcommand{\emptypage}{\newpage \thispagestyle{empty} $\;$ \newpage}

\usepackage{lscape}

\begin{document} \pagenumbering{roman}
	
	\topmargin 1.5cm \clearpage \thispagestyle{empty}

\newcommand{\jMAtitle}{Analog Coding Frame-work
	}
\newcommand{\jAuthor}{Marina Haikin}
\newcommand{\jSubmitDate}{June, 2018}

\begin{center}

\Large{
	\bf TEL AVIV UNIVERSITY
}

\large
The Iby and Aladar Fleischman Faculty of Engineering\\
The Zandman-Slaner School of Graduate Studies\\
the Department of Electrical Engineering - Systems\\
  \vspace{3.5cm}
\Huge  {\bf   \jMAtitle}\\
  \vspace{4.5cm}
  \large{A thesis submitted toward the degree of \\Master of Science in Electrical and Electronic Engineering}\\
  \vspace{0.4cm}
  by \\
  \vspace{0.4cm}
  \Large {\bf \jAuthor}

  \vspace{0.4cm}

  \vspace{3 cm}

  \Large {\jSubmitDate}
\end{center}

\newpage
\emptypage

\thispagestyle{empty}

\begin{center}

\Large{
\bf TEL AVIV UNIVERSITY
}

\large
The Iby and Aladar Fleischman Faculty of Engineering\\
The Zandman-Slaner School of Graduate Studies\\
the Department of Electrical Engineering - Systems\\
  \vspace{1.3cm}
  \Huge {\bf \jMAtitle}\\
  \vspace{1.3cm}
  \large{A thesis submitted toward the degree of \\Master of Science in Electrical and Electronic Engineering}\\
  \vspace{0.3cm}
  by \\
  \vspace{0.3cm}
  \Large {\bf \jAuthor}

  \vspace{3cm}

  \normalsize

  This research was carried out in The School of Electrical Engineering \\The Department of Electrical
  Engineering - Systems,\\
  Tel-Aviv University\\

  \vspace{1cm}

  Supervisors: \\
  {\bf Prof. Ram Zamir}\\
  {\bf Dr. Matan Gavish}\\
  \vspace{3cm}
  \large {\jSubmitDate}
\end{center}

\newpage
\topmargin 0cm

	\onehalfspace
	
	\newpage
	\emptypage
	
	\thispagestyle{empty}

\vspace{1.5cm}

\begin{center}
\textbf{Acknowledgements}
\end{center}

\vspace{2cm}

First I would like to express my sincere gratitude to my advisors Prof. Ram Zamir and Dr. Matan Gavish.
I thank Rami for the continuous support during this long and meaningful period, guidance, patience, and immense knowledge. His intuition and creativity led to many promising directions.
I would like to thank Matan Gavish for his contribution from a different discipline, the statistical perspective, and his enthusiasm and encouragement. 
This work involved both information theoretic aspects and statistics, a fascinating intersection that I had the opportunity to explore thanks to my advisors.
Through our joint work, I learned a lot from their ideas, methodologies, and tips for a good science as well as professional writing.

Finally, I am very grateful to my dear family for their patience and support over the years and intense periods.
Special thanks to my beloved husband Lev, for his encouragement and help with our kids (who joined us during this period of M.Sc study :))

	\newpage
	\emptypage
	
	\begin{abstract}

	Analog coding is a low-complexity method to combat erasures,
	based on linear redundancy in the signal space domain.
	Previous work examined "band-limited discrete Fourier transform (DFT)"
	codes for Gaussian channels with erasures or impulses.
	We extend this concept to source coding with "erasure
	side-information" at the encoder and show that the performance of band-limited DFT
	can be significantly improved using irregular spectrum, and more
	generally, using equiangular tight frames (ETF).
	
	Frames are overcomplete bases and are widely used in mathematics, computer science, engineering, and statistics since they provide a stable and robust decomposition. 
	Design of frames with favorable properties of random subframes is motivated in variety of applications, including code-devision multiple access (CDMA), compressed sensing and analog coding.	
	
	We present a novel relation between deterministic frames and random matrix theory. We show empirically that the MANOVA ensemble offers a universal description of the spectra of randomly selected subframes with constant aspect ratios, taken from deterministic near-ETFs. 
	Moreover, we derive an analytic framework and bring a formal validation for some of the empirical results, specifically that the asymptotic form for the moments of high orders of subsets of ETF agree with that of MANOVA. 
	
	Finally, when exploring over-complete bases, the Welch bound is a lower bound on the root mean square cross correlation between vectors.
	We extend the Welch bound to an erasure setting, in which a reduced frame, composed of a random subset of Bernoulli selected vectors, is of interest. The lower bound involves moment of the reduced frame, and it is tight for ETFs and asymptotically coincides with the MANOVA moments. This result offers a novel perspective on the superiority of ETFs over other frames.
	
\end{abstract}

	\doublespace
	
	\newpage
	\emptypage

	\tableofcontents \singlespace \listoffigures \singlespace \listoftables

	\chapter{Introduction}
		\label{chapter:Introduction}
		\pagenumbering{arabic}
		
\section{Motivations, Objectives and Contributions}
Information theory offers fundamental limits for data compression and transmission. Coding theory studies the properties of codes for the purpose of designing efficient and reliable data transmission methods - data compression or error control. This research was motivated by the problem of source coding with distortion side information (SI) at the encoder. Such a scenario might be interesting in the context of perceptual distortion measures for audio coding or damaged sensors. While proposing and analyzing an analog coding approach using over complete bases, named frames, a whole new field of frame study gradually became the focus of this research.

Consider encoding a source $X$ under a side-information dependent distortion measure $d(x,\hat{x},s)$, where the side information $S$ is statistically independent of the source $X$ and is available only at the encoder.
It is shown in \cite{martinian2008source} that if an optimal conditional
distribution $p(\hat{x}| x,s)$, that achieves the conditional rate-distortion function (RDF) of $x$ given $s$, satisfies $I(S;\hat{X})=0$,
then the rate-distortion performance is the same as if
$S$ was available also at the decoder.
Specifically, this condition holds for the case of an ``erasure distortion measure''
$d(x,\hat{x},s) = s \cdot d(x,\hat{x})$, for $s \in \{0,1\}$,
where only source samples for which $S=1$ are ``important''.
The optimal rate distortion function can be achieved via random coding at an exponential complexity. It requires from the encoder a joint-typicality encoding based on the relevant samples. Hence, in practice, other approaches are necessary. 

For the lossless case a Reed-Solomon (RS) decoder-encoder scheme, suggested in \cite{martinian2008source}, achieves the optimal information-theoretical solution.
Such scheme offers a deterministic, as well as lower complexity coding in a reduced dimension. 
Our goal was to explore the possibility of using a structured, non-random scheme for the setup of lossy coding, specifically by means of analog coding. 

Analog coding as considered in \cite{wolf1983redundancy} is the "analog analogy" of the RS solution (without switching the roles of encoder-decoder). For erasure correction, it creates an analog redundancy by means of band-limited discrete Fourier transform (DFT) interpolation.
In \cite{haikin2016analog}
we examine the analog coding paradigm
for the dual setup of a source with erasure SI
at the encoder. It decouples the analog part of "erasure side information" utilization and the digital component of quantization. 
Preforming an analog dimension reduction enables to deal separately with scalar or vector quantization, in a reduced dimension.
However, it is not clear what structure of transform should be used. Using the DFT matrix naively, as proposed in \cite{wolf1983redundancy}, leads to significant loss in achievable rate.
The band-limited DFT, which is a special case of a frame, suffers from a severe signal amplification caused by the transform at the encoder.
The excess rate of analog coding above the RDF is associated with the energy of the inverse
of submatrices of the frame, where each submatrix corresponds to a
possible erasure pattern.
One of the central motivations of this work is to construct frames such that this performance loss is minimized. Deterministic constructions are often preferred for the sake of implementation simplicity. 
We show that by selecting the DFT frequencies from a {\em difference set},
or more generally, by using equiangular tight frames (ETF),
we minimize the excess rate over all possible frames (although do not achieve the RDF).

Letting $F = [f_1|...|f_n]$ denote the $m$-by-$n$ frame matrix, $f_1,...,f_n$ are frame vectors in the $m$ dimensional space for $n\geq m$.
A tight frame is a frame for which the ratio between the norm of an expanded vector by a frame, $\|F'x\|^2$, and its original norm, $\|x\|^2$, is constant and does not depend on $x$`. A uniform tight frame (UTF) includes also a unit-norm restriction on all frame vectors. Finally, in ETFs the absolute cross-correlation of all pairs of frame vectors is constant. The Welch Bound is a lower bound on the root mean square (rms), and the maximum cross correlation between $n$ unit-norm frame vectors.
UTFs and (unit-norm) ETFs satisfy with equality the rms and maximum Welch bound, respectively. 

A major part of this thesis deals with formalizing the evident superiority of ETFs and proving the surprising observation that random subsets of ETF-like deterministic frames have MANOVA spectrum.
Suppose we draw a random subset of $k$ columns out of $n$ from a frame matrix $F$, where $k$ and $m$ are proportional
to $n$. 
Consider the distribution
of singular values of the $k$-subset matrix.
For a variety of important ETFs and tight non-ETFs, 
we observe in \cite{haikin2017random}
that, for large $n$, 
the singular values can be precisely described by a known probability
distribution: Wachter's MANOVA (multivariate ANOVA) spectral
distribution, a phenomenon that was previously known only for
two types of random frames \cite{farrell2011limiting}. 
In terms of convergence to this limit,
the $k$-subset matrix from all of these frames is shown to be empirically
indistinguishable from the classical MANOVA (Jacobi) random
matrix ensemble. Thus, the MANOVA ensemble
offers a universal description of the spectra of randomly selected
$k$ subframes taken from deterministic frames. 
The same universality phenomena is
shown to hold for notable random frames as well.
This description enables exact calculations of properties of solutions for
systems of linear equations based on a random choice of $k$ frame vectors out of
$n$ possible vectors, and has a variety of implications for erasure coding,
compressed sensing, and sparse recovery.  
When the aspect ratio $m/n$ is small, the MANOVA spectrum tends to the well
known Mar\u cenko-Pastur distribution of the singular values of a Gaussian
matrix, in agreement with previous work on highly redundant frames.
These results are empirical, but they
are exhaustive, precise and fully reproducible.

In the purpose of an analytical support for our results and an analysis of properties of ETF subsets, we explore the moments of unit-norm frames.  
The Welch rms lower bound can be viewed as a lower bound on the second moment of $F$,
namely on the trace of the squared Gram matrix $(F'F)^2$.
In the erasure setting considered we in \cite{haikin2018frames}, a reduced frame, composed of a random subset of Bernoulli selected vectors, is of interest.
We extend the Welch bound to this setting and present the
{\em erasure Welch bound} on the expected value of the Gram matrix of the reduced frame.
Interestingly, this bound generalizes to the $d$-th order moment of $F$.
We provide simple, explicit formulae for the generalized bound for $d=2,3,4$, which is the sum of the $d$-th moment of Wachter’s classical MANOVA distribution
and a vanishing term (as $n$ goes to infinity with $\frac{m}{n}$ held constant).
The bound holds with equality if (and for $d = 4$
only if) $F$ is an ETF.
Our results offer a novel perspective on the superiority of ETFs over other
frames in a variety of applications, including spread spectrum
communications, compressed sensing and analog coding.

As for the asymptotic form for the moments of subsets of ETF, we also found it for the orders of $d=5,6$, and verified that they agree with that of MANOVA. Furthermore, we developed a recursive procedure which allows to continue to higher order moments. A complete computation of all the moments will provide formal validation for some of the empirical results reported in \cite{haikin2017random}, and specifically, that the singular values of random subsets of an ETF asymptotically follow Wachter's MANOVA distribution.

The performance of analog coding \cite{haikin2016analog,ITA17} relies on yet another figure of merit of frame subsets, namely the harmonic-to-arithmetic means ratio of the singular values of the subframe covariance matrix. This quantity is equivalent to the first inverse moment.
With the understanding of the spectral properties of subsets of ETFs, we were able to compute the inverse moment of the MANOVA distribution and to give results on the asymptotic performance of analog erasure coding (both for channel and source coding).

Extension of the Erasure Welch Bounds to higher order moments and $d=-1$ would establish that an ETF is the most robust frame under inversion of subsets and the best candidate for the goal of analog coding with erasures. 
\\
\\

\noindent {\bf To summarize the main contributions of this work}
\begin{enumerate} 
	\item Extending the concept of analog coding from a channel with noise and erasures to source coding with "erasure side-information" at the encoder.
	\begin{itemize} 
	\item Further extending it by suggesting a redundant sampling, i.e. coding in a larger dimension than the inevitable amount of important samples, as well as by the use of general frame.
	\item Showing that the performance of band-limited DFT for analog coding
	can be significantly improved using irregular spectrum, and more
	generally, using ETFs.
	\end{itemize}
	\item Providing overwhelming empirical evidence for universal convergence phenomena of random subsets of deterministic frames. 
	\begin{itemize} 
		\item The Wachter's MANOVA distribution is the universal limiting spectral distribution for the typical $k$-submatrix ensemble of a variety of deterministic frames. 
		\item Presenting a simple method for approximate computation (with
		known and good approximation error) of spectral functionals of $k$-submatrix
		ensemble for a variety of random and deterministic frames.
	\end{itemize}
	\item Analyzing the performance of the analog coding scheme for both random i.i.d frames and ETFs (or more generally every frame with MANOVA spectral limiting density of subsets).
	\begin{itemize} 
		\item Derivation of the first inverse moment of the MANOVA distribution. 
		\item Comparison between the asymptotic gaps of the achievable rates using different frames from the optimal rate.
	\end{itemize}
	\item Providing analytical evidence and proofs for the spectral density of random subsets of deterministic ETFs, as well as their superiority over other frames. 
	\begin{itemize} 
		\item Extending the Welch bound to an erasure setting and proving that it is tight for ETFs and asymptotically equal to moments of the MANOVA distribution.
		\item Developing a recursive algorithm for calculating asymptotic moments of an ETF subset, and showing that the first $6$ moments coincide with those of MANOVA.
	\end{itemize}
	
\end{enumerate} 
	
\section{Thesis Outline}
The rest of this thesis is organized as follows.
Chapter \ref{chapter:AC} explores a scheme for analog coding with erasures.
Section \ref{sec:AC_ISIT} contains the work that was presented in ISIT 2016. This work presents a scheme for analog coding of a source with erasures and shows the benefit of using ETFs. Section \ref{sec:ChannelCoding} shortly projects similar ideas and analysis on the channel coding scenario.
Section \ref{sec:inverse} analytically confirms this benefit for both source an channel coding with erasures. Based on the insight, presented later, that ETFs are closely related to the MANOVA ensemble, it provides the derivation of the inverse moments of the MANOVA distribution as well as asymptotic performance analysis.
In Chapter \ref{chapter:PNAS} we bring empirical study which shows that random subsets of structured deterministic frames, and in particular ETFs, have MANOVA spectra. This novel connection was published in PNAS in June 2017. A detailed supporting information for this work is available at \cite{SI}.
Chapter \ref{chapter:Moments} deals with the moments of random subsets of frames. Section \ref{secEWB} provides an extended Welch bound for an erasure setting which is achieved with equality for ETFs (will be presented in ISIT 2018). Section \ref{sec:asyMoments} suggests a method for calculating asymptotic moments of ETFs, and not surprisingly these moments coincide with those of the MANOVA density.
We finally conclude the thesis and discuss potential future work in Chapter~\ref{chapter:Conclusions}.
In Appendix \ref{app:app2} a brief background on matrices can be found: decompositions, metrics and relation to eigenvalues. 

\section{List of Papers}
The following papers have been published by the author of this thesis during her MSc studies:
\begin{enumerate}
	\item M. Haikin and R. Zamir, ``Analog coding of a source with erasures'', In \emph{IEEE International Symposium on Information Theory (ISIT)}, 2016 IEEE, Barcelona, Spain, pp. 2074--2078, July 2016. (Reference \cite{haikin2016analog}).
	\item M. Haikin, R. Zamir, and M. Gavish, ``Random subsets of structured deterministic frames have {MANOVA} spectra'', \emph{Proceedings of the National Academy of Sciences}, pp. 201700203, June 2017. (Reference \cite{haikin2017random}).
	\item M. Haikin, R. Zamir, and M. Gavish, ``Frame moments and {W}elch bound with erasures'', Accepted to ISIT 2018. (Reference \cite{haikin2018frames}).
\end{enumerate}

	\chapter{Analog Coding with Erasures}
	\label{chapter:AC}
In this chapter we suggest a scheme for analog coding with erasures and investigate its performance based on properties of various frames.
The first section is taken from our ISIT 2016 paper \cite{haikin2016analog}. It characterizes a source coding problem with erasures, and brings conjectures and observations on the superiority of ETFs for this application. A formal justification for this superiority is brought in later sections and chapters.
The second section describes a symmetric problem of channel with noise and erasures.
The third section computes the performance of coding schemes based on ETFs. The performance depends on the inverse moment of the ETF, a derivation of which is also brought in this section.

\section{Analog Coding of a Source with Erasures} \label{sec:AC_ISIT}

Consider an i.i.d source sequence ${\textbf{x}=(x_1,..,x_n)}$ from a normal distribution $\mathcal{N}(0,\sigma^2_x)$. The encoder has information regarding the indices of $k$ important samples. Denote by $\textbf{s}=(s_1,..,s_n)$, $s_i\in\{0,1\}$, the vector of this side information. The decoder must reconstruct an ${n}$-dimensional vector ${\bf \hat{x}}$ where only the values of samples dictated by ${\bf s}$ matter, while at the non-important samples the distortion is zero:
\begin{equation} \label{eqDistortion}
D(x,\hat{x},s) = 
\begin{cases}
(\hat{x}-x)^2,& \text{if } s= 1 \text{ (important) } \\
0,              & \text{if } s=0 \text{ (not important) }.\\
\end{cases}
\end{equation}
In \cite{martinian2008source} it is shown that in this setting of "erasure" distortion, the encoder side information is sufficient, and the RDF is equal to that in the case where the side information is available to both the encoder and decoder: 
\begin{equation} \label{eqTheoreticalRD}
R(D) = \frac{p}{2}\log\bigg(\frac{\sigma^2_x}{D}\bigg).
\end{equation}
Here ${R}$ is the rate per source sample, ${D}$ is the average distortion at the important samples, and ${p=\frac{k}{n}}$ represents the probability of important samples. This rate can be achieved by a "digital" coding scheme; i.e. an "${n}$"-dimensional random code with joint-typicality encoding, at an {\it exponential} cost in complexity \cite{martinian2008source}.

In this section we explore the following low complexity "interpolate and quantize" analog coding scheme:
\begin{equation} \label{IQscheme}
\textbf{T}_{enc} \rightarrow Q \rightarrow \textbf{T}_{dec}
\end{equation}
and its achievable rate for different transforms ${\textbf{T}_{enc}}$ and ${\textbf{T}_{dec}}$. Here, ${\textbf{T}_{enc}}$ is an ${n:m}$ "dimension reduction" linear transformation that depends on ${\bf s}$, ${\textbf{T}_{dec}}$ is an ${m:n}$ "interpolation" linear transformation that is independent of ${\bf s}$, for some $m$, ${n\ge m\ge k}$, and ${Q(\cdot{})}$ denotes quantization. Typically we consider a constant ratio of important samples i.e ${k\approx n/2}$ (e.g. $S\sim {\rm Bernoulli}(p=\frac{k}{n})$ process) and are interested in the asymptotic performance (${n\rightarrow \infty}$).

One motivation for the analog coding scheme comes from the solution given in \cite{martinian2008source} to a lossless version of this problem. In this setting, the encoder uses the Reed Solomon (RS) decoding algorithm to correct the erasures and determine the ${k}$ information symbols. It then transmits these symbols to the decoder at a rate of  ${\frac{k}{n}\log(J)}$ bits per sample, where ${J}$ is the size of the source alphabet. To reconstruct the source, the decoder uses the RS encoding algorithm to get the ${n}$ reconstructed samples, that coincide with the source at the ${k}$ non-erased samples, as desired. We can view the RS decoder as a system which performs {\it interpolation} of the erased source signal.

Such an approach could be extended to a continuous source, if we first quantize the important samples to  ${J}$ levels and then apply the RS code solution above.  However, this "quantize and interpolate" solution is limited to scalar quantization. In contrast, the scheme in (\ref{IQscheme}) reverses the order of quantization and interpolation and therefore it is not limited to scalar quantization. However, the interpolation step (${\textbf{T}_{enc}}$) typically suffers from a {\it signal amplification} phenomenon. This is the main issue we deal with as it results in an increase in rate. 

Our problem formulation is dual to Wolf's paradigm of analog channel coding, in which transform techniques are exploited for coding in the presence of impulse noise \cite{wolf1983redundancy}. Wolf's scheme decouples impulse correction - by analog means - and additive white Gaussian noise (AWGN) protection - by digital means. The impulse-pattern dependent transform at the decoder introduces {\it noise amplification} for a general impulse pattern. In our case, the digital component is the quantizer, which is responsible for the lossy coding. The transform at the encoder  causes signal amplification whose severeness depends on the pattern of important samples. 

The main question which we explore is whether analog coding can asymptotically achieve the optimum information-theoretical solution (\ref{eqTheoreticalRD}). And even if not, what are the best tranforms ${\textbf{T}_{enc}}$ and ${\textbf{T}_{dec}}$ in (\ref{IQscheme}). Our preliminary results are unfortunately negative: the coding rate of the scheme in (\ref{IQscheme}) is strictly above the RDF, even for the best transforms, and even if we let the dimension $n$ go to infinity. \footnote{The result were preliminary in 2016. As of today, a much more profound study (empirical and analytical) confirms and extends these results. }

${\textbf{T}_{dec}}$ can be considered as a frame used for dimension expansion after the dimension reduction performed at the encoder.
Several works explored frames which are good for other applications. In compressed sensing, for example, most commonly the goal is to maximize the spectral norm for all sub-matrices \cite{candes2008restricted}. In \cite{rath2004frame}, \cite{vaezi2013systematic} frames for coding with erasures are introduced  but they are tolerant only to specific patterns. In \cite{holmes2004optimal} ETFs are analyzed but only for small amount of erasures.

The main contributions introduced in this section are the asymptotic point of view - concentration properties and universality of different structured frames, and how they compare to random i.i.d transforms; the redundant sampling (${m>k}$) approach; and the empirical observation that some ETFs are optimum (or at least local minimizers) in the sense of average signal amplification over all erasure patterns.
Subsection~\ref{SystemCharacterization} describes the analog coding scheme in detail. Subsection~\ref{TransformOptimization} analyzes the performance of a random i.i.d transform which turns out to be better than a low pass DFT frame, while Subsection~\ref{IrregularSpectrum} explores the superior approach of difference-set spectrum, random spectrum and general ETFs. 
\subsection{System Characterization }\label{SystemCharacterization}
\begin{figure}[htbp]
	\centering
	\hspace{-1.5em}
	\def\svgscale{3}
	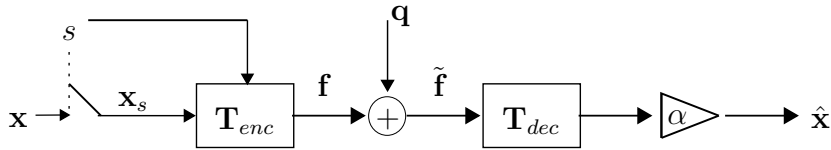
	\caption{Analog coding scheme. ${\textbf{x},\hat{\textbf{x}}}$ are ${n}$-dimensional vectors, ${\textbf{x}_s}$ is ${k}$-dimensional and ${\textbf{f},\tilde{\textbf{f}}}$ and ${\textbf{q}}$ are ${m}$-dimensional. $\textbf{T}_{enc} = \textbf{B}_s$, $\textbf{T}_{dec} = \textbf{A}$}.
	\label{modelFig}
\end{figure}

We begin with defining the system model. 
A {\it Transform Code} is characterized by a "universal" transform at the decoder and pattern dependent transform at the encoder. 
Let ${\bf A}$ be the ${n\times m}$ matrix representation of a frame with ${n}$ ${m}$-dimensional elements as rows, where ${\frac{k}{m}\triangleq\beta}$ is the redundant-sampling ratio, which varies in the range ${p\le\beta\le1}$.\footnote{This is unlike the conventions in frame theory in which the frame elements are column vectors. Note also that the $\beta$ notation from \cite{haikin2016analog} is replaced by $\frac{1}{\beta}$ for consistency with the notations which evolved during this research.} ${\bf A}$ will serve as a transformation applied at the decoder, (${\textbf{T}_{dec}}$ in (\ref{IQscheme})), independent of the pattern of important samples, as the side information is not available to the decoder. The pattern ${\bf s}$ of the important samples defines which ${k}$ rows of ${\bf A}$ contribute to meaningful values.
The corresponding rows define the ${k\times m}$ transform ${\textbf{A}_s}$.
Denote by ${\textbf{B}_s}$ the ${m\times k}$ transform applied at the encoder, (nonzero part of ${\textbf{T}_{enc}}$ in (\ref{IQscheme})), to the vector ${\textbf{x}_s}$ of important samples. ${\textbf{f}=\textbf{B}_s\textbf{x}_s}$ is the vector of transformed samples. As illustrated in Figure~\ref{modelFig}, we use a white additive-noise model for the quantization of the ${m}$ transformed samples, e.g. entropy coded dithered quantization (ECDQ), \cite{mashiach2013noise}, which is blind to the locations of the important samples. As we recall, in the no-erasure case this additive-noise model can achieve the RDF with Gaussian noise (corresponding to large dimensional quantization) and Wiener filtering \cite{berger1971rate}. 
\\
The reconstructed is given by ${{\bf \hat{x}}=\textbf{A}\cdot {\bf\tilde{f}}}$, thus the $k$
reconstructed important samples are
\begin{equation} \label{Model}
{\bf \hat{x}}_s=\alpha\textbf{A}_s{\bf\tilde{f}}=\alpha\textbf{A}_s(\textbf{f}+\textbf{q})=\alpha\textbf{A}_s\textbf{B}_s\textbf{x}_s+\alpha\textbf{A}_s\textbf{q}
\end{equation}
where ${\bf\tilde{f}}$ is the quantized version of the transformed samples and ${\bf q}$ is a white quantization noise with variance ${\sigma^2_{q}}$, independent of ${\bf {x}}_s$.
We can deal separately with the choice of ${\textbf{B}_s}$ and ${\alpha}$. The encoder applies ${\textbf{B}_s}$ such that ${\textbf{A}_s\textbf{B}_s=\textbf{I}}$ and ${\alpha}$ is a Wiener coefficient.

Let ${\|\textbf{A}_s\|^2}$ denote the squared Frobenius norm of the matrix ${\textbf{A}_s}$ normalized by the number of rows ${{\|\textbf{A}_s\|^2}=\frac{1}{k} \|\textbf{A}_s\|_F^2=\frac{1}{k}\sum_{i=1}^{k}\|{\textbf{A}_s}_i\|^2}$, where ${\|{\textbf{A}_s}_i\|}$ is the ${\it l_2}$ norm of the ${i}$'th row. 
\subsubsection{\bf Rate - Distortion Derivation}
Since the decoder is blind to the transform, the rate is given by that of a white input with the same average variance \cite{lapidoth1997role}. The rate per sample for a pattern ${\textbf{s}}$ is therefore the mutual information in an AWGN channel with a white Gaussian input: 
\begin{equation} \label{RateAllRes1}
R=\frac{m}{n}\frac{1}{2}\log\bigg(1+\frac{\frac{1}{m}E\|\textbf{f}\|^2}{\sigma^2_{q}}\bigg)
\end{equation}
\begin{equation} \label{RateAllRes2}
=\frac{m}{n}\frac{1}{2}\log\bigg(1+\frac{\sigma^2_x}{\sigma^2_{q}}\|\textbf{B}_s\|^2\bigg)
\end{equation}
where to obtain (\ref{RateAllRes2}) we substitute the average variance of the transformed samples for a white source $x$:
\begin{equation} \label{VarFredund}
\frac{1}{m}E\|\textbf{f}\|^2=\frac{1}{m}\sum_{i=1}^{m}\sigma^2_{f_i}=\frac{1}{m}\sum_{i=1}^{m}\|{{\textbf{B}}_s}_i\|^2\sigma^2_x=\|\textbf{B}_s\|^2\sigma^2_x.
\end{equation}
For a given ${\textbf{A}_s}$, the matrix ${\textbf{B}_s}$ that minimizes the expected ${\it l_2}$ norm of ${\bf f}$ in (\ref{RateAllRes1}) is the pseudo-inverse  ${\textbf{B}_s=\textbf{A}_s'(\textbf{A}_s\textbf{A}_s')^{-1}}$, hence\footnote{${(\ )'}$ denotes the conjugate transpose.} 
\begin{equation} \label{PointScore}
\|\textbf{B}_s\|^2=\frac{1}{m}\|\textbf{B}_s\|_F^2=\frac{1}{m}\tr(\textbf{B}_s'\textbf{B}_s)=\frac{1}{m}\tr(\big((\textbf{A}_s\textbf{A}_s')^{-1}\big)'\cancel{\textbf{A}_s\textbf{A}_s'(\textbf{A}_s\textbf{A}_s')^{-1}})=\frac{1}{m}\tr((\textbf{A}_s\textbf{A}_s')^{-1}).
\end{equation}

We shall later see that the heart of the problem is the signal amplification caused by the factor ${\|\textbf{B}_s\|^2}$ in (\ref{RateAllRes2}).
The case of ${m>k}$ is referred to as "redundant sampling", where more samples are quantized than the important ones. The motivation for this is the existence of more robust transforms in the sense of signal amplification even at the cost of some extra transmissions.

For convenience, we normalize the transform ${\bf A}$ to have unit-norm rows,
${\|{\textbf{A}}_i\|=1}$, for ${i=1,...,n}$, so that each sample of the additive quantization noise term in (\ref{Model}) has variance ${{\|\textbf{A}_s}_i\|^2\sigma^2_{q}=\sigma^2_{q}}$. 
The variance of each sample of ${\textbf{A}_s\textbf{B}_s\textbf{x}_s}$ is ${\sigma^2_x}$. As a result of the Wiener estimation the distortion is:
\begin{equation} \label{D}
D\triangleq E\bigg\{\frac{1}{k}\sum_{i=1}^{n}D(x_i,\hat{x}_i,s_i) \bigg\}=\frac{1}{k}\Ev\|\bf \hat{x}_s-\textbf{x}_s\|^2=\frac{\sigma^2_x\sigma^2_{q}}{\sigma^2_x+\sigma^2_{q}}.
\end{equation}
Combining (\ref{RateAllRes2}),(\ref{PointScore}) and (\ref{D}) we can relate the rate and distortion of the scheme for a specific pattern ${\bf s}$:
\begin{equation} \label{Rate7}
R=\frac{m}{n}\frac{1}{2}\log\bigg(1+\frac{\frac{1}{m}\tr((\textbf{A}_s\textbf{A}_s')^{-1})(\sigma^2_x-D)}{D}\bigg)
\end{equation}

We define the excess rate of the scheme as ${\delta \triangleq R-R(D)}$:
\begin{equation} \label{ExcessRate2}
\delta(\beta,\gamma,\eta_s)=\frac{k}{n}\frac{1}{2}\big[\frac{1}{\beta}\log(\eta_s\gamma+(1-\eta_s))-\log(\gamma)\big]
\end{equation}
\begin{equation} \label{ExcessRate3}
\simeq\frac{1}{\beta} \frac{p}{2}\log(\eta_s)+(\frac{1}{\beta}-1)\frac{p}{2}\log(\gamma)
\end{equation}
where ${\gamma=\frac{\sigma^2_x}{D}}$ is the signal-to-distortion ratio (SDR), 
\begin{equation} \label{IE}
\eta_s=\frac{1}{m}\tr((\textbf{A}_s\textbf{A}_s')^{-1})
\end{equation}
is the inverse energy (IE) of a pattern ${\textbf{s}}$, which is related to harmonic mean of the eigenvalues of ${\textbf{A}_s\textbf{A}_s'}$, and ${\simeq}$ is true for high resolution (${\gamma\gg1}$).
We also define ${\rho}$ as the mean logarithmic inverse energy (MLIE) of the frame ${\bf A}$:
\begin{equation} \label{MLIE}
\rho=\frac{1}{{n \choose k}}\sum_{s}\frac{m}{n}\frac{1}{2}\log(\eta_s)
\end{equation}
i.e. the average (over all possible patterns of ${k}$ important samples) excess rate above the RDF caused by signal amplification.
As we will see, for "good" transforms this average becomes asymptotically the typical case.

In the high resolution case, the excess rate (\ref{ExcessRate3}) is composed of two terms, one as the result of signal amplification and the second as a result of ${m-k}$ extra samples transmission. 
Note that for fixed (${n,k,\beta}$) minimizing the excess rate ${\delta}$ over the transform $\bf A$ is equivalent to minimizing its MLIE ${\rho}$.

\subsubsection{\bf Side Information Transmission}\label{SI}
It is most natural to compare the proposed system with the alternative  naive approach of transmitting the side information regarding the locations of the important samples. Pattern transmission requires ${\frac{1}{n}\log({n \choose k})}$ bits per input sample, which is ${H_b(p)}$ bits in the limit ${n\to \infty}$.

\subsection{Transform Optimization} 
\label{TransformOptimization}
\subsubsection{\bf (\Romannum{1}) Band Limited Interpolation }
\label{BLinterpolation}
\begin{figure}[ht]
	\begin{center}
		\includegraphics[scale=.49]{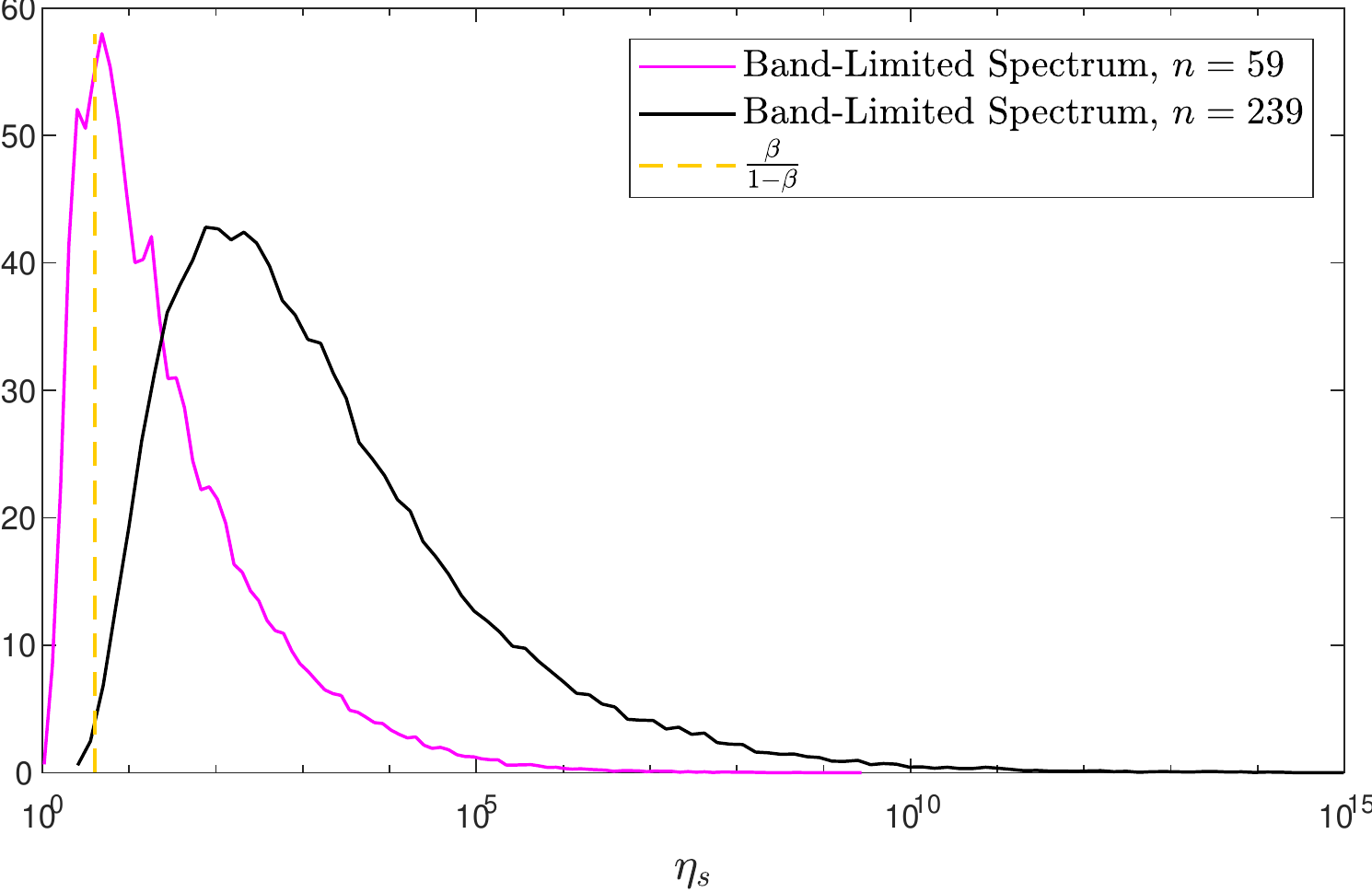}
		\caption{Logarithmic histogram of ${\eta_s}$, ${m=\lfloor\frac{n}{2}\rfloor}$, ${\beta=0.8}$.}
		\label{figBL_IE}
	\end{center}
\end{figure}

The most basic frame includes ${m}$ consecutive rows of the IDFT. Without loss of generality, the transform matrix ${\bf A}$ consists of the first ${m}$ columns of an IDFT matrix, meaning that the reconstructed samples are part of a band limited (lowpass) temporal signal with the quantized DFT coefficients as its spectrum (${m}$ lower frequencies). Such a transform ${\bf A}$ forms a "full spark frame" - every subset of ${m}$ rows of ${\bf A}$ is independent, i.e ${\textbf{A}_s\textbf{A}_s'}$ is full rank and invertible for every pattern ${\bf s}$ \cite{alexeev2012full}. However, it is not robust enough to different patterns. The intuitive reason for that is that, although the source samples are i.i.d, the low-pass model forces slow changes between close samples. Thus, it is good for a uniform sampling pattern, but for most other patterns it suffers from signal amplification that causes a severe loss in rate-distortion performance \cite{mashiach2013noise}. Asymptotically almost every sub-matrix ${\textbf{A}_s}$ is ill-conditioned and the IE ${\eta_s}$ (\ref{IE}) is unbounded even for redundant sampling (${\beta<1}$).
Figure~\ref{figBL_IE} shows the IE distribution for band-limited interpolation
and a random pattern ${\bf s}$. The dashed line at ${\frac{\beta}{1-\beta}}$ is the asymptotic theoretical value achieved by a random i.i.d. transform;
see further in this subsection. 
We can see that the IE diverges as ${n}$ grows.
\subsubsection{\bf Signal Amplification} \label{Signal Amplification}  
The following "Inversion-Amplification Lemma" describes the condition for an optimal transform.
\begin{lemma} \label{lemma1Amp}
	{\em The IE ${\eta_s}$ in (\ref{IE}) of any ${k\times m}$ matrix ${\textbf{A}_s}$, s.t. ${\|{\textbf{A}_s}_i\|=1}$, is lower bounded as ${\eta_s\ge \frac{k}{m}}$, with equality iff ${\textbf{A}_s\textbf{A}'_s= \textbf{I}}$.}
\end{lemma}

\begin{proof}
	Denote by ${\{\lambda_i\}_{i=1}^{k}}$ the eigenvalues of ${\textbf{A}_s\textbf{A}'_s}$.
	\begin{align}
	&1=\frac{1}{k}\tr(\textbf{A}_s\textbf{A}'_s)=\frac{1}{k}\sum_{i=1}^{k}\lambda_{i}\ge \frac{1}{\frac{1}{k}\sum_{i=1}^{k}\frac{1}{\lambda_{i}}}=\frac{1}{\frac{1}{k}\tr((\textbf{A}_s\textbf{A}'_s)^{-1})}
	\nonumber \\
	& \Rightarrow \;
	\frac{1}{m}\tr((\textbf{A}_s\textbf{A}'_s)^{-1})=\beta\frac{1}{k}\tr((\textbf{A}_s\textbf{A}'_s)^{-1})\ge\beta.	
	\nonumber
	\end{align}
	where the inequality follows from the arithmetic-harmonic mean inequality, with equality iff all the eigenvalues are equal, i.e ${\textbf{A}_s\textbf{A}'_s= \textbf{I}}$.
\end{proof}

For ${\beta=1}$, Lemma~\ref{lemma1Amp} becomes ${\|\textbf{A}^{-1}_s\|\ge1}$, with equality iff ${\textbf{A}_s}$ is unitary. Thus for a non-unitary transform the signal is amplified by the factor ${\|\textbf{A}^{-1}_s\|}$.

\subsubsection{\bf (\Romannum{2}) Random i.i.d Transforms } \label{RandomTransform}   
For "digital" coding, we know that random i.i.d codes are optimal. Thus, a natural approach is to investigate the asymptotic performance of a random transform. Consider a matrix ${\textbf A}$ whose entries are i.i.d Gaussian random variables with variance ${\frac{1}{m}}$. For any ${k\times m}$ sub-matrix ${{\textbf{A}_s}}$, ${\lim_{k\to \infty}\|{\textbf{A}_s}_i\|=1}$ almost surely.

\noindent{\bf Amplification Analysis}
We bring here two results which show that for ${m=k}$ ($\beta=1$), random i.i.d transform is definitely bad in the sense of amplification. 
The proof of both of these results is based on characterization of the minimum eigenvalue of a random matrix.

\begin{lemma} \label{RandomAmp1Lemma}
With a square complex random matrix a 'non-amplifying' transformation cannot be achieved and w.p.1 the amplification diverges for large dimensions:
\begin{equation} \label{RandomAmp1}
\lim_{k\to \infty}P[\frac{1}{k}\tr((\textbf{A}_s\textbf{A}_s')^{-1})\ge \zeta]=1,\ \ \ \forall \zeta\ge0
\end{equation}
\end{lemma}
\begin{proof}
	From \cite{tulino2004random} we know that if $\bf H$ is a standard complex $k\times k$ Gaussian matrix then its minimum singular value $\sigma_{min}(\textbf{H})$ satisfies 
	\begin{equation}\label{sigma_min}
	\lim\limits_{k\to \infty}P[k\sigma_{min}(\textbf{H})>x]=e^{-x-\frac{x^2}{2}.} 
	\end{equation}
	According to the normalization convention of $\bf A$, we can assume $\textbf{A}_s=\frac{1}{\sqrt{k}}\textbf{H}$, and $\sigma_{min}^2(\textbf{A}_s)=\frac{1}{k}\sigma_{min}^2(\textbf{H})$. The trace of interest can be bounded:
	\begin{equation}
\frac{1}{k}\tr((\textbf{A}_s\textbf{A}_s')^{-1})\ge \frac{1}{k}\sigma_{max}\big((\textbf{A}_s\textbf{A}_s')^{-1}\big)=\frac{1}{k}\frac{1}{\sigma_{min}^2(\textbf{A}_s)}=\frac{1}{\sigma_{min}^2(\textbf{H})}.
	\end{equation}
	Thus,
	\begin{equation}
	P\big[\frac{1}{k}\tr((\textbf{A}_s\textbf{A}_s')^{-1})\ge \zeta\big]\ge P\big[\frac{1}{\sigma_{min}^2(\textbf{H})}\ge \zeta\big]=P\big[k\sigma_{min}(\textbf{H})\le \frac{k}{\sqrt{\zeta}}\big]=1-P\big[k\sigma_{min}(\textbf{H})\ge \frac{k}{\sqrt{\zeta}}\big]
	\nonumber
	\end{equation}
	\begin{equation}
	\Rightarrow \lim\limits_{k\to \infty}P\big[\frac{1}{k}\tr((\textbf{A}_s\textbf{A}_s')^{-1})\ge \zeta\big]\ge 1- \lim\limits_{k\to \infty}e^{-\frac{k}{\sqrt{\zeta}}-\frac{k^2}{2\zeta}}=1
	\end{equation}
	
\end{proof}
\begin{lemma} \label{RandomAmp2Lemma}
For  ${k \to \infty}$ the divergence rate can be bounded as follows:
\begin{equation} \label{RandomAmpBound}
\frac{k^2}{2\pi e} \le E\bigg[\frac{1}{k}\tr((\textbf{A}_s\textbf{A}_s')^{-1})\bigg]\le \frac{k^3}{2\pi e}
\end{equation}
\end{lemma}
\begin{proof}
	Based on \eqref{sigma_min} we see that for large enough $k$ the distribution of $k\sigma_{min}(\textbf{H})$ is independent of $k$, thus $\Ev\big[k\sigma_{min}(\textbf{H})\big]=c$, $c=$constant.
	\begin{equation}\label{sigma_min2}
	\Ev \bigg[\sigma_{max}\big((\textbf{A}_s\textbf{A}_s')^{-1} \big)\bigg] \le \Ev \bigg[\tr((\textbf{A}_s\textbf{A}')^{-1}_s)\bigg]\le k\Ev \bigg[\sigma_{max}\big((\textbf{A}_s\textbf{A}_s')^{-1} \big)\bigg]
	\end{equation}
	\begin{equation}\label{sigma_min3}
	\Ev \bigg[\sigma_{max}\big((\textbf{A}_s\textbf{A}_s')^{-1} \big)\bigg] = \frac{k}{\Ev \big[\sigma_{min}^2(\textbf{H})\big]}=\frac{k^3}{c^2}.
	\end{equation}
	From \eqref{sigma_min2}, \eqref{sigma_min3} we have
	\begin{equation}\label{trace_bounds}
	\frac{k^2}{c^2} \le \Ev \bigg[\frac{1}{k}\tr((\textbf{A}_s\textbf{A}')^{-1}_s)\bigg]\le \frac{k^3}{c^2}.
	\end{equation}
	We can evaluate the constant $c$ using \eqref{sigma_min}:
	\begin{align}\label{constant_c}
	\Ev\big[k\sigma_{min}(\textbf{H})\big]&=\int_{-\infty}^{\infty}x\frac{\partial}{\partial x}\bigg(1-e^{-x-\frac{x^2}{2}}\bigg) dx=\int_{-\infty}^{\infty}x(1+2x)e^{-x-\frac{x^2}{2}} dx\\&=-xe^{-x-\frac{x^2}{2}}\bigg\rvert^{\infty}_{-\infty}+\int_{-\infty}^{\infty}e^{-x-\frac{x^2}{2}} dx=\sqrt{2\pi e}\int_{-\infty}^{\infty}\frac{1}{\sqrt{2\pi}}e^{-\frac{(x+1)^2}{2}} dx\\&=\sqrt{2\pi e}.
	\end{align}
	Substituting \eqref{constant_c} into \eqref{trace_bounds} we get \eqref{RandomAmpBound}.
\end{proof}

For ${m>k}$ the amplification is finite. As random matrix theory shows, \cite{tulino2004random}, if ${\textbf{H}}$ is an ${k\times m}$ random matrix with i.i.d entries of variance ${\frac{1}{k}}$ and ${\frac{k}{m}\to \beta, \beta<1}$, then 

\begin{equation} \label{RandomMatrix2}
\lim_{k\to \infty}\frac{1}{k}\tr((\textbf{H}\textbf{H}')^{-1})=\frac{\beta}{1-\beta} \; \; a.s.
\end{equation}
A ${k\times m}$ sub-matrix ${\textbf{A}_s}$ has element variance of ${\frac{1}{m}}$. Denote ${\textbf{H}=\sqrt{\frac{m}{k}}\textbf{A}_s}$, which has element variance of ${\frac{1}{k}}$: 
\begin{displaymath}
\frac{1}{m}\tr((\textbf{A}_s\textbf{A}_s')^{-1})=\frac{1}{k}\tr((\textbf{H}\textbf{H}')^{-1})
\end{displaymath}
\begin{equation} \label{RandomAmp3}
\Rightarrow \lim_{k\to \infty}\eta_s=\frac{\beta}{1-\beta}. \,\,\,\,\,\ (\text{where} \,\,\ \frac{k}{m}\to \beta)
\end{equation}
Note that \eqref{RandomAmp3} cannot be used with {$\beta=1$}. Lemmas \ref{RandomAmp1Lemma} and \ref{RandomAmp2Lemma} deal with the asymptotic behavior of the square case.

\noindent{\bf Comparison to the SI Transmission Benchmark (in Subsection~\ref{SI})}
Substituting (\ref{RandomAmp3}) as the IE in (\ref{ExcessRate2}) we get the following expression for the excess rate using a random transform: 
\begin{equation} \label{RateLoss}
\lim\limits_{n\to\infty}\delta=\frac{k}{n}\frac{1}{2}\bigg[\frac{1}{\beta}\log\bigg(\frac{\beta}{1-\beta}\gamma+1-\frac{\beta}{1-\beta}\bigg)-\log(\gamma)\bigg].
\end{equation}
For some scenarios this outperforms the naive side information transmission. 

\begin{figure}[ht]
	\begin{center}
		\includegraphics[scale=.49]{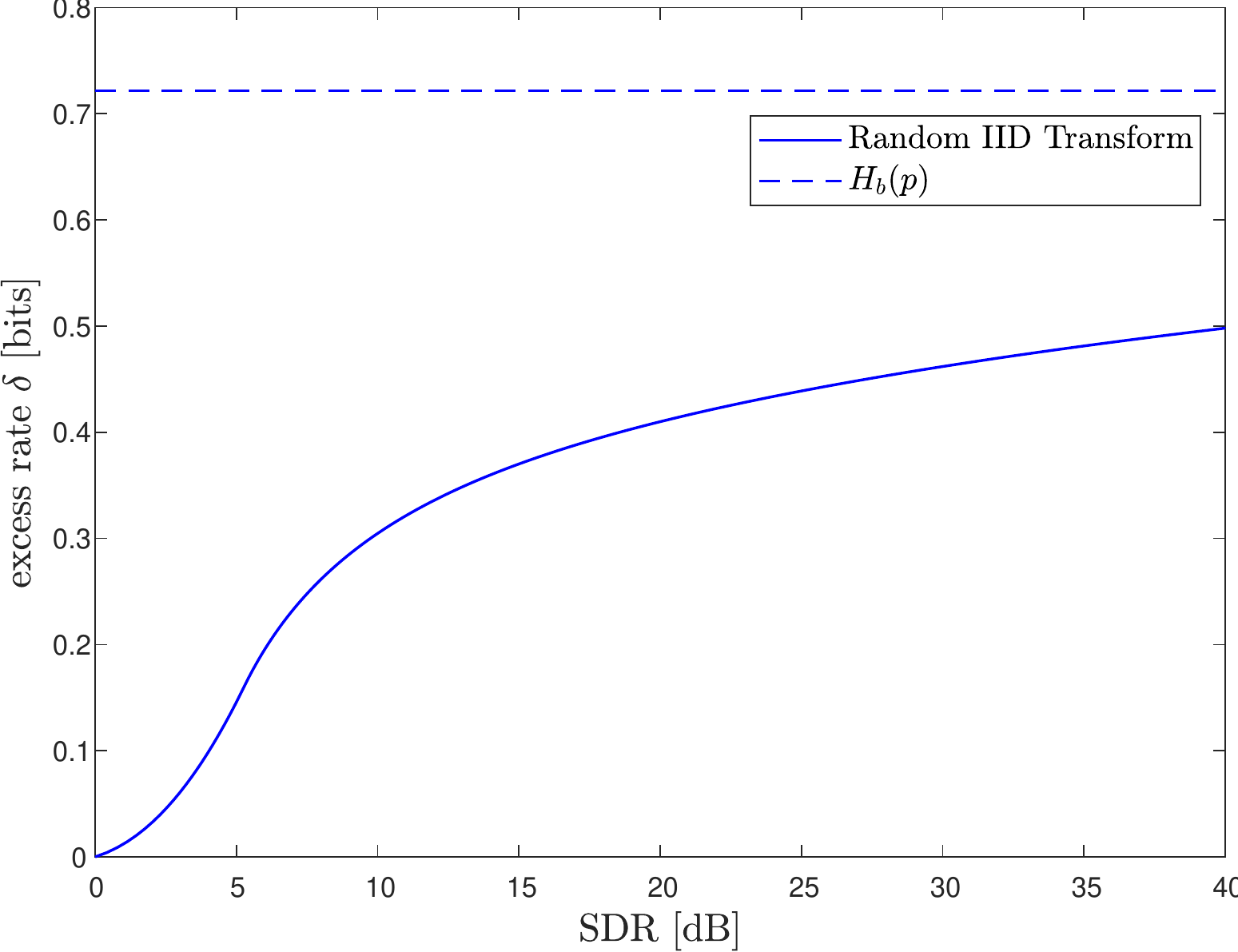}
		\caption{Rate loss for ${\frac{k}{n}=\frac{1}{5}}$.}
		\label{figRateLossVsSDR}
	\end{center}
\end{figure}

Figure~\ref{figRateLossVsSDR} shows the asymptotic excess rate above (\ref{eqTheoreticalRD}) for random transform with optimal ${\beta}$ for each SDR compared to the cost of SI transmission.
In the limit of high SDR the expression for the excess rate (for the best choice of ${\beta}$) takes the following form:
\begin{equation} \label{RateLossHighSDR}
\delta=\frac{k}{n}\frac{1}{2}\log(\ln(\gamma)),
\end{equation}
which goes (very slowly) to ${\infty}$. Nevertheless, for reasonably high SDR there is an advantage to the random matrix approach relative to the benchmark.
Analysis of ${\beta}$ which minimizes the rate loss in (\ref{RateLoss}) and the proof of (\ref{RateLossHighSDR}) appears in Subsection \ref{CR}. 

\begin{figure}[ht]
	\begin{center}
		\includegraphics[scale=.6]{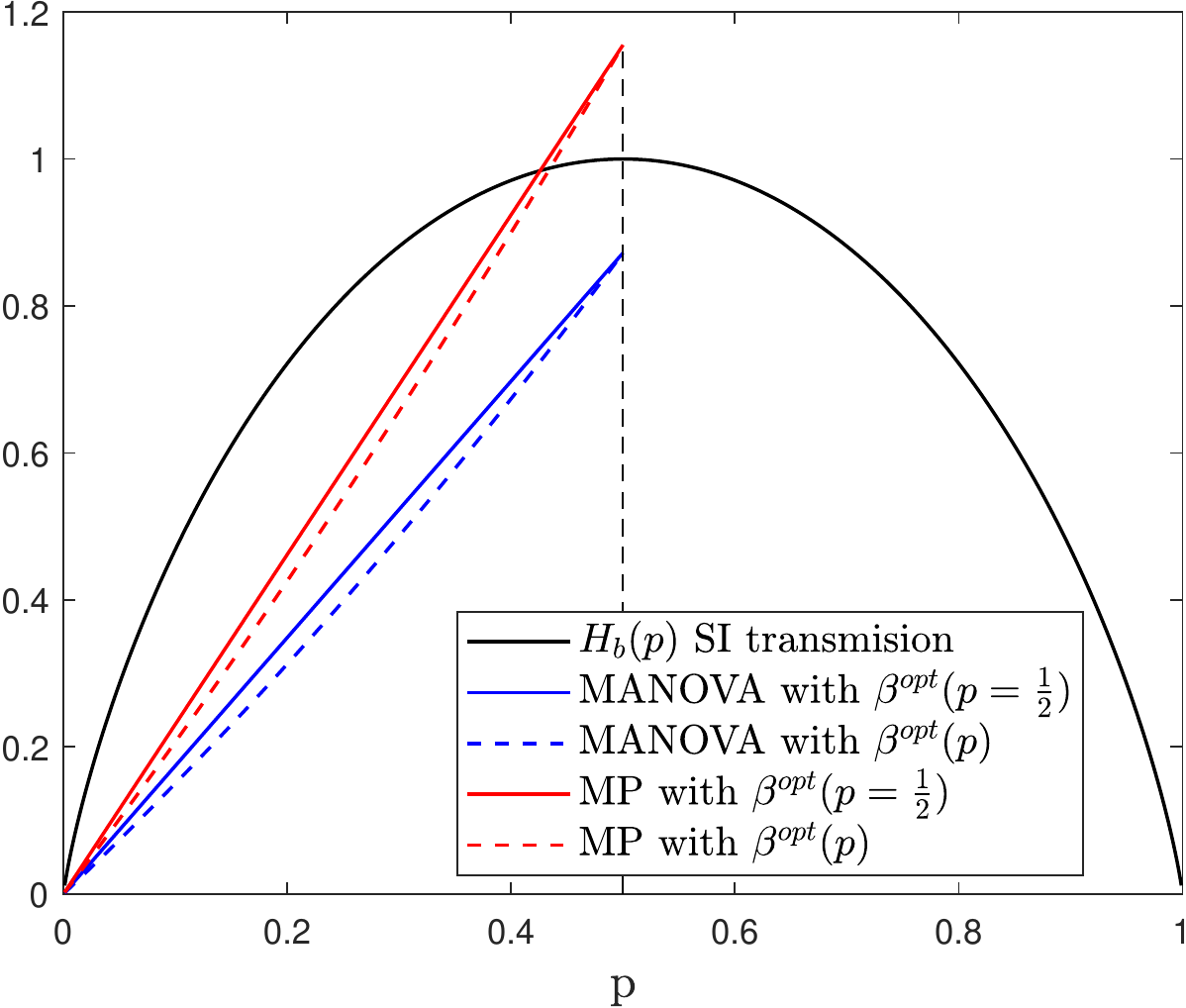}
		\caption{Redundancy above Shannon, SDR=30dB.}
		\label{RedundancyAboveShannon}
	\end{center}
\end{figure}

Figure \ref{RedundancyAboveShannon} shows that below some value of $p$ even the random i.i.d performs better than the benchmark of SI transmission. The solid red line is the excess rate using a random transform and the optimal $\beta$ for $p=\frac{1}{2}$. Using an optimal $\beta$ for each $p$ must be better and the dashed line schematically demonstrates it. The blue line corresponds to another transform which is better already for $p=\frac{1}{2}$ and will be discussed in the next subsection.
\subsubsection{\bf (\Romannum{3}) Irregular Spectrum } 
Special case to which we devote the next subsection.
\subsection{Irregular Spectrum } \label{IrregularSpectrum}     
As we saw in Subsection~\ref{TransformOptimization}, a band-limited DFT frame suffers from high signal amplification for non-uniform erasure patterns. In contrast, the signal amplification of a random frame is invariant to the erasure pattern. We can increase the robustness of a DFT-based frame to the erasure pattern by selecting an irregular "symmetry breaking" set of frequencies.
Thus, the encoder performs interpolation to a signal with irregular spectrum. 

Every set of $m$ frequencies (columns) from the $n\times n$ IDFT matrix, generates an $n\times m$  frame ${\bf A}$.
For a general spectral patten the corresponding frame is not necessarily full spark (for a general ${n}$). But for prime ${n}$, Chebotarev's theorem guarantees that for every spectral choice and every pattern of important samples, the matrix ${\textbf{A}_s}$ is full rank \cite{stevenhagen1996chebotarev}. We thus restrict the discussion to prime ${n}$'s when exploring the DFT transform.  

\subsubsection{\bf Difference-Set Spectrum} \label{DSS}  
For small dimensions it is possible to exhaustively check all spectrum choices and look for the one with the best worst case or average amplification (logarithmic IE). It turns out that the best spectrum consists of frequencies from a {\it difference set} (DS), forming the so called difference-set spectrum (DSS). 

{\it Definition:} an ${m}$ subset of ${\mathbb{Z}_n}$ is a ${(n,m,\lambda)}$ difference set if the distances (modulo ${n}$) of all distinct pairs of elements take all possible values ${1,2,..,n-1}$ exactly ${\lambda}$ times. The three parameters must satisfy the relation
\begin{equation}\label{DSrelation}
\lambda (n-1)=m(m-1). 
\end{equation}
Difference sets are known to exist for some pairs of ${(n,m)}$. In the numerical examples in this chapter we shall use the case of prime ${n}$ and ${m\approx \frac{n}{2}}$. Specifically, we consider a {\it Quadratic Difference Set} \cite{xia2005achieving} with the following parameters:
\begin{equation}\label{QuadraticDS}
\bigg(n, m=\frac{n-1}{2}, \lambda=\frac{n-3}{4}\bigg), \;\;\;\; n-prime.
\end{equation}
This DS can be constructed by a cyclic sub group ${\langle g\rangle}$, for some element ${g}$ from the multiplicative group of ${\mathbb{Z}_n}$. 

An ${n\times m}$ DSS transform ${\textbf{A}}$ is constructed from the ${m}$ columns of an ${n\times n}$ IDFT matrix, that correspond to indices from a difference set. The normalization of the IDFT is such that ${\|\textbf{A}_i\|=1}$, i.e the absolute value of the elements is ${\frac{1}{\sqrt{m}}}$. 

\subsubsection{\bf Random Spectrum and the MANOVA Distribution} \label{RandomSpectrum}  

Interestingly, asymptotically in this setup, a {\it random spectrum} achieves similar performance as the DSS. Recall that the IE \eqref{IE} is determined by the eigenvalue distribution (Lemma~\ref{lemma1Amp}). Farrell showed in \cite{farrell2011limiting} that for a random spectrum, the limiting empirical eigenvalue distribution of ${\textbf{A}_s\textbf{A}_s'}$, for a randomly chosen ${\textbf s}$, converges almost surely to the limiting density of the Jacobi ensemble. The Jacobi ensemble corresponds to the eigenvalue distribution of MANOVA matrices - random matrices from multi-variate distribution \cite{erdHos2013local}. With our notations the MANOVA distribution is equal to:
\begin{equation}\label{Jacobi}
f^{M}(x)=\frac{\sqrt{(x-r_-)(r_+-x)}}{2\pi \beta x(1-\frac{m}{n}x)}\cdot I_{(r_-,r_+)}(x),
\end{equation}
\begin{equation}\label{JacobiExtrimalValues}
r_\pm=\bigg(\sqrt{(1-\frac{m}{n})\beta}\pm\sqrt{1-\frac{m}{n}\beta}\bigg)^2.
\end{equation}

For the examples with ${\frac{m}{n}\rightarrow\frac{1}{2}}$, thus:
\begin{equation}\label{ManovaScore}
\lim_{n\to \infty}\eta_s=\beta E_{f^M}(X^{-1})=\beta\int\frac{1}{x}f^M(x)dx
\end{equation}
\begin{equation}\label{ManovaScore2}
=\int_{1-c}^{1+c}\frac{\sqrt{c^2-(x-1)^2}}{\pi x^2(2-x)}dx, \;\;\; c=\sqrt{\beta\bigg(2-\beta\bigg)}.
\end{equation}

Figure~\ref{figIE_rate} (on the left) shows the histogram of ${\eta_s}$ over randomly chosen patterns for a large dimension in the case of a random i.i.d transform and a DSS transform, as well as random spectrum transform.   
We see that the IE of a random i.i.d transform concentrates on ${\frac{\beta}{1-\beta}}$ (\ref{RandomAmp3}). For DSS/random spectra, almost all patterns (sub-matrices) are equivalent, and the IE concentrates on a lower value which fits the MANOVA density based calculation (\ref{ManovaScore2}). The ideal lower bound (of Lemma~\ref{lemma1Amp}) is also presented.
The advantage of these structured transforms lies in their better eigenvalue distribution.

\begin{figure}[ht]
	\begin{center}
		\includegraphics[scale=.49]{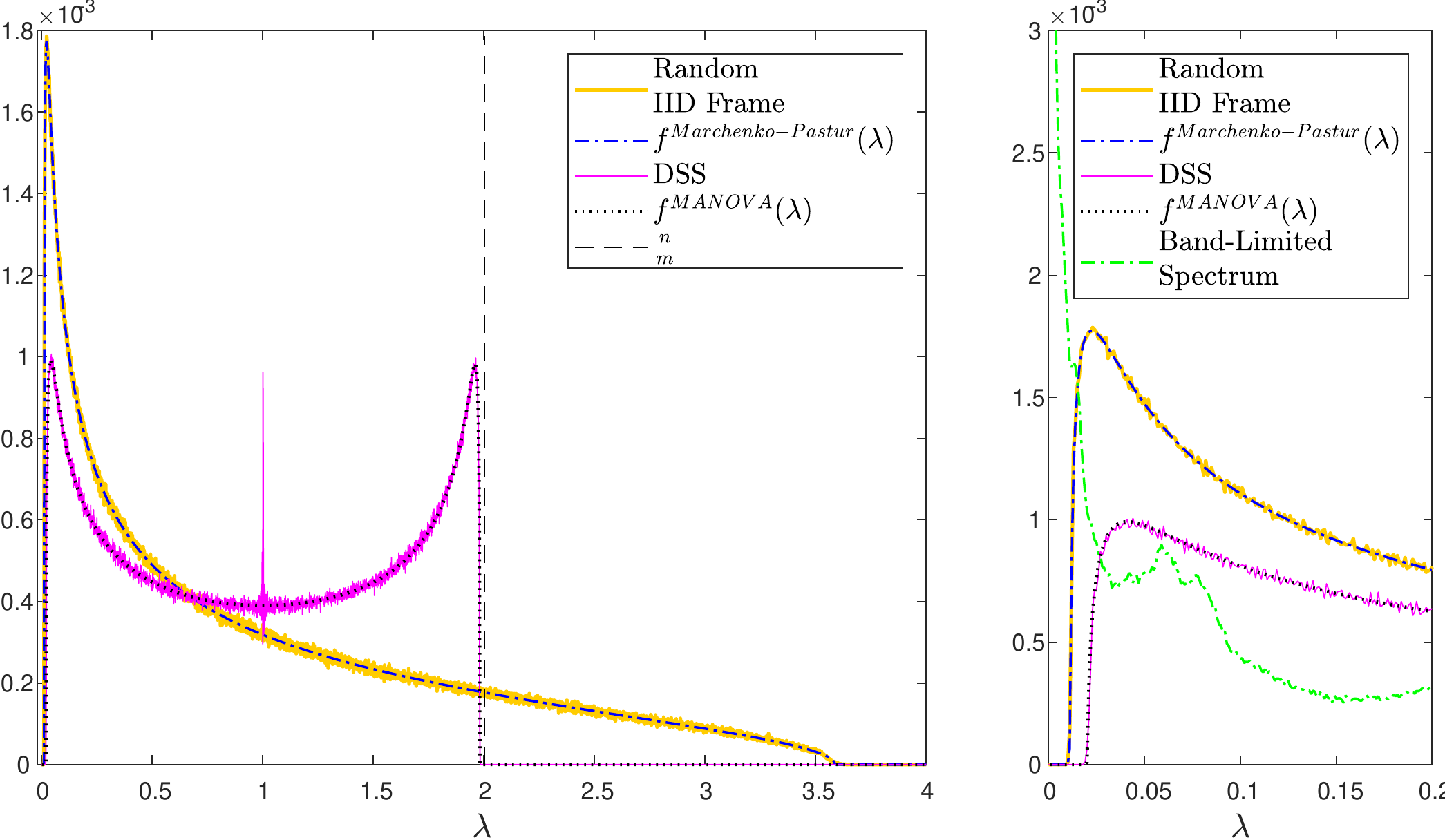}
		\caption{Eeigenvalue distribution of ${\textbf{A}_s\textbf{A}_s'}$, ${n = 947, \beta = 0.8}$.
			The graph on the right hand side zooms in into the behavior near zero.}
		\label{figEigenValues}
	\end{center}
\end{figure}

Figure~\ref{figEigenValues} shows the empirical distribution of the eigenvalues of ${\textbf{A}_s\textbf{A}_s'}$ for different transforms. Figure~\ref{figEigenValues} shows also the theoretical limiting eigenvalue density of an i.i.d random matrix (Mar\u cenko-Pastur) \cite{tulino2004random} and the MANOVA distribution (\ref{Jacobi}).
Observe the concentration property of the eigenvalue empirical distribution and of the IE ${\eta_s}$ (\ref{IE}); namely with high probability these random functionals are close to their mean value. It is evident that DSS also fits the asymptotic MANOVA density of a random spectrum.
Observe also that the minimal eigenvalue of a random i.i.d transform is closer to zero than that of DSS and thus its contribution to the IE amplification is more significant (see the zoom in graph on the right). As ${\beta}$ increases, the extremal eigenvalues move towards the edges (0 and ${\frac{n}{m}}$), and the minimal eigenvalue becomes the most dominant for the IE. For ${\beta =1}$, the support of the density function approaches zero, and as a result the IE diverges
and there is no concentration to a finite value. Note that in band limited spectrum this is the case even for ${\beta < 1}$. 

\subsubsection{\bf Equiangular Tight Frames} \label{ETFs}  

It turns out that a DSS spectrum is a special case of an {\it equiangular tight frame} (ETF) or a {\it maximum-Welch-boud-equality codebook} (MWBE) {\cite{xia2005achieving}}. Moreover, we observe that asymptotically many different ETFs are similar in terms of their ${\eta_s}$ distribution. 

An ${n\times m}$ ETF transform ${\textbf{A}}$ is defined as a uniform tight frame (i.e., it satisfies ${\textbf{A}'\textbf{A}=\frac{n}{m}\textbf{I}_m}$) such that the absolute value of the correlation between two row vectors is constant for all pairs and equal to the Welch bound:
\begin{equation}\label{WBwc3}
|\textbf{a}_l\textbf{a}'_{l'}| = \sqrt{\frac{n-m}{(n-1)m}}=cos(\theta),\;\;\;\;\forall l\not= l'.
\end{equation}
The matrix ${\textbf{A}\textbf{A}'}$ is Hermitian positive semi-definite whose diagonal elements are 1 and whose off-diagonal elements have equal absolute value ${cos(\theta)}$ as in (\ref{WBwc3}). It has ${m}$ eigenvalues equal to ${\frac{n}{m}}$ (same as in ${\textbf{A}'\textbf{A}}$) and the rest ${n-m}$ eigenvalues are zero.
For any ${k\le m}$ rows of ${\textbf{A}}$ (induced by the important samples pattern ${s}$) ${\textbf{A}_s\textbf{A}_s'}$ is positive semi-definite. The absolute value of the off-diagonal elements of the ${k\times k}$ matrix ${\textbf{A}_s\textbf{A}_s'}$ is also ${cos(\theta)}$ but its eigenvalue spectrum is induced by the subset of the element's phases. The distribution of this spectrum is the main issue of interest when exploring the IE ${\eta_s}$ (Figure~\ref{figEigenValues}).

\begin{figure}[!htp]
	\begin{center}
		\includegraphics[scale=.49]{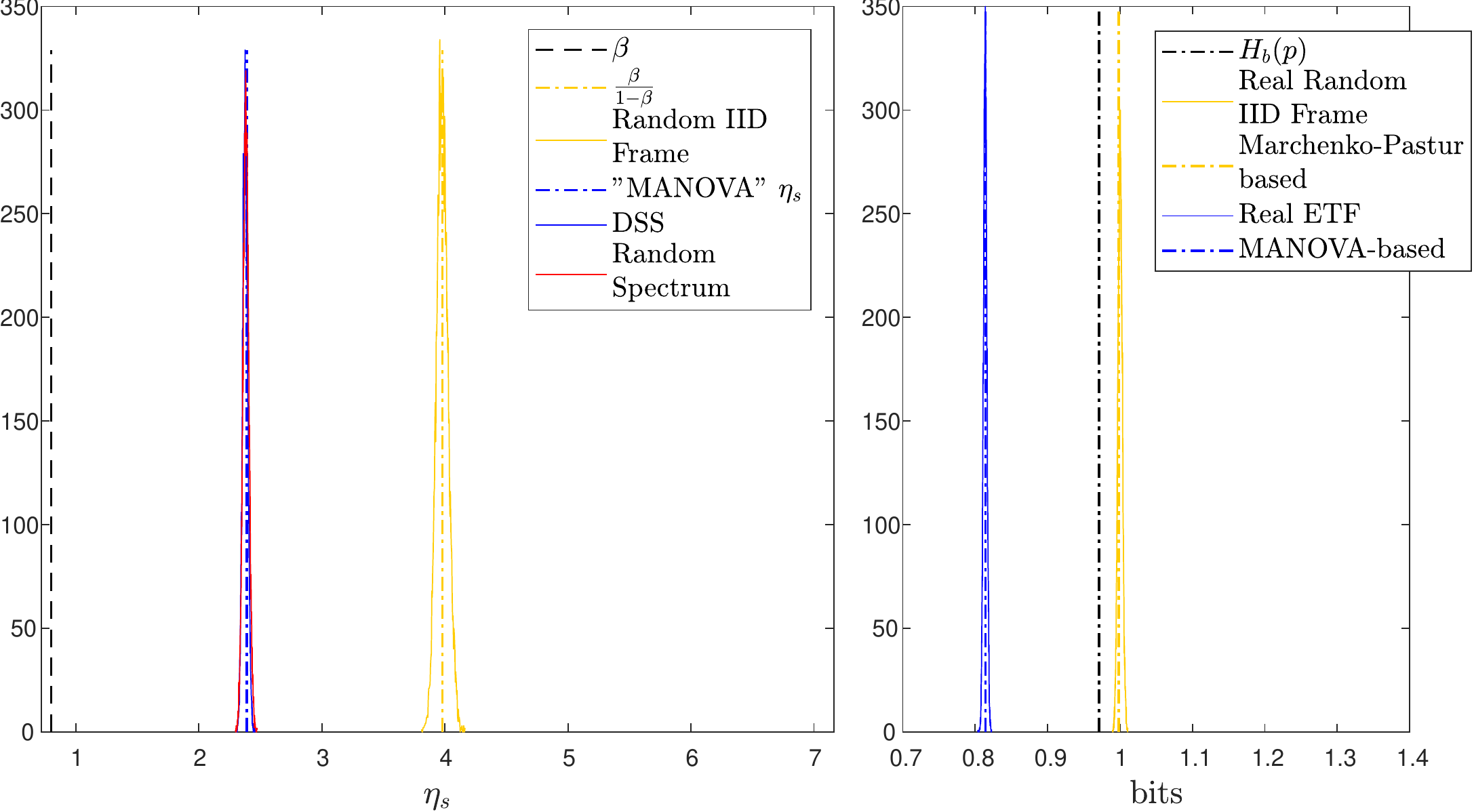}
		\caption{Left: Histogram of the inverse energy ${\eta_s}$ for ${n = 947, \beta = 0.8}$. 
			Right: Histogram of ${\delta}$, ${n = 1902, m = \frac{n}{2}, \beta = 0.8, \text{SDR = 30dB}}$.}
		\label{figIE_rate}
	\end{center}
\end{figure}

While DFT-based transforms assume a complex source, ETFs allow us to consider also real valued sources, which is our original motivation. As stated before, other types of ETFs mentioned above achieve similar IE distribution.  
Figure~\ref{figIE_rate} (on the right) shows the histogram of ${\delta}$, as defined in (\ref{ExcessRate3}), for a real random i.i.d transform as well as Paley's real ETF \cite{paley1933orthogonal}, in high SDR. 
We can see that for this setup the rate of a random i.i.d transform exceeds that of SI transmission $H_b(p)$, but a scheme based on a real-valued ETF achieves a lower rate.

Finally, for given ${\frac{k}{n}}$ and ${\frac{m}{n}}$ values, we observe that many different frames, not just ETFs, are asymptotically equivalent and share a similar MANOVA eigenvalue distribution. In \cite{haikin2017random} we further study this asymptotic behavior for a gallery of structured frames, including different ETFs but not only. (As far as we know, there are no previous results on the asymptotic spectra of submatrices of any deterministic frame other than a random choice of columns from DFT or Haar matrices.) 
We conjecture that these frames are asymptotically optimal for analog coding of a source with erasures. Moreover, we have a strong evidence that for every dimension where an ETF exists, it is optimal in the sense of the average excess rate caused by signal amplification; i.e, it minimizes the MLIE (\ref{MLIE}) over all possible ${(n,m)}$ frames. 

In particular, we verified for specific dimensions that DSS and real ETF (based on Paley’s construction of symmetric conference matrices) are local minimizers of the MLIE. The minimization is constrained due to the unit norm row constraint. Thus, the following term is added to the Lagrangian: ${L=\rho + \sum_{u=1}^{n}\lambda_u(\sum_{v=1}^{m}|\textbf A_{u,v}|^2-1)}$. The elements of ${\rho}$ which depend on ${\textbf A_{i,j}}$ are those for which ${\textbf s}$   includes sample ${i}$ in the pattern of important samples. Such patterns produce ${\textbf{A}_{s^{i}}}$ which consists of row ${i}$ and ${n-1 \choose k-1}$ other rows of ${\textbf A}$. Denote by ${r_{s^{i}}}$ the location of row ${i}$ with respect to the ${k}$ rows corresponding to pattern ${\textbf s^{i}}$. We use the following matrix derivative result: ${\frac{\partial}{\partial \textbf{X}}\tr(\textbf{X}^{-1})=-(\textbf{X}^{-2})^T}$.  The derivative ${\frac{\partial}{\partial \textbf A_{i,j}}L}$ of the Lagrangian with respect to ${\textbf A_{i,j}}$, is:
\begin{equation}\label{ScoreDerivative}
\frac{1}{{n \choose k}}\frac{1}{n\ln(2)}\sum_{s^i=1}^{{n-1 \choose k-1}}\frac{\sum_{t=1}^{k}(\textbf{A}_{s^{i}}\textbf{A}_{s^{i}}')^{-2}_{r_{s^{i}},t}(\textbf{A}_{s^{i}})_{t,j}}{\frac{1}{m}\tr((\textbf{A}_{s^{i}}\textbf{A}_{s^{i}}')^{-1})}-2\lambda_i\textbf A_{i,j}.
\end{equation}
Note that for simplicity we assumed here a real frame, but the result is true also for complex frames.
Numerically, substituting different normalized ETFs for various dimensions ${\frac{\partial}{\partial \textbf A_{i,j}}L}$ is zeroed for all ${i,j}$. 

\section{Analog Coding for an AWGN Channel with Erasures} \label{sec:ChannelCoding}
A symmetric problem to the source coding with erasures is the channel with noise and erasure coding. 
When $x$, the channel input, in additive white Gaussian noise (AWGN) $z$  with variance $\sigma_z^2$, passes through an erasure memoryless channel with probability $1-p$   for erasure, the output is:
\begin{equation} \label{y}
y = 
\begin{cases}
x+z,&  \text{ w.p. } p \\
0,              & \text{ w.p. }1- p.\\
\end{cases}
\end{equation}
Tulino at al. analyzed in \cite{tulino2007gaussian} the capacity of this channel:
\begin{equation} \label{CapacityTulino}
C=\frac{p}{2}\log(1+{\rm SNR})
\end{equation}
where SNR is the signal to noise ratio.
Wolf suggested in \cite{wolf1983redundancy} an analog coding scheme for an equivalent impulse channel $y=x+z$ where
\begin{equation} \label{z}
z\sim
\begin{cases}
\text{ AWGN},&  \text{ w.p. } p \\
\text{ Impulse},              & \text{ w.p. }1- p.\\
\end{cases}
\end{equation}
We can apply an exactly symmetric scheme based on a general frame, when now the encoder and the decoder change roles. The transform at the encoder is the frame matrix $\bf A$, the decoder applies the pseudo inverse of the subframe which corresponds to the non-erased pattern $\textbf{B}_s=(\textbf{A}'_s\textbf{A}_s)^{-1}\textbf{A}'_s$.
The equivalent AWGN channel is thus $\tilde{y}=\tilde{x}+\tilde{z}$:
\begin{figure}[htbp]
	\centering
	\hspace{-1.5em}
	\def\svgscale{2.4}
	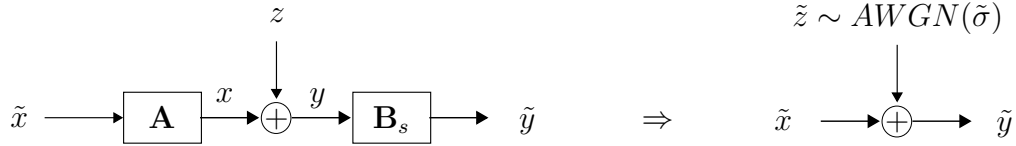
	\caption{Analog channel coding scheme. ${x,y}$ are ${n}$-dimensional vectors, ${\textbf{B}_s}$ is applied on the ${k}$ non-erased components of $y$ and ${\tilde{x},\tilde{y}}$ are ${m}$-dimensional. $k\ge m$.}
	\label{modelFigC}
\end{figure}\\
Due to unit norm normalization, the signal doesn't suffer from amplification, but there is noise amplification which according to \eqref{PointScore} is $\tilde{\sigma}^2=\frac{1}{m}\tr((\textbf{A}'_s\textbf{A}_s)^{-1})\sigma_z^2$.
The signal to noise ratio of the equivalent channel is 
\begin{equation} \label{SNReq}
\widetilde{\rm SNR}=\frac{\rm SNR}{\frac{1}{m}\tr((\textbf{A}'_s\textbf{A}_s)^{-1})}
\end{equation}
where SNR is the true (physical) signal to noise ratio of the channel.
In such scheme we pass $m$ samples in $n$ channel uses thus the achievable rate is 
\begin{equation} \label{Capasityeq}
\tilde{C}=\frac{m}{n}\frac{1}{2}\log (1+\widetilde{\rm SNR})=\frac{m}{k}C\left(\frac{\rm SNR}{\frac{1}{m}\tr((\textbf{A}'_s\textbf{A}_s)^{-1})}\right).
\end{equation}
We choose $m<pn$ in order to reconstruct $\tilde{x}$.

\section{Capacity and Rate Distortion of Equiangular Erasure Coding}\label{sec:inverse}
An analog coding scheme based on frames reveals the main measure of interest, the signal or noise amplification. As has been empirically shown in \cite{haikin2017random} for a variety of deterministic frames, the limiting eigenvalues distribution of subsets of a collection of frames is the MANOVA density which is superior in many aspects, including analog coding performance. Both empirical evidence and an analytical analysis will be further presented in Chapters \ref{chapter:PNAS} and \ref{chapter:Moments}. In this section we use the MANOVA limiting density for the evaluation of this amplification and its implication on analog coding performance. This serves as an analytical confirmation of the superiority which was empirically observed in \cite{haikin2016analog} (Subsection \ref{IrregularSpectrum}). In this section we stick to 
the convention that frame element are column vectors and not row vectors as assumed in previous sections and in Chapter \ref{chapter:PNAS}.
Let $A$ be an $m$-by-$k$ submatrix of the  $m$-by-$n$ frame matrix $F$ constructed by taking $k= p\cdot n$ out of $n$ columns of $F$ at random, or by taking each column with probability $p$
(same as removing the non-zero columns of $X$ \eqref{X}). 
Denote the aspect ratio of the frame $\gamma \triangleq \frac{m}{n}$, and the aspect ratio of the subset $\beta \triangleq \frac{k}{m}$,
where asymptotically, $\frac{k}{n} \to p$ almost surely in both cases. 
\subsection{The Inverse Moment}
\begin{figure}[!htb]
	\centering
	\includegraphics[width=3.7in]{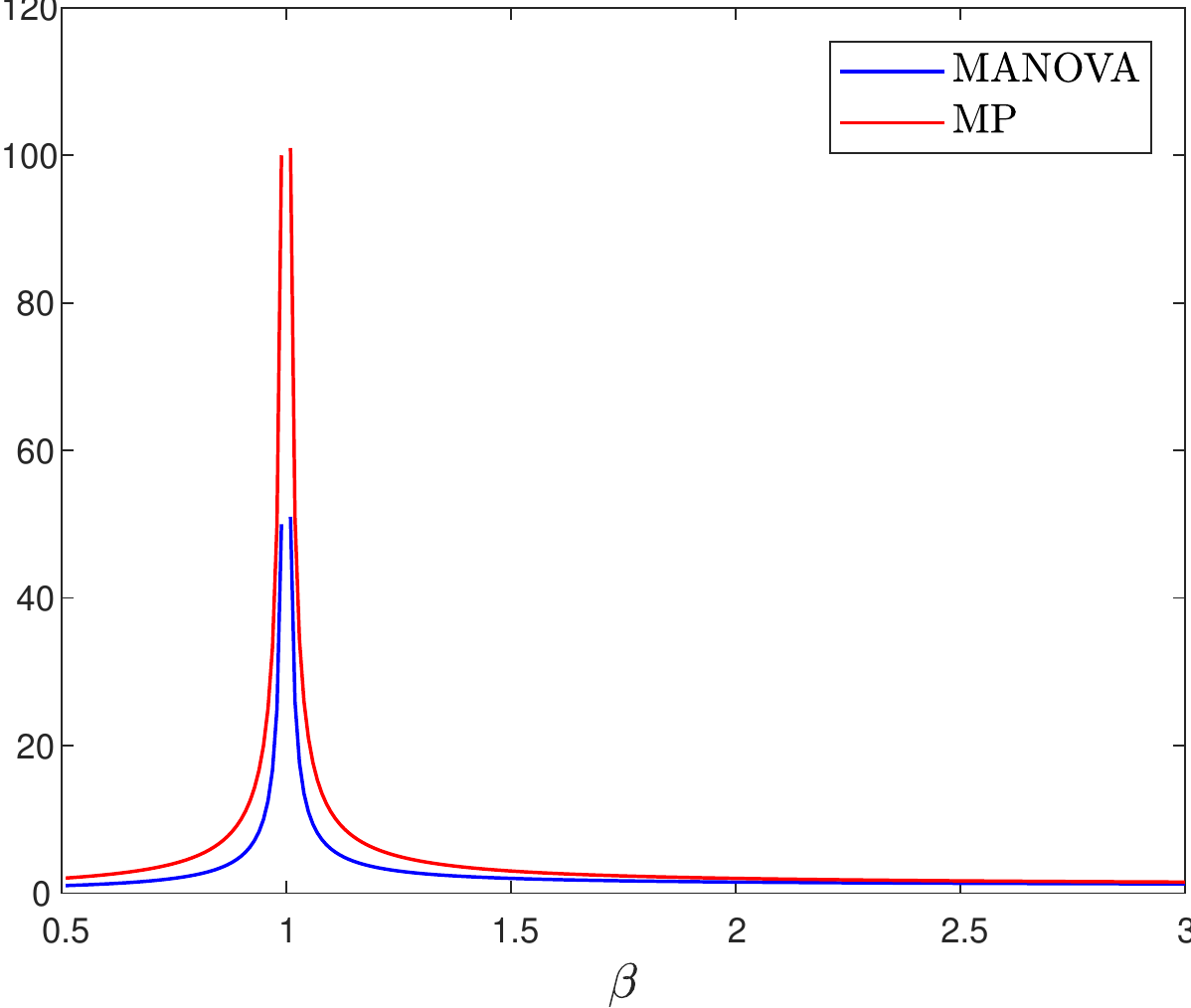}
	\caption{Amplification $\Lambda(\beta,p)$ 
		for $p=0.5$. 
	}
	\label{fig:AMP}
\end{figure}

\begin{figure}[!htb]
	\centering
	\includegraphics[width=3.7in]{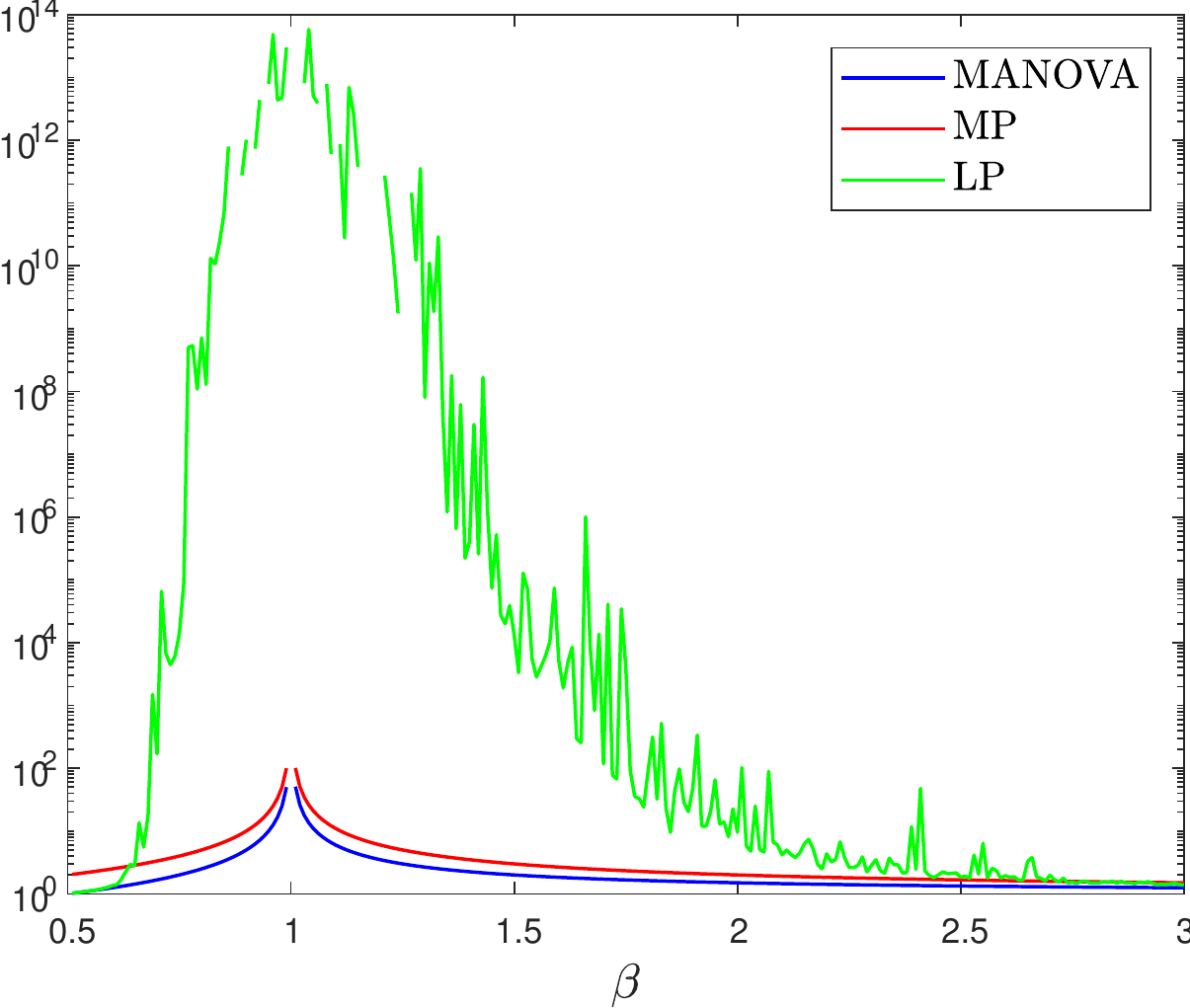}
	\caption{Amplification $\Lambda(\beta,p)$ 
		for $p=0.5$, logarithmic scale. 
	}
	\label{fig:AMP2}
\end{figure}

We denote $G=A'A$ if $m>k$ or $G=AA'$ if $k>m$ and define $r$ as the dimension of $G$,
\begin{equation} \label{r_dim}
r=\min(k,m).
\end{equation}
$\lambda_1, ... , \lambda_r$ are the $r$ non-zero eigenvalues of $A'A$ (or $AA'$), i.e, the $r$ eigenvalues of $G$. We define the measure of amplification as the normalized "$-1$ moment" of the Gram matrix $G$:
\begin{align}
	\Lambda(\beta,p)= \frac{\frac{1}{r}\sum_{i=1}^{r}\frac{1}{\lambda_i}}{\left(\frac{1}{r}\sum_{i=1}^{r}\lambda_i\right)^{-1}}
\end{align}
It is the Arithmetic-to-Harmonic Means Ratio (AHMR) which is obviously larger or equal to one. If the energy of the submatrix is one (i.e, $\frac{1}{r}\sum_{i=1}^{r}\lambda_i=1$), the inverse moment is given by $\frac{1}{r}\tr(G)^{-1}=\Lambda(\beta,p)$.
In the applications of source and channel coding with erasures this measure is responsible for the signal and noise amplification, respectively, and equals to
\begin{align} \label{amplification_CC_SC}
	\Lambda(\beta,p)=
	\begin{cases}
		\frac{1}{k}\tr(A'A)^{-1},& \beta<1  \text{ (Source coding) } \\
		\beta\frac{1}{m}\tr(AA')^{-1},          &   \beta>1  \text{ (Channel coding) }.
	\end{cases}
\end{align} 
This measure is motivated by the inverse energy definition in \eqref{IE}. For source coding $\Lambda(\beta,p)={\rm IE}\frac{1}{\beta}$.
Note that in \cite{haikin2016analog}, we showed that ${\rm IE}\ge \beta$, thus $\Lambda(\beta,p)\ge 1$.
The first order asymptotic moment of $(AA')^{-1}$, in case of i.i.d frame, is shown in \cite{tulino2004random} and the resulting asymptotic amplification is:
\begin{align} \label{InverseMP}
	\lim\limits_{n\to \infty}\Lambda_{MP}(\beta,p)=
	\begin{cases}
		\frac{1}{1-\beta},& \beta<1  \text{ (Source coding) } \\
		\frac{\beta}{\beta -1},          &   \beta>1  \text{ (Channel coding) }\\
	\end{cases}
\end{align}
To the best of our knowledge, our derivation of the asymptotic "inverse moment" of the MANOVA density is novel.
\begin{lemma} \label{lemmaInverse}
	For a frame with the MANOVA limiting eigenvalue distribution, and a minimal
	eigenvalue that converges almost surely to the edge of the support of this
	distribution:
	\begin{align}\label{lemmaInverseEq}
		\lim\limits_{n\to \infty}\Lambda_{MANOVA}(\beta,p)=
		\begin{cases}
			\frac{1-p}{1-\beta},& \beta<1  \text{ (Source coding) } \\
			\frac{\beta-p}{\beta-1},          &   \beta>1  \text{ (Channel coding) }\\
		\end{cases}
	\end{align}
\end{lemma}
\begin{proof}
	We shall derive the asymptotic moment of $A'A$ (or $AA'$) using its connection to the $\eta$ transform.
	The $\eta$-transform of a nonnegative random variable $Y$ is defined as
	\begin{equation}\label{etaTransform}
		\eta_Y(z) = \Ev_Y\left[\frac{1}{1+zY}\right].
	\end{equation}
	When speaking of $\eta$ transform of a positive semi-definite matrix, the expectation is over its limiting eigenvalue distribution.
	From \cite{tulino2004random}
	\begin{equation}\label{lim eta}
		\lim\limits_{n\to \infty}\frac{1}{r}\tr(G^{-1})=\lim\limits_{z\to \infty}z\eta_G(z).
	\end{equation}
	In \cite{farrell2011limiting}, Farrell showed that "Random Spectrum" frames have MANOVA limiting distribution. He derived the $\eta$ and Stieltjes transforms for the matrix $\tilde{G}=\tilde{A}\tilde{A}'$, where $\tilde{A}$ is a $F_n=DFT(n)$ matrix with rows
	and columns set to zero with probability $s$ and $t$ respectively, for $n\to \infty$. 
	We can use this result to compute the $\eta$-transform of $G$, using the fact that it shares the same MANOVA density as the "shrinked" $\tilde{G}$. Note that asymptotically, $\frac{r}{n}\to \min(1-t,1-s)=1-\max(s,t)$.
	Denote the limiting empirical distribution of eigenvalues of $\tilde{G}$ by $\tilde{f}_{s,t}$.
	When $Y$ stands for a random variable with density as the limiting distribution of the eigenvalues of the matrix $\tilde{G}$, the $\eta$-transform of $\tilde{G}$ converges almost surely to the asymptotic $\eta$-transform: 
	\begin{equation}\label{asymptoticEta}
		\tilde{\eta}_{s,t}(z) = \frac{1+(s+t)z+\sqrt{1+(2(s+t)-4st)z+(s-t)^2z^2}}{2(1+z)}
	\end{equation}
	Note that $\eta(z=0)=1$ and $\lim\limits_{z\to \infty}\eta_Y(z)=\Pr(Y=0)$. Thus, the asymptotic fraction of zero eigenvalues of $\tilde{G}$ is: $\lim\limits_{z\to \infty}\tilde{\eta}_{s,t}(z)=\frac{(s+t)+|s-t|}{2}=\max(s,t)$, which is the asymptotic fraction of erased rows or columns (the largest).\\
	The matrix $A$ is obtained by removing zero rows and columns.
	By removing the atom at 0 in $\tilde{f}_{p,q}$ and proper normalization, we can compute the $\eta$-transform of $G$.
	\begin{align}
		&\tilde{\eta}_{s,t}(z)=\Ev\bigg[\frac{1}{1+z\lambda}\bigg]=\int \frac{1}{1+z\lambda}\tilde{f}_{s,t}(\lambda)d\lambda \nonumber
		=\int \frac{1}{1+z\lambda}(\max(s,t)\delta(\lambda)+(1-\max(s,t))f_{s,t})d\lambda
		\nonumber\\
		& = \;
		\max(s,t)+(1-\max(s,t))\eta_{s,t}(z)
		\nonumber
	\end{align}
	\begin{equation}\label{etaNormalized}
		\Rightarrow \eta_{s,t}(z) = \frac{\tilde{\eta}_{s,t}(z)-\max(s,t)}{1-\max(s,t)}
	\end{equation}
	Now, no zero eigenvalues are expected in the limiting spectral density of $G$ and indeed: $\lim\limits_{z\to \infty}\eta_{s,t}(z)=\frac{\max(s,t)-\max(s,t)}{1-\max(s,t)}=0$, $\eta_{s,t}(z=0)=\frac{1-\max(s,t)}{1-\max(s,t)}=1$.
	So we are ready to compute the limit in \eqref{lim eta} (this limit for $\tilde{\eta}_{s,t}(z)$ is unbounded):
	\begin{align}
		&\lim\limits_{z\to \infty}z\eta_{s,t}(z) \nonumber \\&=\lim\limits_{z\to \infty}\frac{z}{1-\max(s,t)}\bigg[\frac{\frac{1}{z}+(s+t)+\sqrt{(s-t)^2+(2(s+t)-4st)\frac{1}{z}+\frac{1}{z^2}}}{2(\frac{1}{z}+1)}-\max(s,t)\bigg]
		\nonumber\\
		& = \;
		\lim\limits_{z\to \infty}\frac{z}{1-\max(s,t)}\bigg[\frac{\frac{1}{z}+(s+t)+|s-t|\bigg(1+\frac{1}{2}\frac{(2(s+t)-4st)}{z(s-t)^2}\bigg)-\max(s,t)2(\frac{1}{z}+1)}{2(\frac{1}{z}+1)}\bigg]
		\nonumber
		\nonumber\\
		& = \;
		\lim\limits_{z\to \infty}\frac{1-2\max(s,t)+\frac{s+t-2st}{|s-t|}}{2(1-\max(s,t))(\frac{1}{z}+1)}=\frac{\max(s,t)-st-\max(s,t)|s-t|)}{(1-\max(s,t))|s-t|}\nonumber \\&=\frac{\max(s,t)(1-\max(s,t))}{(1-\max(s,t))|s-t|}
		\nonumber
	\end{align}
	\begin{equation}\label{limTrace}
		\Rightarrow \lim\limits_{z\to \infty}z\eta_{s,t}(z) = \frac{\max(s,t)}{|s-t|}
	\end{equation}
	Applying our notations, 
	\begin{equation}\label{notations}
		1-t=p=\frac{k}{n}=\beta\gamma, \;\;\;\;
		1-s=\frac{m}{n}=\gamma.
	\end{equation}
	We consider unit-norm frames, thus the DFT submatrix should be multiplied by $\frac{\sqrt{n}}{\sqrt{m}}=\frac{1}{\sqrt{\gamma}}$.
	$\Rightarrow$ For a unit norm frame with MANOVA spectrum of random subsets, the average of the inverse of eigenvalues is:
	\begin{equation}\label{scoreAmpManova}
		\lim\limits_{n\to \infty}\frac{1}{\min(k,m)}\tr(G^{-1})= \gamma \frac{n-\min(k,m)}{|k-m|}
	\end{equation}
	For channel coding we use \eqref{amplification_CC_SC} and \eqref{scoreAmpManova} with $k\ge m$ 
	\begin{equation} \label{ChannelCodingAmpMANOVA}
		\Lambda_{MANOVA}^{CC}(\beta,p)=\beta\frac{1}{m}\tr(AA')^{-1}=\beta \gamma \frac{s}{s-t}=\frac{\beta-\beta \gamma}{\beta -1}=\frac{\beta-p}{\beta -1}.
	\end{equation}
	For source coding we use \eqref{amplification_CC_SC} and \eqref{scoreAmpManova} with $m\ge k$ 
	\begin{equation} \label{SourceCodingAmpMANOVA}
		\Lambda_{MANOVA}^{SC}(\beta,p)=\frac{1}{k}\tr(A'A)^{-1}= \gamma \frac{t}{t-s}=\frac{1-\beta \gamma}{1-\beta}=\frac{1-p}{1-\beta}.
	\end{equation}
\end{proof}
In Figure~\ref{fig:AMP} we can see the asymptotic amplifications of i.i.d frame \eqref{InverseMP} and frames with MANOVA limiting distribution \eqref{lemmaInverseEq}.
In Figure~\ref{fig:AMP2} a comparison to low-pass (LP) frame (band limited interpolation) is presented (for LP frame, an average value of 100 realizations). It is apparent that for $\beta$ around 1, the LP amplification is exploding. Note that for small $\beta \to p$, $m\to n$ and LP is equivalent to ETF with $n=m$ (i.e. a unitary matrix). 

\subsection{Capacity and Rate Distortion}\label{CR}
\begin{figure}[!htb]
	\centering
	\includegraphics[width=4.5in]{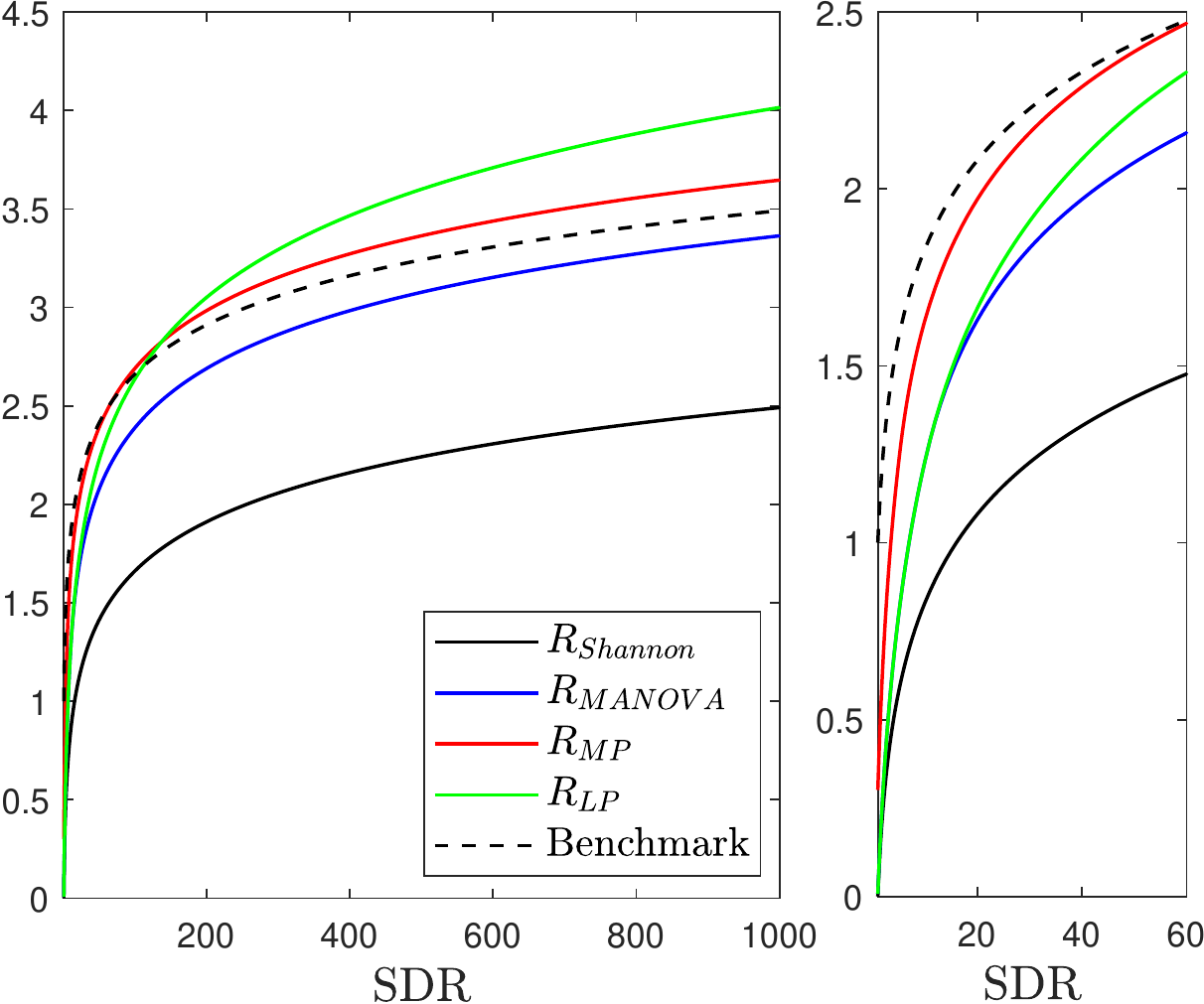}
	\caption{Rate at optimum $\beta$, $p=0.5$. 
	}
	\label{fig:R}
\end{figure}

In this subsection we explore the behavior of the achievable rates using different frames and bring analytical evaluation in extreme SNR and SDR.
According to \eqref{Rate7}
\begin{equation} \label{RateAccurate}
\tilde{R}=\frac{1}{\beta}\frac{p}{2}\log(1+(y-1)\beta \Lambda^{SC}(\beta)),
\end{equation}
where $y$ is the SDR.
In high SDR the rate-distortion function of source coding with erasures using analog codes is: 
\begin{equation} \label{Rate}
\tilde{R}=\frac{1}{\beta}R(y\beta \Lambda^{SC}(\beta)),
\end{equation}
where $R(y) = \frac{p}{2}\log(y)$.
Figure~\ref{fig:R} shows the rate of analog source coding using 3 different frames, compared to the optimal RDF and to the benchmark of SI transmission ($R(y)+H_b(p)$). Note that  in the low SDR, as optimum $\beta$ decreases, the LP frame becomes almost unitary full DFT, just like ETF with $\gamma \to 1$.

\begin{figure}[!htb]
	\centering
	\includegraphics[width=4in]{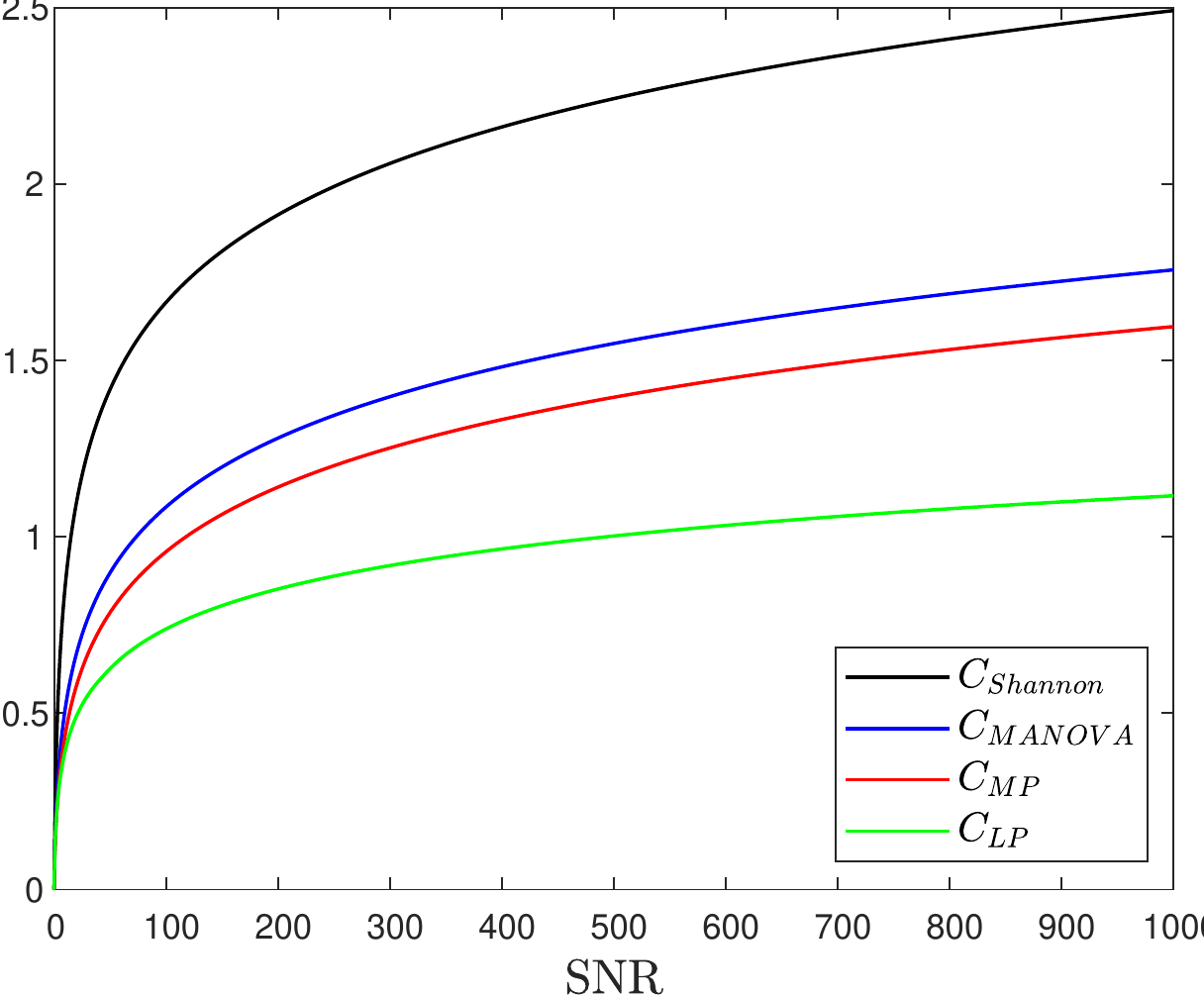}
	\caption{Capacity at optimum $\beta$, $p=0.5$. 
	}
	\label{fig:C}
\end{figure}

\begin{figure}[!htb]
	\centering
	\includegraphics[width=4in]{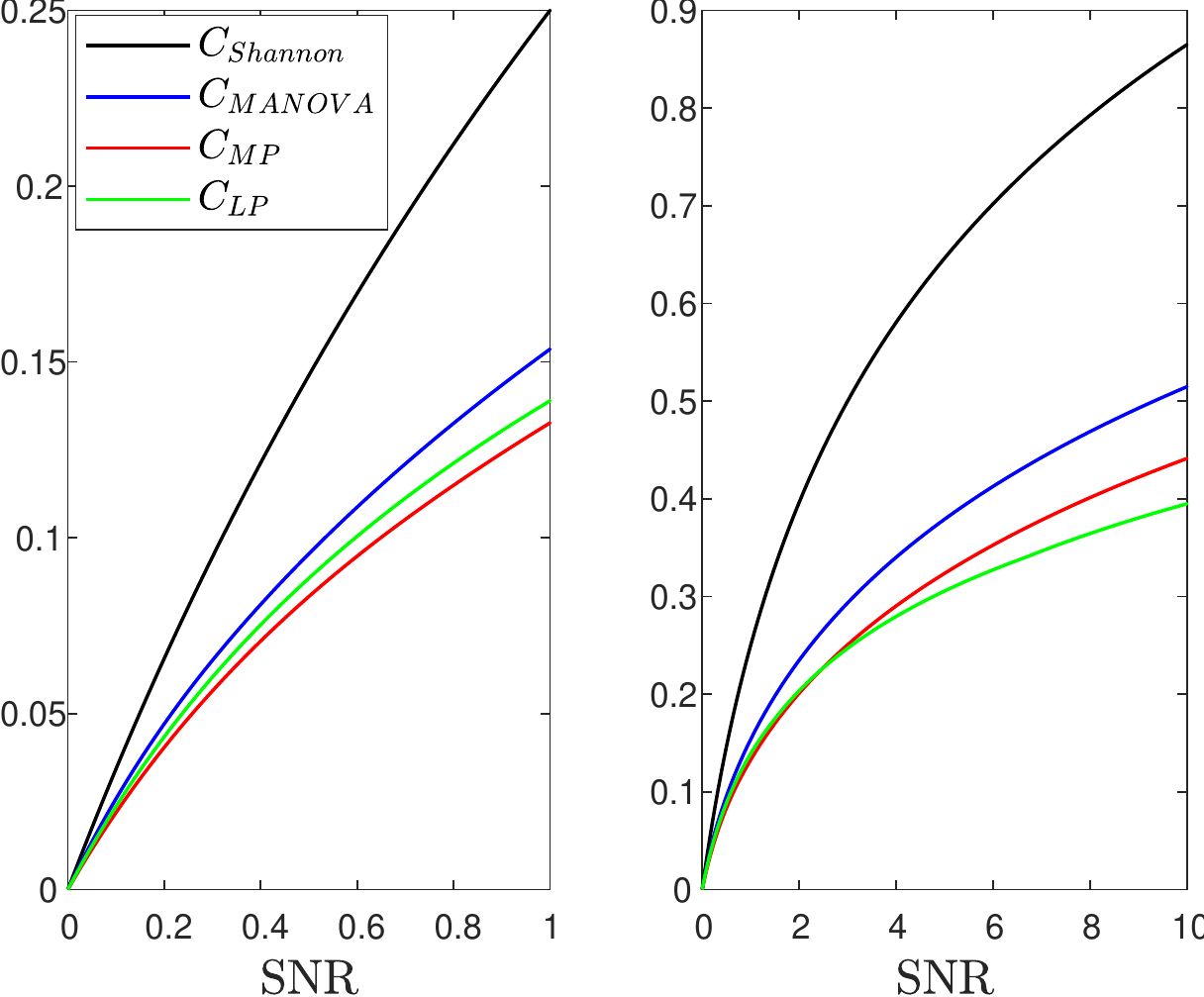}
	\caption{Capacity at optimum $\beta$, $p=0.5$, low SNR. 
	}
	\label{fig:Czoom}
\end{figure}

Similar and even simpler analysis of the capacity of erasure channel with analog coding, which is given in Section \ref{sec:ChannelCoding} and \cite{ITA17}, yields:
\begin{equation} \label{Capacity}
\tilde{C}=\frac{1}{\beta}C(y\frac{\beta}{\Lambda^{CC}(\beta)}),
\end{equation}
where $y$ is the SNR and $C(y) = \frac{p}{2}\log(1+y)$.
Figure~\ref{fig:C} shows the capacity of analog channel coding using 3 different frames, compared to optimal Shannon capacity. Figure~\ref{fig:Czoom} zooms in to a low SNR. 

\begin{lemma} \label{lemmaRC}
	In high SNR/SDR regime the gap from Shannon capacity/rate distortion function, for both i.i.d frame (MP density) and ETF (MANOVA density) is:
	\begin{equation} \label{GapC}
		\Delta^{CC} = \tilde{C}-C = -\frac{p}{2}\log\log y +O(1)
	\end{equation}
	\begin{equation} \label{GapR}
		\Delta^{SC} = \tilde{R}-R = \frac{p}{2}\log\log y +O(1)
	\end{equation}
	and the gaps between these two schemes are:
	\begin{equation} \label{GapCdiff}
		\Delta^{CC}_{MANOVA}-\Delta^{CC}_{MP} = -\frac{p}{2} \log(1-p)
	\end{equation}
		\begin{equation} \label{GapRdiff}
		\Delta^{SC}_{MANOVA}- \Delta^{SC}_{MP}= \frac{p}{2}\log(1-p)
	\end{equation}
\end{lemma}
Note that $\Delta^{SC}$ is the $\delta$ \eqref{ExcessRate3} discussed in Section \ref{sec:AC_ISIT}.

\begin{proof}
	Substituting the expression of the amplification of an i.i.d frame \eqref{InverseMP} in \eqref{Rate} we have:
	\begin{equation} \label{RateIID}
		\tilde{R}_{MP}=\frac{1}{\beta}R\left(y\frac{\beta}{1-\beta}\right)=\frac{p}{2}\frac{1}{\beta}\log\left(y\frac{\beta}{1-\beta}\right)=\frac{p}{2}(x+1)\log\left(\frac{y}{x}\right), 
	\end{equation}
	where $y=\text{SDR}, \; x=\frac{1}{\beta}-1$.
		The derivative with respect to $x$ is:
	\begin{align}
		&\frac{\partial\tilde{R}_{MP}}{\partial x}=\frac{p}{2}\left(\log\left(\frac{y}{x}\right)-(x+1)\frac{x}{y\ln2}\frac{y}{x^2}\right)
		=\frac{p}{2}\left(
		\log\left(\frac{y}{x}\right)-\left( 1+\frac{1}{x} \right)\frac{1}{\ln2}\right).	
		\nonumber
	\end{align}
	For high SDR, i.e. $y \to \infty$, $x$ must go to zero, thus, $\frac{1}{x\ln2}+\log x \approx \log y$ $\Rightarrow$ $ x \approx \frac{1}{\ln y}$.
	For high SDR the optimum $\beta$ behaves like
	\begin{equation} \label{ bestbetaR}
		\beta\approx \left( 1+\frac{1}{\ln y} \right)^{-1}\approx  1-\frac{1}{\ln y}.
	\end{equation}
	The rate in high SDR is:
	\begin{equation} \label{MinRateIID2}
		\tilde{R}_{MP} \approx \frac{p}{2}\left(1+\frac{1}{\ln y}\right)\left(\log y + \log \ln y\right) 
	\end{equation}
	And the gap from the RDF is:
	\begin{equation} \label{gapR}
		\Delta^{SC}_{MP}=\tilde{R}_{MP} -R = \frac{p}{2} \log \log y + \frac{p}{2\ln 2} +\frac{p}{2} \log \ln 2+ O(\frac{\log \log y}{\log y}) 
	\end{equation}
	We repeat similar analysis with substitution of the derived amplification for the MANOVA case in Lemma \ref{lemmaInverse}.
	\begin{equation} \label{RateMANOVA}
		\tilde{R}_{MANOVA}=\frac{1}{\beta}R\left(y\frac{\beta(1-p)}{1-\beta}\right)=\frac{p}{2}\frac{1}{\beta}\log\left(y\frac{\beta(1-p)}{1-\beta}\right)=\frac{p}{2}(x+1)\log\left(\frac{y(1-p)}{x}\right), 
	\end{equation}
	where $y=\text{SDR}, \; x=\frac{1}{\beta}-1$.
	The derivative with respect to $x$ is:
	\begin{align}
		&\frac{\partial\tilde{R}_{MANOVA}}{\partial x}=\frac{p}{2}\left(\log\left(\frac{y(1-p)}{x}\right)-(x+1)\frac{x}{y\ln2}\frac{y}{x^2}\right)
		=
		\frac{p}{2}\left(\log\left(\frac{y(1-p)}{x}\right)-\left( 1+\frac{1}{x} \right)\frac{1}{\ln2}\right).	
		\nonumber
	\end{align}
	Again, for high SDR the optimum $\beta$ behaves like
	\begin{equation} \label{ bestbetaR}
		\beta\approx \left( 1+\frac{1}{\ln y} \right)^{-1}\approx  1-\frac{1}{\ln y}
	\end{equation}
	The rate in high SDR is:
	\begin{equation} \label{MinRateIID2}
		\tilde{R}_{MANOVA} \approx \frac{p}{2}\left(1+\frac{1}{\ln y}\right)\left(\log (y(1-p)) + \log \ln y\right) 
	\end{equation}
		And the gap from the RDF is:
	\begin{equation} \label{gapR}
		\Delta^{SC}_{MANOVA} = \tilde{R}_{MANOVA} -R = \frac{p}{2} \log \log y+ \frac{p}{2\ln 2} +\frac{p}{2} \log \ln 2+ \frac{p}{2}\log(1-p) + O(\frac{\log \log y}{\log y})
	\end{equation}
	The gap between Mar\u cenko-Pastur and MANOVA is thus:
	\begin{equation} \label{gapR_MPvsMANOVA}
		\Delta^{SC}_{MANOVA}- \Delta^{SC}_{MP}= \frac{p}{2}\log(1-p)
	\end{equation}
	
	Now we turn to evaluation of asymptotic rates in channel coding. Substituting the expression of the amplification of an i.i.d frame \eqref{InverseMP} in \eqref{Capacity} we have:
	\begin{equation} \label{CapacityIID}
		\tilde{C}_{MP}=\frac{1}{\beta}C(y(\beta-1))=\frac{p}{2} \frac{1}{\beta} \log(1+y(\beta-1))=\frac{p}{2} \frac{1}{x+1} \log(1+y x), 
	\end{equation}
	where $y=\text{SNR}, \; x=\beta-1$.
	\begin{align}
		&\frac{\partial\tilde{C}_{MP}}{\partial x}=\frac{p}{2}\left(-\log(1+y x)\frac{1}{(x+1)^2}+\frac{1}{(x+1)}\frac{y}{1+y x}\frac{1}{\ln2}\right)\\&
		=
		\frac{p}{2}\frac{1}{(x+1)^2}\left( \frac{y x+y}{1+y x}\frac{1}{\ln2}-\log (1+y x) \right).	
		\nonumber
	\end{align}
	For high SNR, $y \to \infty$, thus the limit $y x \to \infty$ must be satisfied (otherwise the derivative is unbounded). In this case the condition for optimum $x$ ($\beta$) is:
	
	\begin{equation} \label{MaxCapacityIID2Condition}
		\frac{x+1}{x\ln2}-\log y x=0
	\end{equation}
	$x \to 0$ $\Rightarrow$ $\frac{1}{x\ln2}-\log x \approx \log y$ $\Rightarrow$ $ x \approx \frac{1}{\ln y}$.
	For high SNR the optimum $\beta$ behaves like 
	\begin{equation} \label{ bestbetaC}
		\beta\approx 1+\frac{1}{\ln y}
	\end{equation}
	The capacity in high SNR is:
	\begin{equation} \label{MaxCapacityIID2}
		\tilde{C}_{MP} \approx \frac{p}{2}\left(1-\frac{1}{\ln y}\right)\left(\log y - \log \ln y\right) 
	\end{equation}
	\begin{equation} \label{gapC}
		\Delta^{CC}_{MP}=\tilde{C}_{MP} -C \approx -\frac{p}{2} \log \log y - \frac{p}{2\ln2}-\frac{p}{2}\log \ln2 + O(\frac{\log \log y}{\log y}) 
	\end{equation}
\end{proof}

For a general frame we can understand how the ratio of capacities behaves in extreme SNR:
\begin{equation} \label{CapacityRatio}
\frac{\tilde{C}}{C}=\frac{\frac{1}{\beta}\log(1+y\frac{\beta}{\Lambda(\beta)})}{\log(1+y)}
\end{equation}

\begin{lemma} \label{lemmaC_highSNR}
	For every frame $F$, if $\forall \beta$ (close to 1) $\Lambda(\beta)$ is finite, the ratio of capacities in high SNR is:
\begin{equation} \label{CratiocHighSNR}
\lim\limits_{{\rm SNR}\to \infty}\frac{\tilde{C}}{C}=1.
\end{equation}
\end{lemma}

\begin{proof}
	The optimal $\beta$ in high SNR regime goes to 1. For $y\to \infty$, 
	\begin{equation} \label{CapacityRatioHighSNR}
	\frac{\tilde{C}}{C}=\frac{\log y + \log \frac{\beta}{\Lambda(\beta)}}{\log y}=1+\frac{\log \frac{\beta}{\Lambda(\beta)}}{\log y}
	\end{equation}
	thus, if $\Lambda(\beta)$ is bounded, the ratio is 1.
\end{proof}

\begin{lemma} \label{lemmaC_lowSNR}
	For every frame $F$, when ${\rm SNR}\to 0$, if $\lim\limits_{\beta \to \infty}\Lambda(\beta)=1$ then:
	\begin{equation} \label{CratiocLowSNR}
\lim\limits_{{\rm SNR}\to 0}\frac{\tilde{C}}{C}=1.
	\end{equation}
\end{lemma}

\begin{proof}
	For every $\beta$, when $y\to 0 \Rightarrow \frac{\tilde{C}}{C}=\frac{\frac{1}{\beta}y\frac{\beta}{\Lambda(\beta)}}{y}=\frac{1}{\Lambda(\beta)}$ ($\log(1+\epsilon)\approx \epsilon$).
	If for $\beta \to \infty$, $\Lambda(\beta)\to 1$, the ratio in high SNR is 1.
\end{proof}
From \eqref{InverseMP} and Lemma \ref{lemmaInverse} follows that this holds for both i.i.d frames and ETFs.


	\chapter{Random Subsets of Structured Deterministic Frames have	MANOVA Spectra}
	\label{chapter:PNAS}
	
	This chapter is taken from our PNAS paper \cite{haikin2017random}. It explores a variety of structured frames and proves empirically that the MANOVA ensemble universally describes the spectra of their random
	subsets.
	Consider a frame $\{\vx_i\}_{i=1}^n \subset \R^\m$ or
	$\C^\m$ and stack the vectors as rows to obtain the $n$-by-$\m$ frame matrix
	$X$.  Assume that $\norm{ \vx_i}_2=1$ (deterministic frames) or $\lim_{n\to
		\infty}\|\vx_i\|=1$ almost surely (random frames).  This chapter studies
	properties of a random subframe $\{\vx_i\}_{i\in K}$, where $K$ is chosen
	uniformly at random from $[n]=\left\{ 1,\ldots,n \right\}$ and $|K|=k\leq n$.
	We let $\Xk$ denote  the $k$-by-$\m$ submatrix of $X$ created by picking only
	the rows $\{\vx_i\}_{i\in K}$; call this object {\em a typical $k$-submatrix of
		$X$}.  We consider a collection of well-known deterministic frames, listed in
	Table \ref{frames:tab}, which we denote by $\cal X$. Most of the frames in $\cal
	X$ are equiangular tight frames (ETFs), and some are near-ETFs.
	
	The results of this chapter suggest that for a frame in $\cal X$ it is possible to calculate
	quantities of the form $\E_K \specstat(\lambda(\Gk))$, where
	$\lambda(\Gk)=(\lambda_1(G_K),...,\lambda_k(G_K))$ is the vector of eigenvalues
	of the $k$-by-$k$ Gram matrix $\Gk=\Xk \Xk'$ and $\specstat$ is a functional of
	these eigenvalues.  As discussed below, such quantities are of considerable
	interest in various applications where frames are used, across a variety of
	domains, including compressed sensing, sparse recovery and erasure coding.
	
	We present a simple and explicit formula for calculating 
	$\E_K \specstat(\lambda(\Gk))$ for a given frame in $\cal X$ and a given spectral 
	functional $\specstat$. Specifically, for the case $k \le m$,
	\[
	\E_K\specstat(\lambda(\Gk)) \approx \specstat\left( f^{MANOVA}_{\beta,\gamma}
	\right)\,, \] 
	where $\beta=k/\m$, $\gamma=\m/n$ and where 
	$f^{MANOVA}_{\beta,\gamma}$ is the density of
	Wachter's classical 
	MANOVA$(\beta,\gamma)$
	limiting distribution \cite{wachter1980limiting}. The fluctuations about this approximate 
	value are given {\em exactly} by
	\begin{eqnarray} \label{main_for:eq}
	\E_K \big| \specstat(\lambda(\Gk)) - \specstat\left( f^{MANOVA}_{\beta,\gamma}
	\right)\big|^2 =   
	C n^{-b} \log^{-a}(n) \,.
	\end{eqnarray}
	While the constant $C$ may depend on the frame, 
	the exponents $a$ and $b$ are {\em universal} and depend only 
	on $\specstat$ and on the aspect ratios $\beta$ and $\gamma$. 
	Evidently, the precision of the MANOVA-based approximation is good, known, and 
	improves as $m$ and $k$ both grow proportionally to $n$.
	
	Formula \ref{main_for:eq} is based on a far-reaching 
	{\em universality hypothesis}: 
	For all frames in $\cal X$, as well as for well-known random frames also listed in
	Table \ref{frames:tab}, we find that
	the spectrum of the typical $k$-submatrix ensemble is indistinguishable 
	from that of the classical MANOVA (Jacobi) random matrix ensemble
	\cite{forrester2010log}
	of the same
	size. 
	(Interestingly, it will be shown that for deterministic ETFs this
	indistinguishably holds in a stronger sense than for deterministic non-ETF
	frames.)
	This universality is not asymptotic, and concerns 
	finite $n$-by-$\m$ frames. However, it does imply that the spectrum of the
	typical $k$-submatrix ensemble converges to 
	a  universal limiting distribution, which is non other than 
	Wachter's  
	MANOVA$(\beta,\gamma)$ limiting distribution \cite{wachter1980limiting}. 
	It also implies that the universal exponents $a$ and $b$
	in \eqref{main_for:eq}
	are previously unknown, universal quantities corresponding to the 
	classical MANOVA (Jacobi) random matrix ensemble.
	
	This brief announcement tests Formula
	\ref{main_for:eq} and the underlying universality hypothesis
	by conducting substantial computer
	experiments, in which a large number of random $k$-submatrices are generated. 
	We  study a large variety of  deterministic frames, both
	real and complex. In addition to the universal object (the MANOVA ensemble)
	itself, 
	we study difference-set spectrum frames,
	Grassmannian frames, real Paley frames, complex Paley frames, quadratic phase
	chirp frames, Spikes and Sines frames, and Spikes and Hadamard frames. 
	
	We report compelling empirical evidence,  systematically documented and
	analyzed, which fully supports the universality hypothesis and
	\eqref{main_for:eq}.  Our results are empirical, but they are exhaustive,
	precise, reproducible and meet the best standards of empirical science. 
	
	For this purpose, we develop a natural
	framework for empirically testing such hypotheses regarding
	limiting distribution and convergence rates of 
	random matrix ensembles. 
	Before turning to deterministic frames, 
	we validate our framework on well-known random frames, including 
	real orthogonal Haar frames, complex unitary Haar frames, real random
	Cosine frames and complex random Fourier frames.
	Interestingly, rigorous proofs that identify the MANOVA distribution as the
	limiting spectral distribution of typical $k$-submatrices can be found in the
	literature for two of these random frames, namely the random Fourier frame
	\cite{farrell2011limiting} and the unitary Haar frame \cite{edelman2008beta}. 
	
	\section{Motivation} \label{sec:motivation} 
	
	Frames can be viewed as an analog counterpart for digital coding. They provide 
	overcomplete representation of signals, adding redundancy and increasing
	immunity to noise. 
	Indeed, they are used in
	many branches of science and engineering for stable signal representation,
	as well as error and erasure correction.
	
	Let $\lambda(G)$ denote the vector of nonzero eigenvalues of $G=X'X$ and let 
	$\lambda_{max}(G)$ and $\lambda_{min}(G)$ denote its max and min,
	respectively. 
	Frames were traditionally designed to achieve frame bounds  $\lambda_{min}(G)$
	as high as possible (resp. $\lambda_{max}(G)$ as low as possible).
	Alternatively, they were designed to
	minimize  {\em mutual coherence} \cite{donoho2006stable,Elad2010},
	the maximal pairwise correlation between any two frame vectors.
	
	In the passing decade it has become
	apparent that neither frame bounds (a global criterion) 
	nor coherence (a local, pairwise criterion) are
	sufficient to
	explain various phenomena related to overcomplete representations, and that
	one should also look at collective behavior of $k$ frame vectors from the
	frame, $2\leq k \leq n$. 
	While different applications focus on different properties of the submatrix
	$\Gk$, most of these properties can be expressed as a function of
	$\lambda(\Gk)$,
	and even just an average of a scalar function of the eigenvalues. Here are a few notable examples. 
	\paragraph{Restricted Isometry Property (RIP).} 
	Recovery of any $k/2$-sparse
	signal $\V{v}\in\R^n$ from its linear measurement $F'\V{v}$ using $\ell_1$
	minimization is guaranteed if the spectral radius of $\Gk - I$, namely,
	\begin{eqnarray} \label{rip_func:eq}
	\specstat_{RIP}(\lambda(\Gk))= 
	\max\{ \lambda_{max}(\Gk)-1 \,,\,
	1-\lambda_{min}(\Gk) \}\,, 
	\end{eqnarray} 
	is uniformly bounded by some $\delta<0.4531$ on all $K\subset[n]$
	\cite{candes2006near, candes2008restricted,foucart2009sparsest}.
	
	\paragraph{Statistical RIP.} Numerous authors have studied a relaxation of the
	RIP condition suggested in \cite{calderbank2010construction}. Define 
	\begin{eqnarray} \label{strip_func:eq}
	\specstat_{StRIP,\delta}(\lambda(\Gk)) =
	\begin{cases} 1 &  \specstat_{RIP}(\lambda(\Gk)) \leq
	\delta \\ 
	0 & otherwise \end{cases}\,.  
	\end{eqnarray}
	Then
	$\E_K \specstat_{StRIP,\delta}(\lambda(\Gk))$ is
	the probability that the RIP condition 
	with bound $\delta$ holds when $X$ acts on a signal
	supported on a random set of $k$ coordinates.
	\paragraph{Analog coding of a source with erasures.} 
	In \cite{haikin2016analog} we  
	considered a typical erasure pattern of
	$n-k$ random samples known at the transmitter, but not the receiver.
	The rate-distortion function  
	of the coding scheme suggested in \cite{haikin2016analog} is determined
	by $\E_K \log(\beta\specstat_{AC}(\lambda(\Gk)))$, with
	\begin{eqnarray} \label{ac_func:eq}
	\specstat_{AC}(\lambda(\Gk))=\frac{1}{k}\tr[(\Gk)^{-1}]/\left(\frac{1}{k}\tr[\Gk]\right)^{-1}\,,
	\end{eqnarray}
	i.e., $\specstat_{AC}(\lambda(\Gk))$ is the arithmetric-to-harmonic means ratio of the eigenvalues (the arithmetric mean is $1$ due to the normalization of frames). This quantity is the signal
	amplification responsible for the excess rate of the suggested coding scheme. 
	%
	\paragraph{Shannon transform.} 
	The quantity 
	\begin{eqnarray} \label{shannon_func:eq}
	\begin{aligned}
	\specstat_{Shannon}(\lambda(\Gk)) 
	&= \frac{1}{k}\log(\det(I+\alpha \Gk))= \frac{1}{k}\tr(\log(I+\alpha \Gk))\,,
	\end{aligned}
	\end{eqnarray}
	which was suggested in \cite{tulino2004random},
	measures the capacity of a linear-Gaussian erasure channel. 
	Specifically, it assumes $y=XX'x+z$,
	(where $x$ and $y$ are the channel input and output)
	followed by $n-k$ random erasures. 
	The quantity $\alpha$ in \eqref{shannon_func:eq} is the 
	signal-to-noise ratio $SNR=\alpha\geq 0$.
	\\
	~
	\\
	\noindent In this work, we focus on {\em typical-case} performance criteria (those that
	seek to optimize $\E_K \specstat(\lambda(\Gk))$ over random choice of $K$) rather than
	{\em worse-case} performance criteria (those that seek to optimize 
	$\max_{K\subset [n]} \specstat(\lambda(\Gk))$, such as RIP). For the remainder of this
	chapter, $K\subset[n]$ will denote a uniformly distributed random subset of size
	$k$. Importantly, $k$ should be allowed to be large, even as large as $\m$. 
	
	For a given $\specstat$, one would like to design frames that optimize $\E_K
	\specstat(\lambda(\Gk))$.  This turns out to be a difficult task; in fact, it is
	not even known how to calculate $\E_K \specstat(\lambda(\Gk))$ for a given frame
	$X$.
	Indeed, to calculate this quantity one effectively has to average $\specstat$
	over  the spectrum $\lambda(\Gk)$ for all $\binom{n}{k}$ subsets $K\subset [n]$.
	It is of little surprise to the information theorist that the first frame
	designs, for which performance was formally bounded (and still not calculated
	exactly), consisted of random vectors
	\cite{candes2006near,nelson2014new,rudelson2008sparse,haviv2017restricted,pfander2013restricted,cheraghchi2013restricted}.
	
	
	\section{Random Frames} \label{sec:random_frames}
	
	When the frame is random, namely when $X$ is drawn from some ensemble of random
	matrices, the typical $k$-submatrix $\Xk$ is also a random matrix. Given a
	specific $\specstat$, rather than seeking to bound $\E_K
	\specstat(\lambda(\Gk))$ for specific $n$ and $\m$, it can be extremely
	rewarding to study the limit of $\specstat(\lambda(\Gk))$ as the frame size $n$
	and $\m$ grow. This is because tools from random matrix theory become available,
	which allow exact asymptotic calculation of $\lambda(\Gk)$ and
	$\specstat(\lambda(\Gk))$, and also because their limiting values are usually
	very close to their corresponding values for finite $n$ and $\m$, even for low
	values of $n$.
	
	Let us consider then a sequence of dimensions $\m_n$ with $\m_n/n=\gamma_n\to \gamma$
	and a sequence of 
	random frame matrices $X^{(n)}\subset \R^{n\times\m_n}$ or $\C^{n\times\m_n}$.
	To
	characterize the collective behavior of $k$-submatrices we choose a sequence
	$k_n$ with $k_n/\m_n=\beta_n\to \beta$ 
	and look at the spectrum
	$\lambda(\Gkn)$ 
	of the random
	matrix 
	$\Xkn$ as $n\to\infty$, where $K_n\subset [n]$ is a randomly chosen 
	subset with $|K_n|=k_n$. Here and below, to avoid cumbersome notation we omit the subscript $n$ and write $m$,$k$ and $K$ for $m_n$,$k_n$ and $K_n$.
	
	A mainstay of random matrix theory is the celebrated convergence of the
	empirical spectral distribution of random matrices, drawn from a certain
	ensemble, to a limiting spectral distribution corresponding to that ensemble. 
	This has indeed been established for three random frames:
	\begin{enumerate} 
		
		\item {\em Gaussian i.i.d frame:} Let $X_{normal}^{(n)}$ have i.i.d normal
		entries with mean zero and variance $1/\m$.  The empirical
		distribution of $\lambda(\Gk)$ famously converges, almost surely in
		distribution, to the Mar\u cenko-Pastur density \cite{marvcenko1967distribution} with parameter
		$\beta$:  
		\begin{equation}
		\label{MPdensity:eq}
		f_{\beta}^{MP}(x)
		=\frac{\sqrt{(x-\lambda^{MP}_-)(\lambda^{MP}_+-x)}}{2\beta\pi x}\cdot
		I_{(\lambda^{MP}_-,\lambda^{MP}_+)}(x),
		\end{equation} 
		supported on $[\lambda^{MP}_-,\lambda^{MP}_+]$ where
		$\lambda^{MP}_\pm = (1\pm \sqrt{\beta})^2$.  Moreover,  almost surely
		$\lambda_{max}(G_{normal}^{(n)}) \to \lambda_+$ and
		$\lambda_{min}(G_{normal}^{(n)}) \to \lambda_-$; in other words, the maximal
		and minimal empirical eigenvalues converge almost surely to the edges of the
		support of the limiting spectral distribution \cite{bai2010spectral}.
		
		\item {\em Random Fourier frame:}
		Consider the
		random Fourier frame, in which the $\m_n$ columns of $X_{fourier}^{(n)}$ are drawn
		uniformly
		at random
		from the columns of the $n$-by-$n$ discrete Fourier transform (DFT) matrix
		(normalized s.t absolute value of matrix entries is $1/\sqrt{\m}$). 
		Farrell \cite{farrell2011limiting} has proved that the 
		empirical distribution of $\lambda(\Gk)$ converges, almost surely in
		distribution,  as $n\to\infty$ and as $m$ and $k$ grow proportionally to $n$, 
		to the so-called MANOVA
		limiting distribution, which we now describe briefly.

		The classical MANOVA$(n,\m,k,\Fc)$ ensemble\footnote{Also known as the
			beta-Jacobi ensemble with beta=$1$ (orthogonal) 
			for $\Fc=\R$, and beta=$2$ (unitary) for $\Fc=\C$.}, with $\Fc\in\left\{ \R,\C
		\right\}$ is the distribution of the random 
		matrix 
		\begin{equation}
		\label{ManovaRandomMatrix}
		\frac{n}{\m}(AA'+BB')^{-\frac{1}{2}}BB'(AA'+BB')^{-\frac{1}{2}}\,,
		\end{equation} 
		where $A_{k\x (n-\m)}, B_{k\x \m}$ are random standard Gaussian
		i.i.d matrices with entries in $\Fc$.  
		Wachter \cite{wachter1980limiting} discovered that, as $k/\m\to \beta \le 1$ and $\m/n\to\gamma$, 
		the empirical spectral
		distribution of the MANOVA$(n,\m,k,\R)$ ensemble converges, almost surely in distribution,
		to the so-called
		MANOVA$(\beta,\gamma)$ limiting
		spectral distribution\footnote{The literature uses the term MANOVA to refer both
			to the random matrix ensemble, which we denote here by MANOVA$(n,\m,k,\Fc)$,
			and to the limiting spectral distribution, which we denote here by
			MANOVA$(\beta,\gamma)$.}, whose density is given by 
		\begin{eqnarray}
		\label{ManovaDensity}
		f_{\beta,\gamma}^{MANOVA}(x)
		=\frac{\sqrt{(x-r_-)(r_+-x)}}{2\beta\pi x(1-\gamma x)}\cdot
		I_{(r_-,r_+)}(x) 
		+ \left(1+\frac{1}{\beta}-\frac{1}{\beta\gamma}\right)^+ \cdot \delta\left(x-\frac{1}{\gamma}\right) \, &&
		\end{eqnarray}
		where $(x)^+ = \max(0,x)$.
		The limiting MANOVA distribution is compactly supported on $[r_-,r_+]$  with
		\begin{equation}
		\label{ManovaDensityExtrimalValues}
		r_\pm=\bigg(\sqrt{\beta(1-\gamma)}\pm\sqrt{1-\beta\gamma}\bigg)^2\,.
		\end{equation} 
		\noindent The same holds for the MANOVA$(n,\m,k,\C)$ ensemble. 
		
		Note that the support of the MANOVA$(\beta,\gamma)$ distribution is smaller than
		that of the corresponding Mar\u cenko-Pastur law for the same aspect ratios. Figure
		\ref{fig:MANOVA_MP} shows these two densities for $\beta=0.8$ 
		and $\gamma=0.5$.
		Nevertheless, as the MANOVA dimension ratio becomes small,
		its distribution tends to the Mar\u cenko-Pastur distribution \eqref{MPdensity:eq},
		i.e., $f^{MANOVA}_{\beta,\gamma}(x) \rightarrow f^{MP}_\beta(x)$ as $\gamma \rightarrow 0$.
		Thus, a highly redundant random Fourier frame behaves like
		a Gaussian i.i.d. frame.
		\begin{figure}[h]
			\centering
			\includegraphics[width=2.7in]{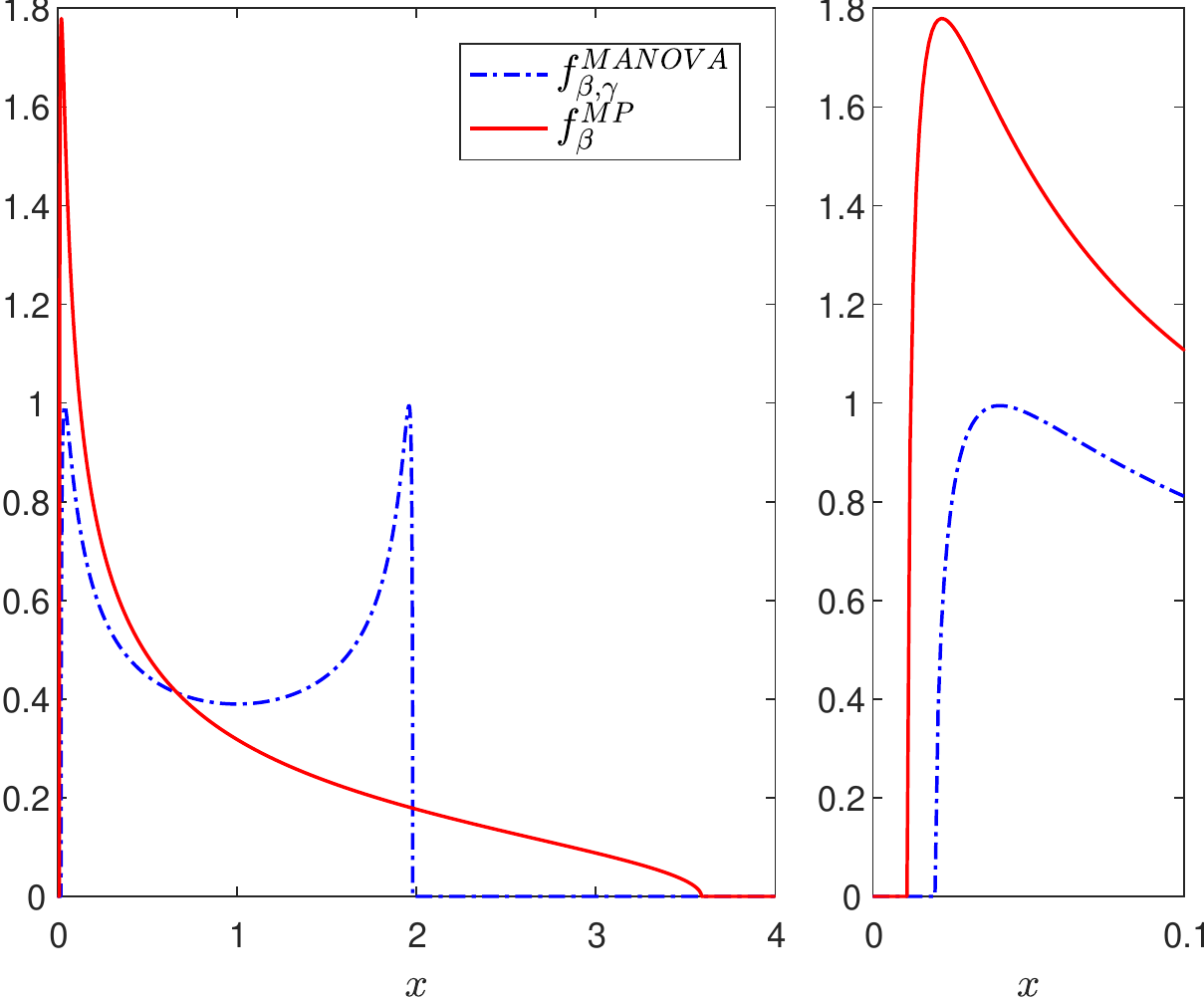}
			\caption{Limiting 
				MANOVA ($\beta=0.8,\gamma=0.5$) 
				and Mar\u cenko-Pastur ($\beta=0.8$) density
				functions. Left: density on the interval $x\in [0,4]$. Right: Zoom in on the
				interval $x\in[0,0.1]$. 
			}
			\label{fig:MANOVA_MP}
		\end{figure}
		
		\item {\em Unitary Haar frame:}
		Let $X_{haar}^{(n)}$ consist of
		the first $\m$ columns of a
		Haar-distributed $n$-by-$n$ unitary matrix normalized by $\sqrt{n/\m}$ (the Haar distribution being the
		uniform distribution over the group of $n$-by-$n$ unitary matrices).
		Edelman and Sutton \cite{edelman2008beta} proved that 
		the empirical spectral distribution of $\lambda(\Gk$) also converges, almost
		surely in distribution, to the
		MANOVA limiting spectral distribution 
		(See also \cite{wachter1980limiting} and the closing remarks of \cite{farrell2011limiting}.)
	\end{enumerate}
	The maximal and minimal eigenvalues of a matrix
	from the MANOVA$(n,\m,k,\Fc)$ ensemble ($\Fc\in\left\{ \R,\C \right\}$) 
	are known to converge almost surely 
	to $r_+$ and $r_-$, respectively \cite{johnstone2008multivariate}.
	While we are not aware of any parallel results for the random Fourier and Haar
	frames,  
	the empirical evidence in this work show that it must be the case.
	
	These random matrix phenomena have practical significance for evaluations of
	functions of the form $\specstat(\lambda(\Gk))$ such as those mentioned above.
	The functions $\specstat_{AC}$ and $\specstat_{Shannon}$, for example, are what \cite{yao2015sample}
	call {\em linear spectral statistics}, namely functions of $\lambda(\Gk)$ that
	may be written as an integral of a scalar function against the empirical measure of $\lambda(\Gk)$.
	Convergence of the empirical
	distribution of $\lambda(\Gk^{(n)})$ to the limiting 
	MANOVA distribution with density $f^{MANOVA}_{\beta,\gamma}$ 
	implies 
	\begin{eqnarray} \label{ac_limit:eq}
	\lim_{n\to\infty} \specstat_{AC}(\lambda(\Gkn^{(n)})) &=& \int \frac{1}{x}
	\,f_{\beta,\gamma}^{MANOVA}(x)dx \\
	\lim_{n\to\infty} \specstat_{Shannon}(\lambda(\Gkn^{(n)})) &=&  \int \log(1+\alpha x)
	\,f_{\beta,\gamma}^{MANOVA}(x)dx \nonumber 
	\end{eqnarray}
	for both the random Fourier and Haar frames; 
	the integrals on the right hand side may be evaluated explicitly.
	Similarly, convergence of 
	$\lambda_{max}(\Gk)$ and $\lambda_{min}(\Gk)$ to $r_+$ and $r_-$ implies, for
	example, that 
	\begin{eqnarray} \label{rip_limit:eq}
	\lim_{n\to\infty} \specstat_{RIP}(\lambda(\Gk^{(n)})) = \max(r_+-1,1-r_-)\,.
	\end{eqnarray}
	
\begin{figure*}[t]
		\centering
		\includegraphics[width=2.1in]{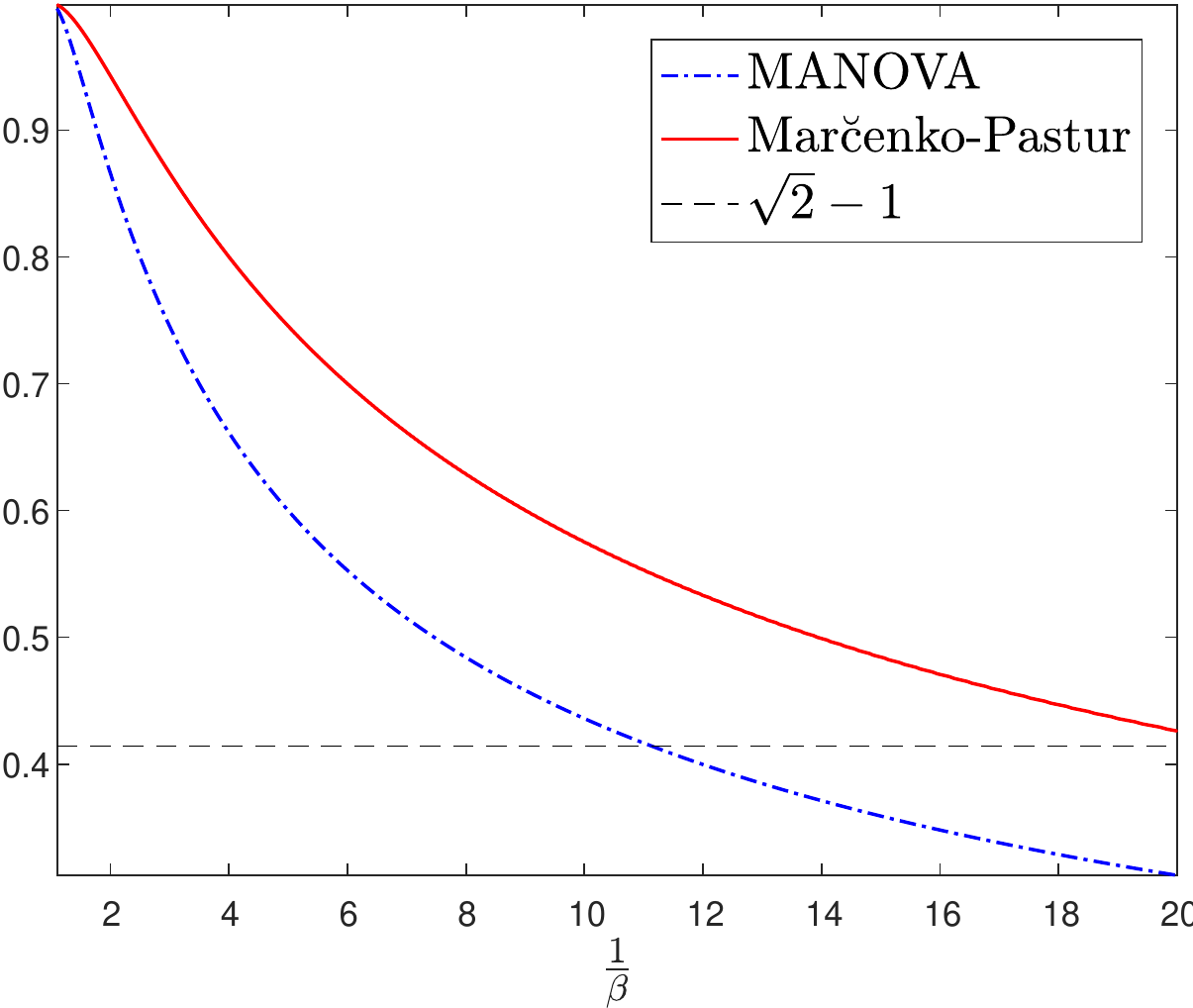}
		\includegraphics[width=2.1in]{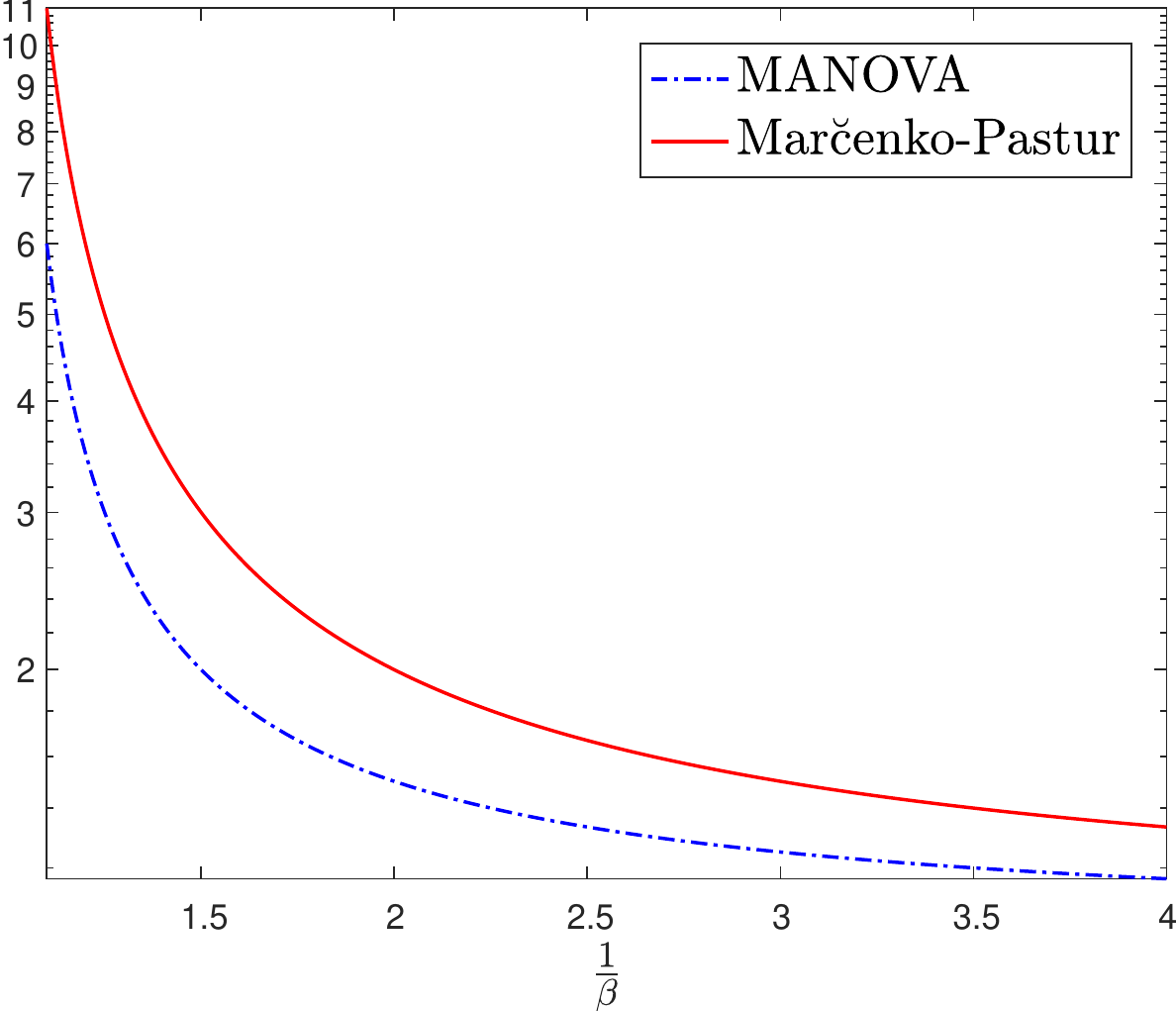}
		\includegraphics[width=2.1in]{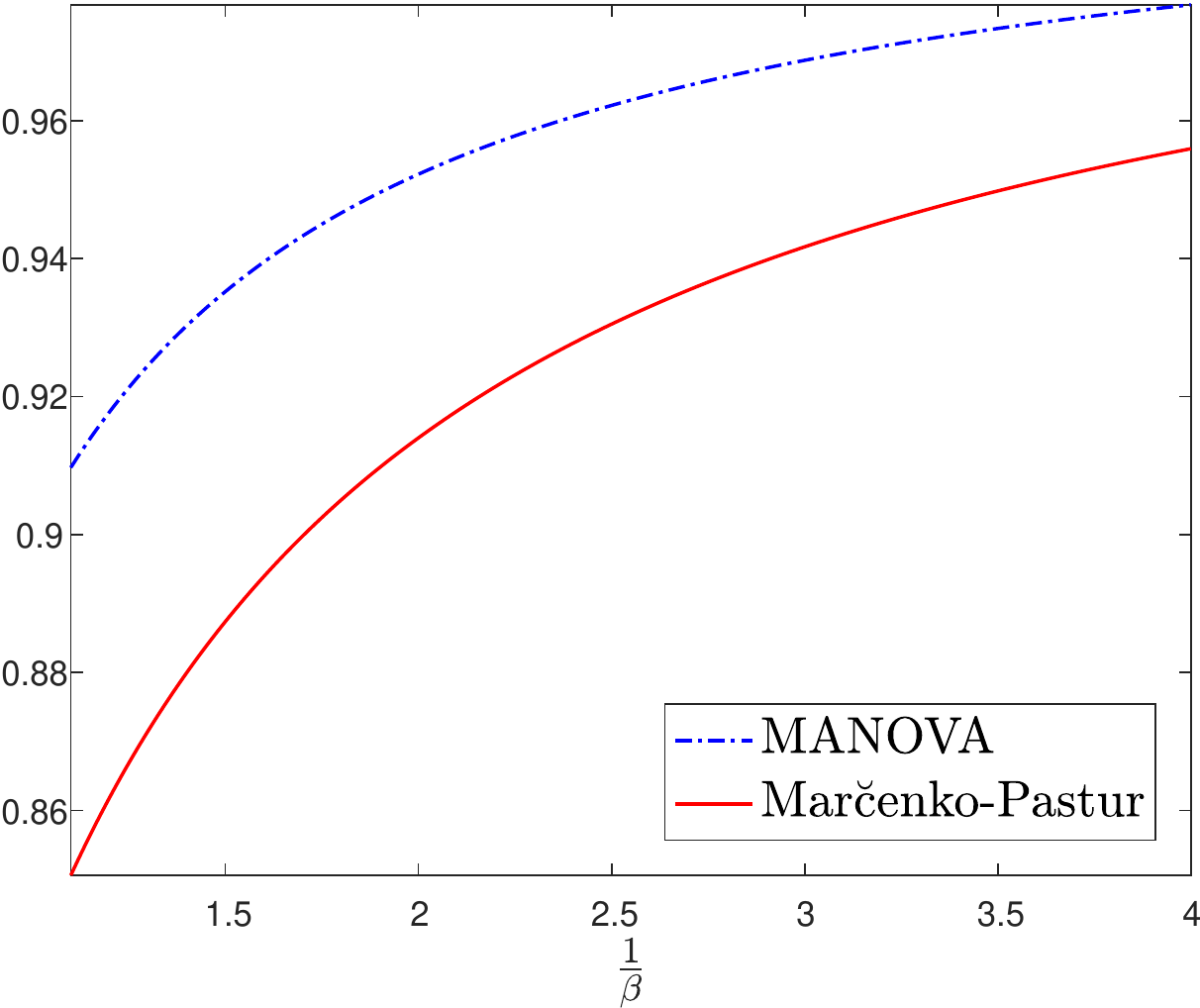}
		\caption{Comparison of limiting values of $\E_K \specstat(\lambda(G_K))$ for the three
			functions $\specstat$ discussed in Motivation Section between the Mar\u cenko-Pastur limiting distribution and the MANOVA
			distribution. Left: $\specstat_{RIP}$ (lower is better). 
			Middle: $\specstat_{AC}$ (lower is
			better). Right: $\specstat_{Shannon}$ (higher is better).
		}
		\label{fig:limiting_F}
	\end{figure*}

	To demonstrate why such calculations are significant,
	we note that Equations \eqref{ac_limit:eq} and \eqref{rip_limit:eq} 
	immediately allow us to compare the Gaussian i.i.d frame
	with the random Fourier and Haar frames, 
	in terms
	of their limiting value of functions of interest.
	Figure \ref{fig:limiting_F} compares the limiting value of $\specstat_{RIP}$,
	$\specstat_{AC}$ and $\specstat_{Shannon}$ over varying values of $\beta=\lim_{n\to \infty}
	k/\m$. The plots 
	clearly demonstrate that frames whose typical
	$k$-submatrix exhibits a MANOVA spectrum, are superior to frames whose typical
	$k$-submatrix exhibits a Mar\u cenko-Pastur spectrum, 
	across the performance measures.

	\section{Deterministic Frames: Universality Hypothesis}
	\label{sec:deterministic_frames}
	
	Deterministic frames, namely
	frames whose design involves no randomness, have so far eluded this kind of
	asymptotically exact analysis. 
	While
	there are results regarding RIP \cite{bandeira2013road,fickus2015group} and statistical RIP
	\cite{calderbank2010construction,gurevich2008incoherent,mazumdar2011general}, for example, of deterministic
	frame designs, they are mostly focused on highly redundant frames ($\gamma \rightarrow 0$)
	and the wide submatrix ($\beta \rightarrow 0$) case, where the spectrum tends
	to the Mar\u cenko-Pastur distribution. Furthermore, nothing analogous, say, to the precise comparisons of Figure
	\ref{fig:limiting_F} exists in the literature to the best of our knowledge. 
	Specifically, no results analogous to \eqref{ac_limit:eq} and \eqref{rip_limit:eq}
	are known for deterministic frames, 
	let alone the associated convergence rates, if any.
	
	In order to subject deterministic frames to an asymptotic analysis, we shift our
	focus from a single frame $X$ to a family of deterministic frames $\{X^{(n)}\}$
	created by a common construction. The frame matrix $X^{(n)}$ is $n$-by-$m$.
	Each frame family determines allowable sub-sequences $(n,m)$; to
	simplify notation, we leave the subsequence implicit and index the frame
	sequence simply by $n$. The frame family
	also determines the aspect ratio limit $\gamma=\lim_{n\to\infty} \m/n$.
	In what follows we also fix a sequence $k$ with $\beta=\lim_{n\to\infty}
	k/\m $,
	and let $K\subset [n]$ denote a uniformly distributed random subset.
	
	\paragraph{Frames under study.} 
	
	The different frames that we studied are listed in Table \ref{FramesTable},
	in a manner inspired by \cite{monajemi2013deterministic}.
	In addition to our deterministic frames of interest 
	(the set ${\cal X}$),
	the table contains also two examples of random frames
	(real and complex variant for each), 
	for validation and convergence analysis purposes.
	\begin{table*}[t]
		\scriptsize
		\centering
		\caption{Frames under study}
		\label{FramesTable}
		\renewcommand{\arraystretch}{0.85}%
		\begin{tabular}{lllllll}
			\toprule
			Label & Name & $\R$ or $\C$ & Natural $\gamma$ & Tight frame &
			Equiangular & References \\
			\midrule
			& & & & &  & \\
			{\bf Deterministic frames} & & & & &  & \\
			DSS & Difference-set spectrum & $\C$  & & Yes & Yes & \cite{xia2005achieving}\\
			GF & Grassmannian frame & $\C$ & $1/2$& Yes & Yes &  \cite[Corollary 2.6b]{strohmer2003grassmannian}
			\\
			RealPF & Real Paley's construction & $\R$ & $1/2$ & Yes & Yes & \cite[Corollary 2.6a]{strohmer2003grassmannian}
			\\
			ComplexPF & Complex Paley's construction & $\C$ & $1/2$ & Yes & Yes & 
			\cite{paley1933orthogonal}
			\\
			Alltop & Quadratic Phase Chirp & $\C$ & $1/L$ & Yes & No &  
			\cite[eq. S4]{monajemi2013deterministic} 
			\\
			SS & Spikes and Sines & $\C$ & $1/2$ & Yes & No & \cite{Elad2010} \\
			SH & Spikes and Hadamard & $\R$  & $1/2$ & Yes & No & \cite{Elad2010}
			~ \\\hline  ~\\
			{\bf Random frames} & & & & &  & \\
			HAAR 
			& Unitary Haar frame & $\C$ & & Yes & No & \cite{farrell2011limiting,edelman2008beta} \\
			RealHAAR 
			& Orthogonal Haar frame & $\R$ & & Yes & No & \cite{edelman2008beta} \\
			RandDFT  
			& Random Fourier transform & $\C$ & & Yes & No & \cite{farrell2011limiting} \\
			RandDCT  
			& Random Cosine transform & $\R$ & & Yes & No &  \\
			
			\bottomrule
		\end{tabular}
		\label{frames:tab}
	\end{table*}
	
	\paragraph{Functionals under study.}
	
	We studied the functionals $\specstat_{StRIP}$ from \eqref{strip_func:eq}, 
	$\specstat_{AC}$ from \eqref{ac_func:eq}, $\specstat_{Shannon}$ from
	\eqref{shannon_func:eq}. In addition, we studied the maximal and minimal
	eigenvalues of $\Gk$, and its condition number:
	\begin{eqnarray*}
		\specstat_{max}(\lambda(\Gk)) &=&  \lambda_{max}(\Gk) \\
		\specstat_{min}(\lambda(\Gk)) &=&  \lambda_{min}(\Gk) \\
		\specstat_{cond}(\lambda(\Gk)) &=&  \lambda_{max}(\Gk) / \lambda_{min}(\Gk)\,. 
	\end{eqnarray*}

\begin{figure}[h]
		\centering
		\includegraphics[width=2.5in]
		{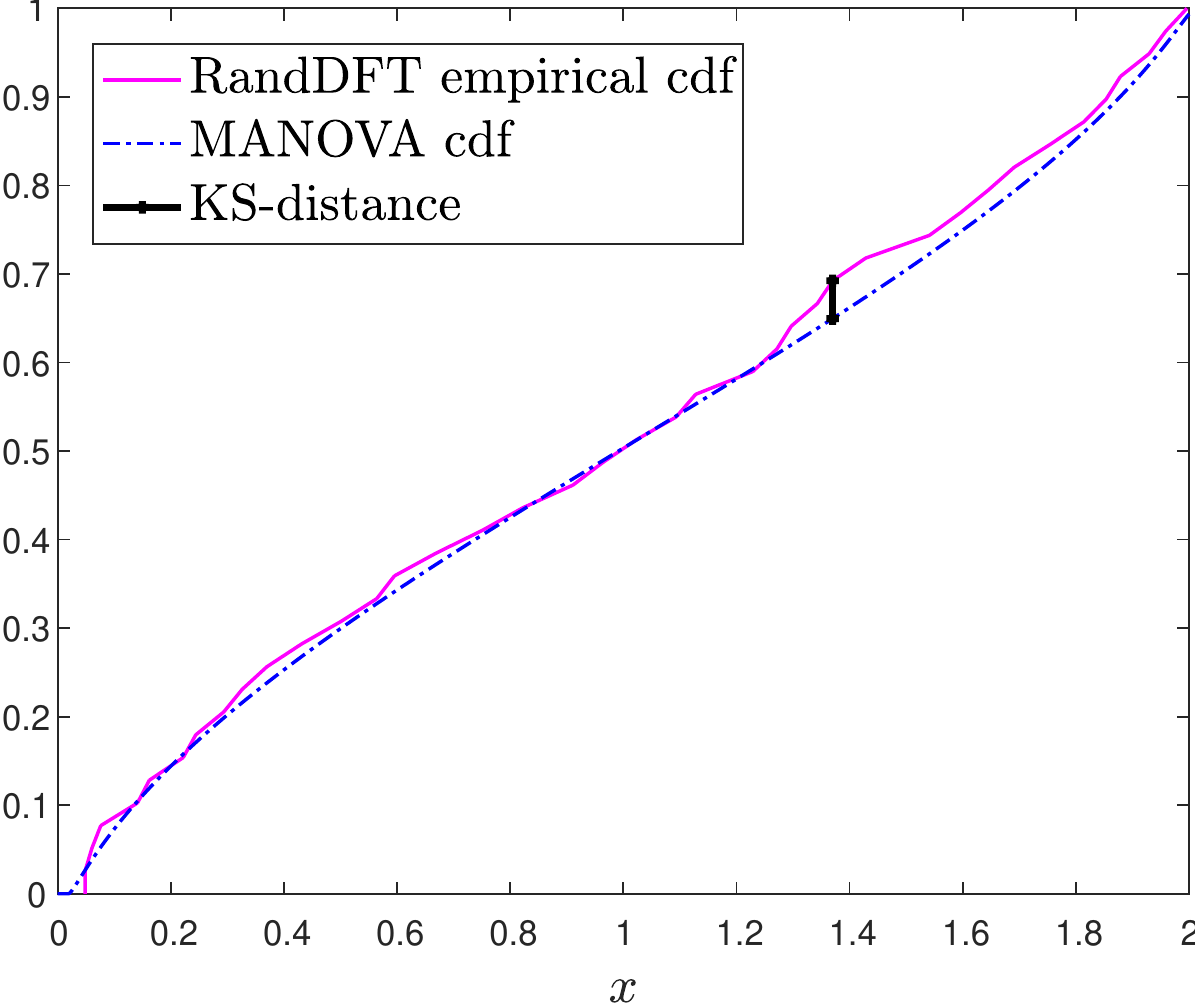}
		\caption{KS-distance of random DFT subframe, $\beta = 0.8$, $\gamma = 0.5$, $n=100$.}
		\label{fig:CDFs}
	\end{figure}

	\paragraph{Measuring the rate of convergence.}
	In order to quantify the rate of convergence of the entire spectrum 
	of the $k$-by-$\m$ matrix 
	$\Xk$, which is a $k$-submatrix of an $n$-by-$\m$ frame matrix $X$, to a limiting
	distribution, we let $F[\Xk]$ denote the
	empirical cumulative distribution function (CDF) of $\lambda(\Gk)$, and 
	let $F^{MANOVA}_{\beta,\gamma}(x) = \intop_{r_-}^x
	f^{MANOVA}_{\beta,\gamma}(x)dx$ denote the CDF of the MANOVA$(\beta,\gamma)$
	limiting distribution. 
	The quantity
	\[
	\Delta_{KS}(\Xk) = \norm{F[\Xk] - F^{MANOVA}_{\beta_n,\gamma_n}}_{KS}\,,
	\]
	where $\norm{\cdot}_{KS}$ is the Kolmogorov-Smirnov (KS) distance between CDFs, 
	measures the distance to the hypothesised limit. Here, 
	$\beta_n=k/m$ and $\gamma_n=m/n$ are the actual aspect ratios for the matrix $\Xk$ at hand.
	As a baseline we use $\Delta_{KS}( Y_{n,\m,k,\Fc})$, where 
	$Y_{n,\m,k,\Fc}$ is a matrix from the MANOVA$(n,\m,k,\Fc)$ ensemble,
	with $\Fc=\R$ if $X_K$ is real and $\Fc=\C$ if complex.
	Figure \ref{fig:CDFs} illustrates the KS-distance 
	between an empirical CDF and the limiting MANOVA CDF.
	
	Similarly, in order to quantify the rate of convergence of a functional
	$\specstat$, the quantity
	
	is the distance between the measured value of $\Psi$ on a given $k$-submatrix
	$\Xk$ and its hypothesised limiting value. 
	For a baseline we can use $\Delta_\specstat(Y_{n,\m,k,\Fc})$,
	with $\Fc=\R$ if $X_K$ is real and $\Fc=\C$ if complex.
	For linear spectral functionals 
	like $\Psi_{AC}$ and $\Psi_{Shannon}$, which may be written as
	$\Psi(\lambda(\Gk))=\int \psi dF[X_K]$ for some kernel $\psi$, we have
	$  \Psi(f^{MANOVA}_{\beta,\gamma}) = \int \psi dF^{MANOVA}_{\beta,\gamma}$. 
	For $\Psi_{RIP}$ that depends on $\lambda_{max}(\Gk)$ 
	and $\lambda_{min}(\Gk)$ we have $\Psi_{RIP}(f^{MANOVA}_{\beta,\gamma}) =
	\max\left\{ r_+-1,1-r_- \right\}$.
	
	\paragraph{Universality Hypothesis.} 
	
	The contributions of this work are based on the following assertions on the typical $k$-submatrix ensemble $\Xk$ corresponding to a frame
	family $X^{(n)}$. This family may be random or deterministic, real or complex. 
	
	\begin{enumerate}
		\item[{\bf H1}] {\em Existence of a limiting spectral distribution.} 
		The empirical spectral distribution of $\Xk^{(n)}$, namely the 
		distribution of $\lambda(\Gk^{(n)})$, converges, as $n\to \infty$, 
		to a compactly-supported limiting distribution; furthermore, 
		$\lambda_{max}(\Gk^{(n)})$ and $\lambda_{min}(\Gk^{(n)})$ converge to the
		edges of that compact support.
		
		\item[{\bf H2}] {\em Universality of the limiting spectral distribution.} 
		The limiting 
		spectral distribution of $\Xk^{(n)}$ is the 
		MANOVA$(\beta,\gamma)$ distribution  \cite{wachter1980limiting} whose density is 
		\eqref{ManovaDensity}. Also 
		$\lambda_{max}(\Gk^{(n)})\to r_+$ and $\lambda_{min}(\Gk^{(n)})\to r_-$
		where $r_\pm$ is given by \eqref{ManovaDensityExtrimalValues}.

		\item[{\bf H3}] {\em Exact power-law rate of convergence for the entire
			spectrum.}
		The spectrum of $\Xk^{(n)}$ converges to the limiting
		MANOVA$(\beta,\gamma)$ distribution
		\begin{eqnarray*} 
			\left(\E_{K_n}\left(\Delta_{KS}(\Xk^{(n)})\right)\right)^2 \searrow 0
		\end{eqnarray*}
		and in fact its fluctuations are given by the law
		\begin{eqnarray} \label{KS_conv:eq}
		Var_{K}(\Delta_{KS}(\Xk^{(n)}))=Cn^{-2b} 
		\end{eqnarray}
		for some constants $C,b$, which may depend on the frame family.
		
		\item[{\bf H4}] {\em Universality of the rate of convergence for the
			entire spectrum of ETFs.}
		For an equiangular tight frame (ETF) family,  
		the exponent
		$b$ in \eqref{KS_conv:eq} is universal and does not depend on the frame. 
		Furthermore, 
		\eqref{KS_conv:eq} also holds, with the same universal exponent,
		replacing $\Gk^{(n)}$ with a same-sized matrix
		from the MANOVA$(n,\m,k,\Fc)$ distribution 
		defined in \eqref{ManovaRandomMatrix}, with $\Fc=\R$ if $X^{(n)}$ is a real
		frame family, and $\Fc=\C$ if complex. 
		In other words, the universal exponent $b$ for ETFs 
		is a property of the MANOVA
		(Jacobi) random matrix ensemble.
		
		\item[{\bf H5}] {\em Exact power-law rate of convergence for functionals.}
		For a ``nice'' functional $\specstat$, the value of
		$\specstat(\lambda(\Gk^{(n)}))$  converges
		to $\specstat(f^{MANOVA}_{\beta,\gamma})$ according to the law
		\begin{eqnarray} \label{func_conv:eq}
		\E_{K}(\Delta_\specstat(\Xk^{(n)})^2)=Cn^{-b}\log^{-a}(n)
		\end{eqnarray}
		for some constants $C,b,a$.
		
		\item[{\bf H6}]  {\em Universality of the rate of convergence for functionals.}
		While the constant $C$ in \eqref{func_conv:eq} may depend on the frame, the
		exponents $a,b$ are universal. \eqref{func_conv:eq} also holds, with the same
		universal exponents, replacing $\Gk^{(n)}$ with a same-sized matrix from the
		MANOVA$(n,\m,k,\Fc)$ ensemble defined in \eqref{ManovaRandomMatrix}, with
		$\Fc=\R$ if $X^{(n)}$ is a real frame family, and $\Fc=\C$ if complex.  In
		other words, the universal exponents $a,b$ are a property of the MANOVA
		(Jacobi) random matrix ensemble.

	\end{enumerate}
	
	\paragraph{Nonstandard aspect ratio $\beta>1$.}
	While the classical MANOVA ensemble and limiting density are not defined for
	$\beta>1$, in our case it is certainly possible to sample $k>m$ vectors from the
	$n$ possible frame vectors, resulting in a situation with $\beta>1$.
	In this situation, the hypotheses above require slight modifications.
	Specifically, the limiting 
	spectral distribution of $\Xk^{(n)}$, for $\beta>1$, is
	\begin{equation}
	\label{ManovaDensityBeta}
	\left(1-\frac{1}{\beta}\right) \delta(x)+f_{\beta,\gamma}^{MANOVA}(x)\,,
	\end{equation}
	where $f_{\beta,\gamma}^{MANOVA}(x)$ is the function (no longer a density) 
	defined in \eqref{ManovaDensity}.
	The rate of convergence of the distribution of nonzero eigenvalues to the
	limiting density
	$\frac{1}{\beta}f_{\frac{1}{\beta},\beta\gamma}^{MANOVA}(\frac{1}{\beta}x)=\beta
	f_{\beta,\gamma}^{MANOVA}(x)$ is compared with 
	the baseline $\beta\cdot Y_{n,k,\m,\Fc}$, where $Y_{n,k,\m,\Fc}$ is a matrix from the MANOVA$(n,k,\m,\Fc)$ ensemble
	(i.e., with reversed order of $k$ and $m$).
	
	
	
	{\small
		\section{Methods} \label{sec:methods}
		
		The software we developed has been permanently deposited in the Data and Code
		Supplement \cite{SDR}.  As many of the deterministic frames under study are only
		defined for $\gamma=0.5$, we primarily studied the aspect ratios
		$(\gamma=0.5,\beta)$ with $\beta\in\{0.3,0.5,0.6,0.7,0.8,0.9\}$.  In addition,
		we inspected all frames under study that are defined for the aspect ratios
		$(\gamma=0.25,\beta=0.6)$ and $(\gamma=0.25,\beta=0.8)$ (all random frames, as
		well as DSS and Alltop).  We also studied nonstandard aspect ratios $\beta>1$ as
		described in the Supporting Information \cite{SI}.  For deterministic frames,
		$n$ took allowed values in the range $(240,2000)$, $(2^5,2^{12})$ for
		Grassmannian and Spikes and Hadamard frames and $(600,4000)$ for DSS frame with
		$\gamma=0.25$.  For random frames and MANOVA ensemble we used dense grid of
		values in the range $(240,2000)$.  Hypothesis testing as discussed below, was
		based on a subset of these values where $n\ge 1000$.  For each of the frame
		families under study, and for each value of $\beta$ and $\gamma$ under study, we
		selected a sequence $(n,\m,k)$.  The values $n$ and $m$ were selected so that
		$m/n$ will be as close as possible to $\gamma$, however due to different aspect
		ratio constrains by the different frames occasionally we had $m/n$ close but not
		equal to $\gamma$.  We then determined $k$ such that $k/\m$ will be as close as
		possible to $\beta$.  For each $n$, we generated a single $n$-by-$\m$ frame
		matrix $X^{(n)}$.  We then produced $T$ independent samples from the uniform
		distribution on $k_n$-subsets, $K[1],\ldots,K[T]\subset[n]$, and generated their
		corresponding $k$-submatrices $X_{K[i]}^{(n)}$ ($1\leq i\leq T$). Importantly,
		all these are submatrices of the same frame matrix $X^{(n)}$.  We calculated
		$\overline{\Delta}^{Var}_{KS}(\Xk^{(n)})=\overline{\Delta^2}_{KS}(\Xk^{(n)})-\overline{\Delta}^2_{KS}(\Xk^{(n)})$,
		the empirical variance of $ \Delta_{KS}(X_{K[i]}^{(n)})$, and
		$\overline{\Delta^2}_{KS}(\Xk^{(n)})$, the average value of $
		\Delta^2_{KS}(X_{K[i]}^{(n)})$ on $1\leq i\leq T$, as a monte-carlo
		approximation  to the left-hand side of \eqref{KS_conv:eq}, variance and MSE
		respectively. For each of the functionals under study, we also calculated $
		\overline{\Delta^2}_\specstat(X_{K[i]}^{(n)})$, the average value of $
		\Delta^2_\specstat(X_{K[i]}^{(n)})$ on $1\leq i\leq T$, as a monte-carlo
		approximation to the left-hand size of \eqref{func_conv:eq}.
		
		Separately, for each triplet ($n$,$m$,$k$) and $\Fc\in\left\{ \R,\C \right\}$ we
		have performed $T$ independent draws from the MANOVA$(n,\m,k,\Fc)$ ensembles
		(\ref{ManovaRandomMatrix}) and calculated analogous quantities
		$\overline{\Delta}^{Var}_{KS}(Y_{n,\m,k,\Fc})$,
		$\overline{\Delta^2}_{KS}(Y_{n,\m,k,\Fc})$ and
		$\overline{\Delta^2}_{\specstat}(Y_{n,\m,k,\Fc})$.

		\paragraph{Test 1: Testing H1--H4.}
		For each of the frames under study and each value of $(\beta,\gamma)$, we computed the KS-distance for $T=10^4$
		submatrices and performed simple linear regression
		of 
		$-\frac{1}{2}\log\left( \overline{\Delta}^{Var}_{KS}(\Xk^{(n)})\right)$ 
		on $\log(n)$ with an intercept. We obtained 
		the estimated linear coefficient $\hat{b}$ as an estimate
		for the exponent $b$, and its standard error $\sigma(\hat{b})$.
		Similarly we regressed 
		$-\frac{1}{2}\log\left( 
		\overline{\Delta}^{Var}_{KS}(Y_{n,\m,k,\Fc})\right)$
		on $\log(n)$ to obtain $\hat{b}_{MANOVA}$ and $\sigma(\hat{b}_{MANOVA})$.  
		We performed Student's
		t-test to test the null hypotheses $b=b_{MANOVA}$ using
		the test
		statistic 
		\[
		t = \frac{\hat{b}-b_{MANOVA}}
		{\sqrt{\sigma(\hat{b})^2+\sigma(\hat{b}_{MANOVA})^2 }}\,.
		\]
		Under the null hypothesis, 
		the test statistic is distributed $t_{(N+N_{MANOVA}-4)}$, 
		where $N$, $N_{MANOVA}$ are the numbers of different values of $n$ for which we have collected
		the data for a frame and the MANOVA ensemble respectively.
		We report the $R^2$ of
		the linear fit; the slope coefficient $\hat{b}$ and its standard error; and the
		p-value of the above t-test.
		We next regressed $-\log \left(\overline{\Delta^2}_{KS}\right)$ on $\log(n)$. 
		Since $\overline{\Delta^2}_{KS} = \left( \overline{\Delta}_{KS}
		\right)^2+\overline{\Delta}_{KS}^{Var}$, a linear fit verifies that 
		$ \left( \overline{\Delta}_{KS}
		\right)^2 \searrow 0$.
		
		\paragraph{Test 2: Testing H5--H6.}
		
		For each of the frames under study, each of the functionals $\specstat$ 
		under study, and 
		each value of $(\beta,\gamma)$, we computed the empirical value of the
		functionals on $T=10^3$ submatrices.
		We first performed linear regression
		of 
		$-\log\left( \overline{\Delta^2}_{\specstat}(Y_{n,\m,k,\Fc})\right)$
		on $\log(n)$ and $\log(\log(n))$ with an intercept, for $\Fc\in\left\{ \R,\C
		\right\}$. 
		Let $a_0$ denote the fitted coefficient for $\log(n)$ and let $b_0$ denote the
		fitted coefficient for $\log(\log(n))$. 
		This step was based on triplets $(n,m,k)$ yielding accurate aspect ratios in
		the range $240\le n\le 2000$.
		We then performed simple linear regression
		of 
		$-\log\left( \overline{\Delta^2}_{\specstat}(\Xk^{(n)};n,\m,k)\right)$
		on 
		$\log(n) + (a_0/b_0)\cdot \log(\log(n))$.
		The estimated linear regression coefficient $\hat{b}$
		is the estimate 
		for the exponent $b$ in \eqref{func_conv:eq}, 
		and $\sigma(\hat{b})$
		is its standard error. We used 
		$\hat{b}\cdot(a_0/b_0)$ as an
		estimate for the exponent $a$ in
		\eqref{func_conv:eq}.
		We proceeded as above to test the null hypothesis $b=b_0$.
		We report the $R^2$ of
		the linear fit; the slope coefficient $\hat{b}$ and its standard error; and the
		p-value of the test above.
		
		\paragraph{Computing.}
		
		To allow the number of monte-carlo samples to be as large as $T=10^4$ 
		and $n$ to be as large as $2000$, we used a large Matlab cluster 
		running on Amazon Web Services.
		We used 32-logical core machines, with 240GB RAM each, 
		which were running several hundred hours in total. 
		The code we executed has been
		deposited \cite{SDR}; it
		may easily be executed for smaller values of $T$ and $n$ on smaller machines.

	}
	
	
	\section{Results} \label{sec:results}
	
	The raw results obtained in our experiments, as well as the analysis results 
	of each experiment, have been deposited with their generating code \cite{SDR}.
	
\begin{figure}[h]
		\centering
		\includegraphics[width=3.0in]
		{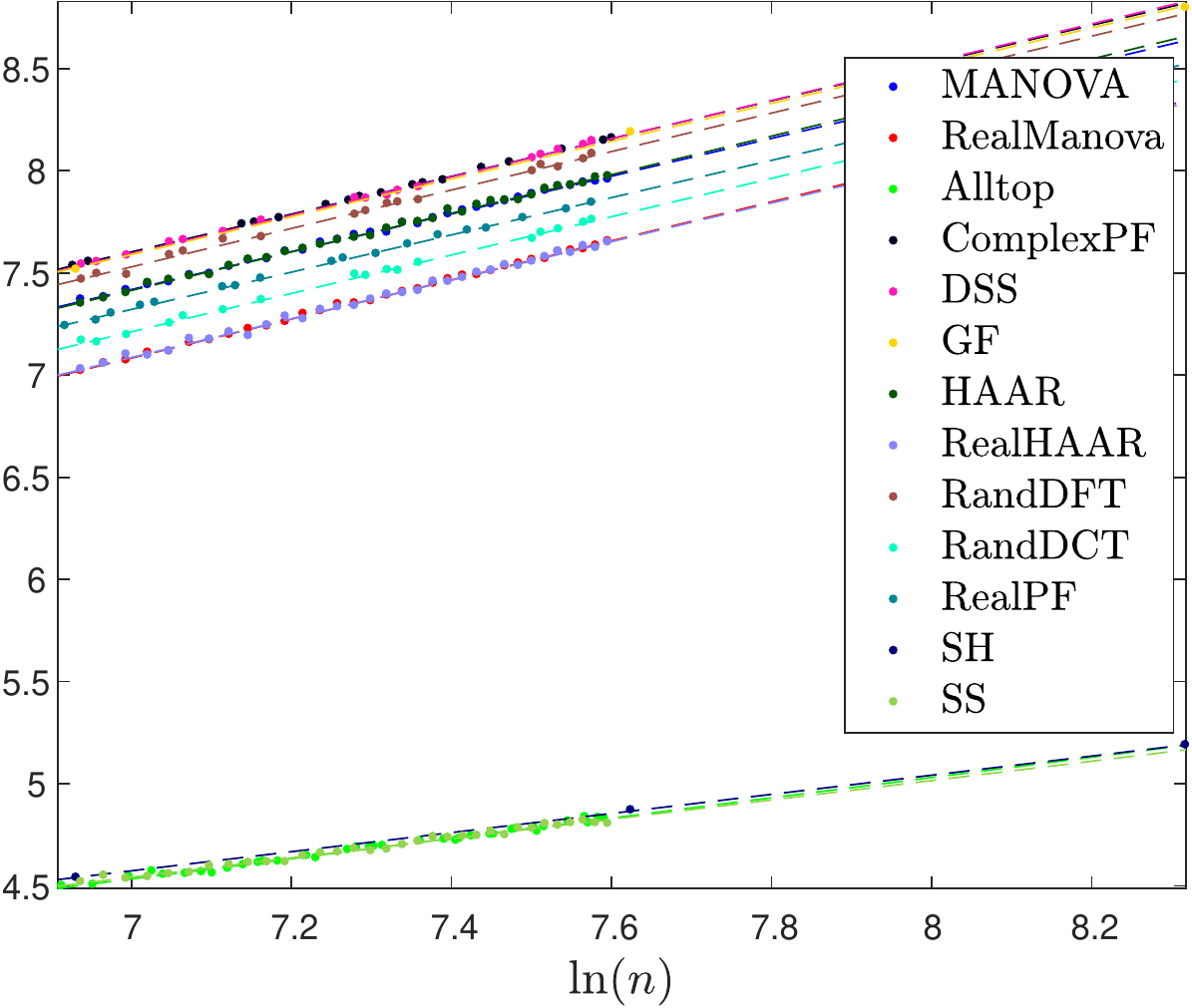}
		\caption{Test 1 for $\gamma=0.5$ and $\beta=0.8$. Plot shows $-\frac{1}{2}\ln Var_{K}(\Delta_{KS}(\Xk^{(n)}))$ over $\ln(n)$. }
		\label{fig:KS1}
	\end{figure}

\begin{figure}[h]
		\centering
		\includegraphics[width=3.0in]{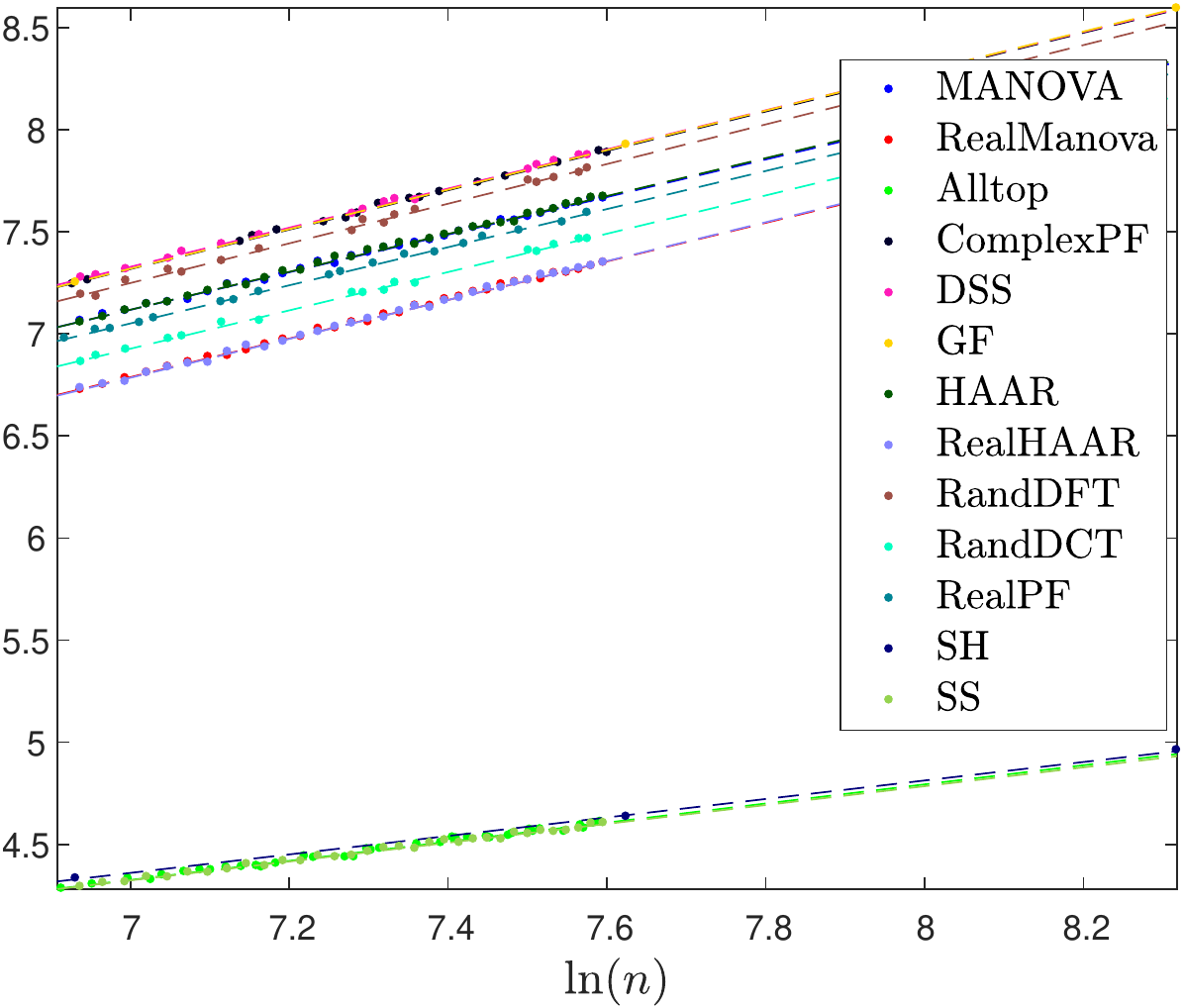}
		\caption{Test 1 for $\gamma=0.5$ and $\beta=0.6$. Plot shows $-\frac{1}{2}\ln Var_{K}(\Delta_{KS}(\Xk^{(n)}))$ over $\ln(n)$. }
		\label{fig:KS2}
	\end{figure}

\begin{figure}[h]
		\centering
		\includegraphics[width=3.0in]{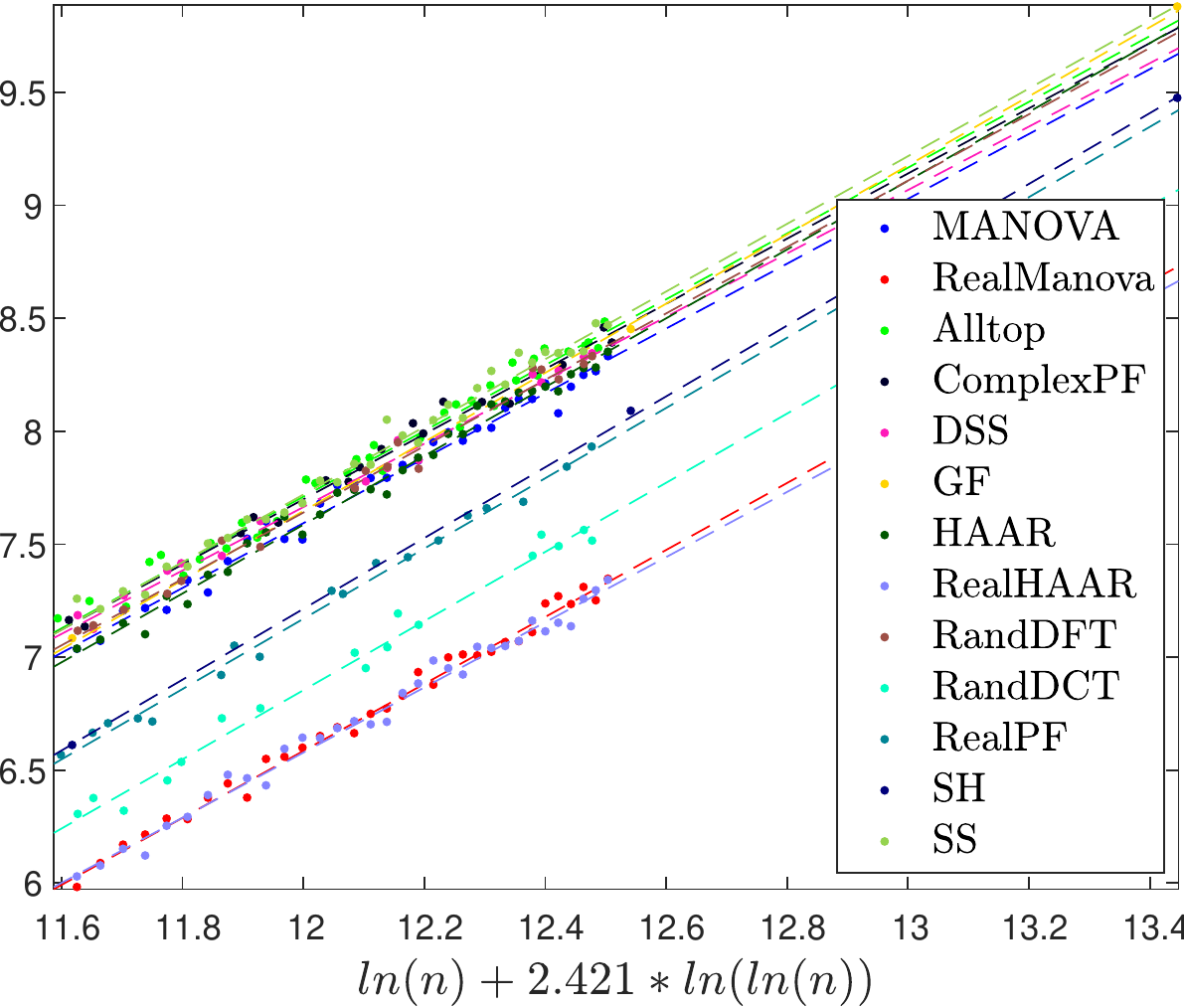}
		\caption{Test 2 for $\specstat_{AC}$, $\gamma=0.5$ and $\beta=0.8$. Plot shows $-\ln\E_{K}(\Delta_\specstat(\Xk^{(n)})^2)$
		}
		\label{fig:f_AC}
	\end{figure}

\begin{figure}[h]
		\centering
		\includegraphics[width=3.0in]{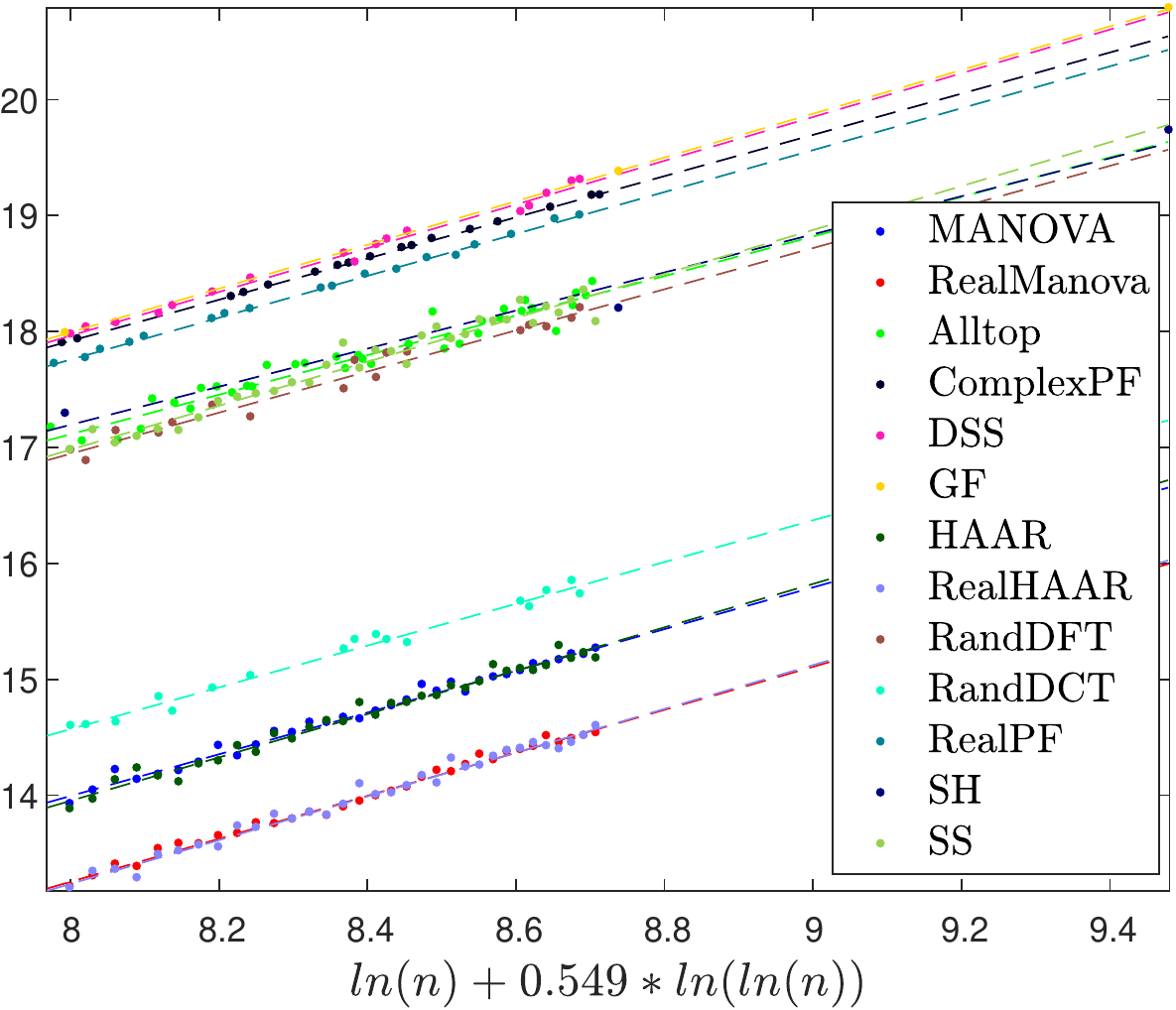}
		\caption{Test 2 for $\specstat_{Shannon}$, $\gamma=0.5$ and $\beta=0.8$. Plot shows $-\ln\E_{K}(\Delta_\specstat(\Xk^{(n)})^2)$
		}
		\label{fig:f_Shannon}
	\end{figure}

	For space considerations, the full documentation of our results is deferred to
	the Supporting Information \cite{SI}.  To offer a few examples, Figure
	\ref{fig:KS1} and Table \ref{tab:KS1} show the linear fit to
	$\overline{\Delta}^{Var}_{KS}$ for  $(\gamma=0.5,\beta=0.8)$.  Figure
	\ref{fig:KS2} shows the linear fit to $\overline{\Delta}^{Var}_{KS}$ for a
	different value of $\beta$, namely  $(\gamma=0.5,\beta=0.6)$.  Figure
	\ref{fig:f_AC} shows the linear fit to $\overline{\Delta}_{\specstat_{AC}}$ for
	$(\gamma=0.5,\beta=0.8)$.  Figure \ref{fig:f_Shannon} and Table
	\ref{tab:f_Shannon} show the linear fit to
	$\overline{\Delta}_{\specstat_{Shannon}}$ for  $(\gamma=0.5,\beta=0.8)$. Similar
	figures and tables for the other values $(\gamma,\beta)$, in particular,
	$(\beta=0.3,\gamma=0.5)$, $(\beta=0.5,\gamma=0.5)$,  $(\beta=0.7,\gamma=0.5)$,
	$(\beta=0.9,\gamma=0.5)$, $(\beta=0.6,\gamma=0.25)$, $(\beta=0.8,\gamma=0.25)$,
	are deferred to the Supporting Information.  Note that in all coefficient
	tables, both those shown here and those deferred to the Supporting Information,
	upper box shows complex frames (with t-test comparison to the complex MANOVA
	ensemble of the same size, denoted ``MANOVA'') and bottom box shows real
	frames (with t-test comparison to the real MANOVA ensemble of the same size,
	denoted ``RealMANOVA'').  In each box, top rows are deterministic frames and
	bottom rows are random frames.  Further note that in plots for Test 2 the
	horizontal axis is slightly different for real and complex frames, as the
	preliminary step described above was performed separately for real and
	complex frames. In the interest of space, we plot all frames over the
	horizontal axis calculated for complex frames.
	
	\paragraph{Validation on random frames.} While our primary interest was in
	deterministic frames, we included in the frames under study random frames. For
	the complex Haar frame and random Fourier frame, convergence of the empirical
	CDF of the spectrum to the limiting MANOVA$(\beta,\gamma)$ distribution has been
	proved in \cite{farrell2011limiting,edelman2008beta}. To our surprise, not only was our framework
	validated on the four random frames under study, in the sense of asymptotic
	empirical spectral distribution, but all universality hypotheses {\bf H1--H6}
	were accepted (not rejected at the 0.001 significance level, with very few
	exceptions).
	
	\paragraph{Test results on deterministic frames.} A tabular summary of our
	results, per hypothesis and per frame under study, is included for convenience
	in the Supporting Information.  Universality Hypotheses {\bf H1--H3} were
	accepted on all deterministic frames.  For {\bf H1--H2}, convergence of the
	empirical spectral distribution to the MANOVA$(\beta,\gamma)$ limit has been
	observed in all cases. For {\bf H3}, the linear fit in all cases was excellent
	with $R^2>0.99$ without exception, confirming the power law in
	\eqref{KS_conv:eq} and the polynomial decrease of $\overline{\Delta^2}_{KS}$
	with $n$.  Universality Hypothesis {\bf H4} was accepted (not rejected) for
	deterministic equiangular tight frames (ETFs) at the 0.001 significance level,
	with few exceptions (see Table \ref{tab:KS1} below, as well as full results and
	summary table in the Supporting Information); it was rejected for deterministic
	non-ETFs. For $\gamma=0.25$, Hypothesis {\bf H4} has also been accepted for the
	Alltop frame, see Supporting Information.  Universality Hypothesis {\bf H5} was
	accepted for all deterministic frames, with excellent linear fits ($R^2>0.97$
	without exception), confirming the power law in  \eqref{func_conv:eq}.
	Universality Hypothesis {\bf H6} was accepted (not rejected) at the 0.001
	significance level (and even 0.05 with few exceptions) for all deterministic
	frames.
	For the reader's convenience, Table \ref{ExpSummary} summarizes the universal
	exponents for convergence of the entire spectrum ({\bf H4}) and the universal
	exponents for convergence of the functionals under study ({\bf H6}), for
	$(\beta,\gamma)=(0.8,0.5)$.  The framework developed in this work readily
	allows tabulation of these new universal exponents for any value of
	$(\beta,\gamma)$.
	We have observed that the universal exponents are slightly sensitive to the
	random seed. However, exact evaluation of this variability requires very
	significant computational resources and is beyond our present scope. Similarly,
	some sensitivity of the p-values to random seed has been observed.

	\begin{table}[h]
		\centering
		\caption{Results of Test 1 for $\gamma=0.5$ and $\beta=0.8$.}
		\small
\begin{tabular}{|l c c c c|}\hline 
Frame & $R^2$& $\hat{b}$ & $SE(\hat{b})$ & p-value \\
 & & & & $b=b_{MANOVA}$\\ \hline 
MANOVA & 0.99828 & 0.92505 & 0.00690 & 1 \\ 
DSS & 0.99858 & 0.93652 & 0.00911 & 0.32089 \\ 
GF & 0.99921 & 0.92474 & 0.02608 & 0.99082  \\ 
ComplexPF & 0.99950 & 0.92454 & 0.00535 & 0.95390  \\ 
Alltop & 0.98906 & 0.49660 & 0.00883 & 9.4651e-47  \\ 
SS & 0.98767 & 0.47354 & 0.00950 & 5.8136e-45  \\ 
\hline 
HAAR & 0.99736 & 0.94421 & 0.00873 & 0.09019  \\ 
RandDFT & 0.99544 & 0.94127 & 0.01644 & 0.36788  \\ 
\hline 
\hline 
RealMANOVA & 0.99873 & 0.95610 & 0.00613 & 1 \\ 
RealPF & 0.99871 & 0.91244 & 0.00821 & 9.7174e-05  \\ 
SH & 0.99989 & 0.46822 & 0.00492 & 6.3109e-35  \\ 
\hline 
RealHAAR & 0.99596 & 0.94456 & 0.01081 & 0.35675  \\ 
RandDCT & 0.99773 & 0.93859 & 0.01156 & 0.18737  \\ 
\hline 
\end{tabular}

		\label{tab:KS1}
	\end{table}
	
	\begin{table}[h]
		\centering
		\caption{Results of Test 2 for $\specstat_{Shannon}$, $\gamma=0.5$ and $\beta=0.8$}
		\small
\begin{tabular}{|l c c c c|}\hline 
Frame & $R^2$& $\hat{b}$ & $SE(\hat{b})$ & p-value \\
 & & & & $b=b_{MANOVA}$\\ \hline 
MANOVA & 0.98721 & 1.79936 & 0.03678 & 1 \\ 
DSS & 0.99110 & 1.88674 & 0.04615 & 0.14551 \\ 
GF & 0.99997 & 1.88548 & 0.01073 & 0.03161  \\ 
ComplexPF & 0.99977 & 1.77783 & 0.00701 & 0.56808  \\ 
Alltop & 0.93841 & 1.70618 & 0.07388 & 0.26297  \\ 
SS & 0.95539 & 1.89501 & 0.07355 & 0.24922  \\ 
\hline 
HAAR & 0.97971 & 1.87082 & 0.04836 & 0.24400  \\ 
RandDFT & 0.96928 & 1.77454 & 0.08157 & 0.78270  \\ 
\hline 
\hline 
RealMANOVA & 0.99202 & 2.05451 & 0.03309 & 1 \\ 
RealPF & 0.99834 & 2.00345 & 0.02045 & 0.19576  \\ 
SH & 0.97850 & 1.81297 & 0.26874 & 0.37904  \\ 
\hline 
RealHAAR & 0.98287 & 2.09078 & 0.04958 & 0.54503  \\ 
RandDCT & 0.98364 & 1.99663 & 0.06648 & 0.43977  \\ 
\hline 
\end{tabular}

		\label{tab:f_Shannon}
	\end{table}
	
	\begin{table*}[t]
		\centering
		\renewcommand{\arraystretch}{0.8}%
		\captionsetup{font=small}
		\caption{Summary of universal exponents for convergence. $\gamma = 0.5$, $\beta = 0.8$, $(\specstat_{S}=\specstat_{Shannon})$.
		}
		\newcommand\T{\rule{0pt}{2.6ex}}       
\newcommand\B{\rule[-1.2ex]{0pt}{0pt}} 

\tiny
\begin{tabular}{l c c c c c c c c c c c c c}  
\toprule 
Frame & $b_{spectrum}$ & $b_{\specstat_{RIP}}$ &$a_{\specstat_{RIP}}$ &$b_{\specstat_{AC}}$ &$a_{\specstat_{AC}}$ &$b_{\specstat_{S}}$&$a_{\specstat_{S}}$ &$b_{\specstat_{max}}$&$a_{\specstat_{max}}$&$b_{\specstat_{min}}$&$a_{\specstat_{min}}$&$b_{\specstat_{cond}}$&$a_{\specstat_{cond}}$ \T \B \\
\hline ~
MANOVA & 0.93 & 1.15 & 2.21 & 1.44 & 3.48 & 1.80& 0.99 & 1.13 & 2.48& 1.00& 3.09 & 1.87& -4.55 \T \\ 
DSS & 0.94 & 1.14 & 2.18 & 1.40 & 3.40 & 1.89& 1.04 & 1.10 & 2.41& 1.00& 3.11 & 1.87& -4.56 \\ 
GF & 0.92 & 1.17 & 2.23 & 1.53 & 3.70 & 1.89& 1.03 & 1.13 & 2.48& 1.04 & 3.22 & 1.95& -4.76\\ 
ComplexPF & 0.92 & 1.13 & 2.17 & 1.44 & 3.49 & 1.78& 0.98 & 1.10 & 2.41& 1.00& 3.12 & 1.87& -4.56 \\ 
Alltop & 0.50 & 1.14 & 2.18 & 1.46 & 3.53 & 1.71& 0.94 & 1.11 & 2.42& 1.01 & 3.13 & 1.86& -4.54\\ 
SS & 0.47 & 1.11 & 2.13 & 1.50 & 3.63 & 1.90& 1.04 & 1.08 & 2.36& 0.98 & 3.06 & 1.83& -4.47\\ 
HAAR & 0.94 & 1.10 & 2.11 & 1.52 & 3.69 & 1.87& 1.03 & 1.09 & 2.37& 1.01& 3.13 & 1.88& -4.59 \\ 
RandDFT & 0.94 & 1.21 & 2.32 & 1.47 & 3.56 & 1.77& 0.97 & 1.11 & 2.42& 1.03& 3.18 & 1.93& -4.70 \\ 
\hline ~
RealMANOVA & 0.96 & 0.87 & 3.58 & 1.26 & 5.21 & 1.27& 5.26 & 0.90 & 3.73& 0.87 & 3.58 & 0.77& 3.17\\ 
RealPF & 0.91 & 0.92 & 3.82 & 1.32 & 5.46 & 1.24& 5.12 & 0.94 & 3.88& 0.94 & 3.88 & 0.81& 3.36\\ 
SH & 0.47 & 0.93 & 3.82 & 1.34 & 5.53 & 1.14& 4.71 & 0.93 & 3.82& 0.93& 3.82 & 0.85& 3.51 \\ 
RealHAAR & 0.94 & 0.86 & 3.54 & 1.23 & 5.07 & 1.29& 5.35 & 0.89 & 3.68& 0.90& 3.73 & 0.79& 3.28 \\ 
RandDCT & 0.94 & 0.99 & 4.08 & 1.30 & 5.38 & 1.24& 5.10 & 0.94 & 3.89& 0.95& 3.93 & 0.82& 3.40 \\ 
\bottomrule 
\end{tabular}

		\label{ExpSummary}
	\end{table*}
	\begin{table*}[h]
		\centering
		\caption{$\Psi(f^{MANOVA}_{\beta,\gamma})\pm\sqrt{\Delta_\specstat(\Xk^{(n)})^2}$ for $\specstat_{AC}$ and DSS frame, $m=\frac{n-1}{2}$, $k=\beta\cdot\m$.
		}
		\scriptsize
\begin{tabular}{|l r r r r r r r|}\hline 
$n$ & 1031 & 1151 & 1291 & 1451 & 1571 & 1811 & 1951 \\ \hline 
RMSE, $\beta=0.8$ & 3$\pm$0.0281 & 3$\pm$0.0253 & 3$\pm$0.0227 & 3$\pm$0.0204 & 3$\pm$0.0189 & 3$\pm$0.0166 & 3$\pm$0.0155  \\ \hline 
RMSE, $\beta=0.6$ & 1.75$\pm$0.0073 & 1.75$\pm$0.0065 & 1.75$\pm$0.0058 & 1.75$\pm$0.0051 & 1.75$\pm$0.0048 & 1.75$\pm$0.0041 & 1.75$\pm$0.0038  \\ \hline 
\end{tabular}

		\label{TabulateAccuracy}
	\end{table*}
	
	\paragraph{Reproducibility advisory.} All the figures and tables in this chapter,
	including those in the Supporting Information, are fully reproducible from our
	raw results and code deposited in the Data and Code Supplement \cite{SDR}.
	
	\section{Discussion}
	\label{sec:discussion}
	
	\subsection{The hypotheses}
	
	Our Universality Hypotheses may be surprising in several aspects: Firstly, the
	frames examined were designed to minimize frame bounds and worse-case pairwise
	correlations. Still it appears that they perform well when the performance
	criterion is based on spectrum of the typical selection of $k$ frame vectors.
	Secondly, under the Universality Hypotheses, all these deterministic frames
	perform exactly as well as random frame designs such as the random Fourier
	frame. Inasmuch as frames are continuous codes, we find deterministic codes
	matching the performance of random codes.  Finally, the Hypotheses suggest an
	extremely broad universality property: many different ensembles of random
	matrices asymptotically exhibit the limiting MANOVA spectrum.
	
	All of the deterministic frames under study satisfy the Universality Hypotheses
	(with Hypothesis {\bf H4} satisfied only for ETFs). This should not give the
	impression that {\em any} deterministic frame satisfies these hypotheses!
	Firstly, because the empirical measures of an arbitrary sequence of frames
	rarely converge (thus violating Hypothesis H1).  Secondly, even if they
	converge, a too-simplistic frame design often leads to concentration of the
	lower edge of the empirical spectrum near zero, resulting in a non-MANOVA
	spectrum and poor performance.  For example, if the frame is sparse, say,
	consisting of some $m$ columns of the $n$-by-$n$ identity matrix, then a
	fraction $(n-m)/n$ of the singular values of a typical submatrix are exactly
	zero. 
	
	The frames under study are all ETFs or near-ETFs, all with favorable frame
	properties.  To make this point, we have included in the Supporting Information
	\cite{SI} study of a low-pass frame, in which the Fourier frequencies included
	in the frame are the lowest ones. This is in contrast with the clever choice of
	frequencies leading to the difference-set spectrum frame (DSS). Indeed the
	low-pass frame does not have appealing frame properties. It's quite obvious from
	the results in the SI, as well as results regarding the closely related random
	Vandermonde ensemble \cite{ryan2009asymptotic}, that such frames do not satisfy
	any of the Universality Hypotheses {\bf H2--H6}.
	
	We note that convergence rates of the form \eqref{KS_conv:eq} and
	\eqref{func_conv:eq} are known for other classical random matrix ensembles 
	\cite{gotze2011rate,gotze2016optimal,chatterjee2004new,meckes2017rates}.
	
	We further note that Hypotheses {\bf H1--H4} do not imply Hypotheses {\bf H5--H6}.
	Even if the empirical CDF converges in KS metric to the limiting
	MANOVA$(\beta,\gamma)$ distribution, 
	functionals which are not continuous in the KS metric do not necessarily
	converge, and moreover no uniform rate of convergence is a-priori implied.

	\subsection{Our contributions}
	
	This work presents a novel, simple method for approximate computation (with
	known and good approximation error) of spectral functionals of $k$-submatrix
	ensemble for a variety of random and deterministic frames, using
	\eqref{main_for:eq}.  Our results make it possible to tabulate these 
	approximate values, creating a useful resource for scientists. 
	As an example, we include Table \ref{TabulateAccuracy}. This is a lookup table
	for the value of the functional $\Psi_{AC}$ on the difference-set spectrum
	deterministic frame family (DSS), listing by values of $n$ and $k$ 
	the asymptotic (approximate) value 
	calculated analytically from the limiting $f^{MANOVA}_{\beta,\gamma}$
	distribution, 
	and the standard approximation error. 
	
	To this end we developed a systematic empirical framework,
	which allows validation of \eqref{main_for:eq} and discovery of the exponents
	there.  Our work is fully reproducible, and our framework is available (along
	with the rest of our results and code) in the Code and Data Supplement
	\cite{SDR}.  In addition, our results provide overwhelming empirical evidence
	for a number of phenomena, which were, to the best of knowledge, previously
	unknown:
	
	\begin{enumerate}
		\item {\bf The typical $k$-submatrix ensemble of deterministic frames is an
			object of interest.}
		While there is absolutely no randomness involved in the submatrix $X_K$
		of a deterministic frame (other than the choice of subset $K$), 
		the typical $k$-submatrix appears to be an ensemble in its own right,
		with properties so far attributed only to random matrix ensembles --
		including a universal, compactly-supported
		limiting spectral distribution and convergence of the maximal (resp.
		minimal) singular value to the upper (resp. lower) edges of the limiting
		distribution.
		
		\item {\bf MANOVA$(\beta,\gamma)$ as a universal limiting spectral distribution.}
		Wachter's MANOVA$(\beta,\gamma)$
		distribution is the limiting spectral distribution of $\lambda(G_K)$, as
		$k/m\to\beta$ and $m/n\to\gamma$, 
		for the typical $k$-submatrix ensemble of deterministic frames 
		(including difference-set, Grassmannian, real Paley, complex Paley,
		quadratic chirp, spiked and sines, and spikes and Hadamard).
		The same is true for real random frames - random cosine transform and
		random Haar.
		
		\item {\bf Convergence of the edge-spectrum.} 
		For all the deterministic frames above, as well as for the random frames 
		(random cosine, random Fourier, complex Haar, real Haar), the 
		maximal and minimal eigenvalues of the $k$-typical submatrix ensemble 
		converge to the support-edges of the MANOVA$(\beta,\gamma)$ limiting 
		distribution. The convergence follows a universal power-law rate.
		
		\item {\bf A definite power-law rate of convergence for the
			entire spectrum of the MANOVA$(n,m,k,\Fc)$ ensemble to its 
			MANOVA$(\beta,\gamma)$ limit,} with different exponents in the real and
		the complex cases.
		
		\item {\bf Universality of the power-law exponents for the entire
			spectrum.} The complex deterministic ETF frames 
		(difference-set, Grassmannian,
		complex Paley)  share the power-law exponents with the
		MANOVA$(n,\m,k,\C)$ ensemble. The same is true for
		the complex random frames (random Fourier and
		complex Haar).
		The complex tight non-equiangular Alltop frame, which can be constructed for various aspect ratios, also share the power-law exponents with the
		MANOVA$(n,\m,k,\C)$ ensemble for $\gamma<0.5$.
		The real deterministic ETF frame 
		(real Paley) 
		shares the exponent with the 
		MANOVA$(n,\m,k,\R)$. The same is true for real random frames 
		(random cosine and real Haar).
		All non-ETFs under study, with $\gamma=0.5$, share different power-law exponents (slower convergence).
		
		\item {\bf A definite power-law rate of convergence for
			functionals} including $\specstat_{StRIP}$, $\specstat_{AC}$ and 
		$\specstat_{Shannon}$.
		
		\item {\bf Universality of the power-law exponents for functionals.} 
		For practically all frames under study, both random and deterministic, 
		the power-law exponents for functionals agree with those of the 
		MANOVA$(n,\m,k,\R)$  (real frames) and 
		MANOVA$(n,\m,k,\C)$ (complex frames).
		
	\end{enumerate}
	\subsection{Intercepts}
	
	Our results showed a surprising categorization of the deterministic and random
	frames under study, according to the constant $C$ in \eqref{KS_conv:eq}, or
	equivalently, according to the intercept (vertical shift) in the linear regression 
	on $\log(n)$. Figure \ref{fig:KS1} and Figure \ref{fig:KS2} 
	clearly show that the regression lines, while having identical slopes (as
	predicated by Hypothesis {\bf H3}), are grouped according to their intercepts
	into the following 
	seven categories:
	Complex MANOVA ensemble and complex Haar (Manova, HAAR);
	Real MANOVA ensemble and real Haar (RealManova, RealHAAR);
	Complex ETFs (DSS, GF, ComplexPF);
	Non-ETFs (SS,SH,Alltop);
	Real ETF (RealPF);
	Complex Random Fourier (RandDFT), and 
	Real Random Fourier (RandDCT).
	
	Interestingly, intercepts of all complex frames are larger (meaning that the
	linear coefficient $C$ in \eqref{KS_conv:eq} is smaller) than those of all
	real frames. Also, the less randomness exists in the frame, the higher the
	intercept: intercepts of deterministic ETFs are higher then those of random
	Fourier and random Cosine, which are in turn higher than those of Haar
	frames and the MANOVA ensembles.

	\subsection{Related work} \label{sec:related}

	Farrell \cite{farrell2011limiting} has
	conjectured that the phenomenon of convergence of the spectrum of typical
	$k$-submatrices to the limiting MANOVA distribution is indeed much broader and
	extends beyond the partial Fourier frame he considered.
	A related empirical study was conducted by Monajemi et al \cite{monajemi2013deterministic}.
	There, the
	authors considered the so-called sparsity-undersampling phase transition in
	compressed sensing. This asymptotic quantity poses a performance criterion for
	frames that interacts with the typical $k$-submatrix $\Xk$ in a manner possibly
	more complicated than the spectrum $\lambda(\Gk)$. The authors investigated
	various deterministic frames, most of which are studied in this work, and
	brought empirical evidence that the phase transition for each of these
	deterministic frames is identical to the phase transition of Gaussian frames.
	Gurevich and Hadani \cite{gurevich2008incoherent} proposed certain deterministic frame
	constructions and effectively proved that the empirical spectral distribution of
	their typical $k$-submatrix converges to a semicircle, assuming
	$k=\m^{1-\varepsilon}$, a scaling relation different than the one considered
	here.
	\cite{applebaum2009chirp} and \cite{babadi2011spectral} also considered deterministic frame designs,
	chirp sensing codes and binary linear codes, with a random sampling.
	In their design the aspect ratios are large (e.g., in \cite{applebaum2009chirp}
	$m \sim k^2$ and $n \sim m^2$), so the spectrum converges to the
	Mar\u cenko-Pastur distribution.
	Tropp \cite{tropp2008conditioning} provided bounds for $\lambda_{max}(\Gk)$ and
	$\lambda_{min}(\Gk)$ when $X$ is a general dictionary.
	Collins \cite{collins2005product} has shown that the spectrum 
	of a matrix model deriving
	from random projections has the same eigenvalue distribution of the MANOVA
	ensemble in finite
	$n$. Wachter \cite{wachter1980limiting} used a connection between the MANOVA ensemble and
	submatrices of Haar matrices to derive the asymptotic spectral distribution
	MANOVA$(\beta,\gamma)$.

	\section{Conclusions}
	
	We have observed a surprising universality property for the $k$-submatrix
	ensemble corresponding to various well-known deterministic frames, as well as to
	well-known random frames. The MANOVA ensemble, and the MANOVA limiting
	distribution, emerge as key objects in the study of frames, both random and
	deterministic, in the context of sparse signals and erasure channels. 
	We hope that our findings will invite rigorous mathematical study of these 
	fascinating phenomena.
	
	In any frame where our Universality Hypotheses hold
	(including all the frames under study here),
	Figure \ref{fig:limiting_F} correctly
	describes the limiting values of $f_{RIP}$, $f_{AC}$ and $f_{Shannon}$ and shows
	that codes based on deterministic frames (involving no randomness and allowing
	fast implementations) are better, across performance measures,
	than i.i.d random codes. 
	
	The empirical framework we proposed in this chapter may be easily applied to new
	frame families $X^{(n)}$ and new functionals $\Psi$, extending our results further and
	mapping the frontiers of the new universality property. 
	In any frame family, and for any functional, where our Universality Hypotheses hold, we
	have proposed a simple, effective method for calculating quantities of the form 
	$\E_K \Psi\left( \lambda(G_K \right))$ to known approximation, which improves
	polynomially with $n$.

	\chapter{Frame Moments}
	\label{chapter:Moments}
In this chapter we analyze the moments of random subsets of frames as a tool for a mathematical study of the phenomena presented in Chapter \ref{chapter:PNAS}. Section \ref{secEWB} is taken from our ISIT 2018 paper, and it develops a lower bound on these moments which surprisingly is achieved with equality for ETFs. Section \ref{sec:asyMoments} derives a recursive method for computation of asymptotic moments of ETFs. This method relies on two conjectures, proofs of which are still part of current work.

\section{Frame Moments and Welch Bound with Erasures} \label{secEWB}

Design of frames or over-complete bases with favorable properties is a thoroughly studied subject in communication, signal processing and harmonic analysis.  In various applications, one is interested in finding over-complete bases
where the favorable properties hold for a random subset of the frame vectors, rather than for the entire frame.

Here are a few examples. In code-devision multiple access (CDMA), spreading sequences with low cross-correlation are preferred; when only a random subset of the users is active, the quantity of interest is the expected cross-correlation within a random subset of the spreading sequences \cite{rupf1994optimum}.
In sparse signal reconstruction from undersampled measurements, the ability to reconstruct the signal crucially depends on 
properties of a subset of the measurement matrix, which corresponds to the non-zero entries of the sparse signal;
for example, if the extreme eigenvalues of the submatrix are bounded, stable recovery is guaranteed \cite{candes2008restricted}. When the support of the sparse vector is random, one is interested in extreme eigenvalues of a random frame subset \cite{calderbank2010construction}.
In analog coding, various schemes of interest require frames, for which the first inverse moment of the 
covariance matrix of a randomly chosen frame subset is as small as possible. This occurs, for example, 
in the presence of source erasures known at the encoder \cite{haikin2016analog},
in channels with impulses \cite{wolf1983redundancy} or with erasures \cite{ITA17},
and in multiple description source coding \cite{mashiach2013sampling}.

A famous result by Welch \cite{welch1974lower} provides a universal lower bound on the mean and maximum value 
of powers of absolute values of inner products (a.k.a cross-correlations) of frame vectors. 
Frames which achieve the Welch lower bound on maximal absolute cross-correlation
are known as equiangular tight frames (ETF). 

Motivated by frame design for various applications, in this paper we show that the Welch bound naturally extends to random frame subsets, such that 
the lower bound is achieved by (and sometimes only by) ETFs. We term this new 
universal lower bound the
{\em Erasure Welch Bound} (EWB) and generalize it to higher-order covariances as well.

As a universal, tight lower bound in frame theory, the EWB is essentially a geometric quantity. Surpringly, 
the EWB itself coincides with a quantity appearing elsewhere in mathematics, namely in random matrix theory.
Below, we prove that the EWB matches the moments of Wachter's classical limiting MANOVA distribution \cite{wachter1980limiting}. 
In a recent paper \cite{haikin2017random} we reported overwhelming empirical evidence
that the covariance matrix of a random frame subset from many well-known ETFs (and near-ETFs) in fact follows the 
Wachter's classical limiting MANOVA distribution. To the best of our knowledge,
the results of this paper are the first theoretical confirmations to the empirical predictions of \cite{haikin2017random}, relating ETFs to Wachter's classical limiting MANOVA distribution and random matrix phenomena.

\subsection{Notation and Setup} \label{definitions}
We consider a unit-norm frame, being an over-complete basis comprising $n$ elements - unit-norm vectors $\bfv_1,\dots,\bfv_n$.
Let $\bF=\{f_{j,i}, \,\,\, j=1\dots m,\,\, i=1\dots n\}$ denote the $m$-by-$n$ frame matrix whose columns are the frame vectors, $\bF=\left[\bfv_1|\cdots|\bfv_n\right]$.
Let us define the vector cross correlation:
\begin{equation} \label{corr}
c_{i_1,i_2}\triangleq<\bfv_{i_1},\bfv_{i_2}> = \bfv_{i_1}'\bfv_{i_2}=\sum_{j=1}^{m}f_{j,i_1}^*f_{j,i_2}
\end{equation}
where 
\begin{equation} \label{c_ii}
c_{i,i} = \|\bfv_i\|^2=1
\end{equation}
by the unit norm property.
The {\em Welch bound} \cite{welch1974lower} lower bounds the mean-square (ms) cross correlation:
\begin{equation} \label{rms WB}
I_{ms}(\bF) \triangleq \frac{1}{(n-1)n}\sum_{i_2\neq i_1}^{n}|c_{i_1,i_2}|^2\ge \frac{n-m}{(n-1)m},
\end{equation}
and it is achieved with equality iff $\bF$ is a Uniform Tight Frame (UTF), i.e.
\begin{equation} \label{UTF}
\bF\bF'=\frac{n}{m}I_m.
\end{equation}
The Welch bound \cite{welch1974lower} implies a bound on the maximum-square cross correlation:
\begin{equation} \label{max WB}
I_{max}(\bF) \triangleq \max_{1\le i_1< i_2\le n}|c_{i_1,i_2}|^2\ge \frac{n-m}{(n-1)m}.
\end{equation}
This stronger lower bound is achieved with equality iff the frame is an Equiangular Tight Frame (ETF), namely, it is UTF \eqref{UTF} and it satisfies 
\begin{equation} \label{ETF2}
|c_{i_1,i_2}|^2={\rm constant}=\frac{n-m}{(n-1)m} \ \ \forall i_1 \neq i_2.
\end{equation}
The unique configuration of ETF, which exists only for some dimensions $m$ and number of vectors $n$, achieves a whole family of lower bounds which are derived below.

Our main object of interest is a submatrix composed of a random subset of the frame vectors, or columns of $\bF$.
Define the "erased" $m$-by-$n$ matrix as
\begin{equation} \label{X}
\bX = \bF\bP,
\end{equation}
where $\bP$ is a diagonal matrix with independent Bernoulli($p$) elements on the diagonal.
In other words, each of the vectors $\bfv_1,...,\bfv_n$ is replaced by a zero vector with probability $1-p$.
%
Let us define the (expected) $d$-th moment of a random subset of $\bF$ as:
\begin{equation} \label{moment d}
m_d \triangleq \frac{1}{n}\Ev\left[ \tr \left((\bX'\bX)^d\right)\right]=\frac{1}{n}\Ev\left[ \tr \left((\bF\bP\bF')^d\right)\right]
\end{equation}
where in the second equality we applied $\tr \left((\bX'\bX)^d\right)=\tr \left((\bX\bX')^d\right)$ and $\bP^2=\bP$. 

Since the trace of a matrix is equal to the sum of its eigenvalues, the argument of expectation in \eqref{moment d} is the $d$-th moment of the empirical eigenvalues distribution of $\bX'\bX$.

The following Lemma deals with the special case of $p=1$, i.e. moments of the \textbf{whole} frame without taking subsets. It is useful for relating the moments definition above to the ms cross correlation in \eqref{rms WB}, and for attaining bounds for $d>1$.  
\begin{lemma} \label{lemma1}
	For any unit-norm frame $\bF$,
	\begin{equation} \label{md bound_p=1}
	\frac{1}{n}\tr \left((\bF\bF')^d \right)\ge\left(\frac{n}{m}\right)^{d-1} 
	\end{equation}
	with equality for $d>1$ iff $\bF$ is a UTF.
\end{lemma}
\begin{proof}
	The trace of the square matrix $\bF\bF'$ is equal to the sum of its eigenvalues $\{\lambda\}_{j=1}^m$. Furthermore, the eigenvalues of $(\bF\bF')^d$ are $\{\lambda^d\}_{j=1}^m$. 
	Using Jensen's inequality for the convex function $(\cdot)^d$, we have:
	\begin{equation} \label{Jensen}
	\frac{1}{m}\sum_{j=1}^{m}\lambda^d_{j}\ge \left( \frac{1}{m}\sum_{j=1}^{m}\lambda_{j}\right)^d
	\end{equation}
	with equality iff all eigenvalues are equal, i.e. $\bF\bF'\propto I_m$, as in \eqref{UTF}. Hence,
	\begin{equation} \label{Jensen tr}
	\Rightarrow \frac{1}{m}\tr \left((\bF\bF')^d \right)\ge \left(\frac{1}{m}\tr(\bF\bF')\right)^d 
	\end{equation}
	and by proper normalization \eqref{md bound_p=1} follows since the argument in the right hand side can be written as:
	\begin{equation} \label{trace FF'}
	\frac{1}{m}\tr(\bF\bF')=\frac{1}{m}\tr(\bF'\bF)=\frac{1}{m}\sum_{i=1}^{n}c_{i,i}=\frac{n}{m}.
	\end{equation}
\end{proof}

Note that for $d=2$, 
\begin{equation} \label{Lemma d=2}
\frac{1}{n}\tr \left((\bF\bF')^2\right)=\frac{1}{n}\sum_{i_1,i_2=1}^{n}|c_{i_1,i_2}|^2=1+\frac{1}{n}\sum_{i_2\neq i_1}^{n}|c_{i_1,i_2}|^2,
\nonumber
\end{equation}
so \eqref{md bound_p=1} becomes 
\begin{equation} \label{WB x}
\frac{1}{n}\sum_{i_2\neq i_1}^{n}|c_{i_1,i_2}|^2\ge \frac{n}{m}-1\triangleq x
\end{equation}
which is the Welch bound \eqref{rms WB} with different normalization.
Therefore, a lower bound on $m_d$ in \eqref{moment d} generalizes the Welch bound in two senses. First, it is a bound on random subsets of $\bF$.
In particular, for $d=2$,
\begin{equation} \label{rms FP m2}
m_2 = \frac{1}{n}\Ev \left[{\sum_{i_1,i_2 \in S} |c_{i_1,i_2}|^2 }\right]
\end{equation}
where $S \subset \{1,...,n\}$ is the random subset of selected indices
(the $i$'s for which $p_{i,i}$ =1).
Second, it is a bound on higher order moments, for $d\ge 2$. \footnote{Our definition is different than the general Welch bound on the powers of the absolute cross-correlations \cite{welch1974lower}.}

\subsection{Main Result}\label{main}
The first moment ($d=1$) of a frame is independent of the choice of $\bF$ since
\begin{equation} \label{m1}
\begin{split}
&m_1 = \frac{1}{n}\Ev\left[ \tr \left(\bX'\bX\right)\right] = \frac{1}{n}\Ev\left[ \sum_{i=1}^{n}\bfv_i'\bfv_ip_{i,i}\right]=\frac{1}{n}\Ev\left[ \sum_{i=1}^{n}p_{i,i}\right]=\frac{1}{n}\sum_{i=1}^{n}\Ev\left[ p_{i,i}\right]=\frac{1}{n}\sum_{i=1}^{n}p = p
\end{split}
\end{equation}
where the third equality is due to \eqref{c_ii}.
To state our main theorem regarding higher order moments, let us define the $d$-th moment of the MANOVA$(\gamma,p)$ density as, \cite{dubbs2015infinite}
\begin{equation} \label{moment d MANOVA}
m^{\rm MANOVA}(\gamma,p,d)\triangleq \min (p,\gamma)\int t^d \,\rho_{p,\gamma}(t)dt
\end{equation}
where $\gamma = \frac{m}{n}$ is the aspect ratio of the frame, and
\begin{equation}	\label{ManovaDensity}
\begin{split}
&\rho_{p,\gamma}(t)=\frac{\gamma \sqrt{(t-r_-)(r_+-t)}}{2\pi t(1-\gamma t)\min (p,\gamma)}\cdot 
I_{(r_-,r_+)}(t) +\left(p+\gamma-1\right)^+/\min (p,\gamma)\cdot \delta(t-\frac{1}{\gamma})
\end{split}
\end{equation}
is Wachter's classical MANOVA desnity \cite{wachter1980limiting}, compactly supported on $[r_-,r_+]$  with
\begin{equation}
\label{ManovaDensityExtrimalValues}
r_\pm=\bigg(\sqrt{\frac{p}{\gamma}(1-\gamma)}\pm\sqrt{1-p}\bigg)^2\,.
\end{equation} 
The non-standard factor $\min (p,\gamma)$ in \eqref{moment d MANOVA} is due to normalization by the full dimension $n$ as defined in \eqref{moment d} for $m_d$, and not by the rank of the subset ($k$ or $m$). 

Using $x=\frac{1}{\gamma}-1$ \eqref{WB x}, let:
\begin{equation} \label{delta EWBd}
\Delta(\gamma,p,d,n)\triangleq
\begin{cases}
0,& d=2,3  \\
p^2(1-p)^2\frac{x^2}{n-1},& d=4.  \\
\end{cases}
\end{equation}
~\\
\begin{theorem}[Erasure Welch Bound of order $d$] \label{th1}
	For any  $m$-by-$n$ unit-norm frame and $d=2,3,4$, the $d$-th moment \eqref{moment d} is lower bounded by
	\begin{equation} \label{moment d bound}
	m_d  \ge 
	m^{\rm MANOVA}(\gamma,p,d)+\Delta(\gamma,p,d,n).
	\end{equation}
	with equality for $d=2,3$ iff $\bF$ is a UTF, and for $d=4$ iff $F$ is an ETF.
\end{theorem}
~\\
The Erasure Welch Bound admits a simple closed form.
We can write the first term in \eqref{moment d bound} for $d=2,3,4$ as
\begin{align} \label{Manova moments}	
m^{\rm MANOVA}(\gamma,p,2) &= p+p^2x\\
\nonumber
m^{\rm MANOVA}(\gamma,p,3) &= p+p^23x+p^3(x^2-x)\\
\nonumber
m^{\rm MANOVA}(\gamma,p,4) &= p+p^26x+p^3(6x^2-4x)
\nonumber
+p^4(x^3-3x^2+x)
\nonumber
\end{align}
where $x$ is defined in \eqref{WB x}.
As for the second term, note that $\Delta(\gamma,p,4,n)\to 0$ as $n\to \infty$. Therefore, the lower bound is asymptotically $m^{\rm MANOVA}(\gamma,p,d)$ for $d=2,3$ and $4$.
This is in line with the empirical results in \cite{haikin2017random}, where we showed empirically that random subsets of ETFs have MANOVA spectra.

We can see from \eqref{delta EWBd} that $\Delta(\gamma,p=1,d,n)=0$, and from \eqref{Manova moments} that $m^{\rm MANOVA}(\gamma,p=1,d)=(x+1)^{d-1}$. Thus for $p=1$ the bound \eqref{moment d bound} becomes $\left(\frac{n}{m}\right)^{d-1}$ and coincides with Lemma \ref{lemma1}.

The second moment case of Theorem \ref{th1} is essentially equivalent to the Welch Bound \eqref{rms WB}. To see why, note that a Bernoulli($p$) selection is asymptotically equivalent to a combinatoric selection with $k = np$. Furthermore, the normalized sum in \eqref{rms WB}, which is the average over all ${n}\choose{2}$ distinct pairs, can be written as the average over all ${n}\choose{k}$ subsets of the average over all ${k}\choose{2}$ pairs within each subset. 

For any $n > k > m$, we can interpret the gap between the Welch bound for an $m$-by-$n$ frame and the Welch bound for an $m$-by-$k$ frame as the penalty in the mean square cross correlation due to randomly choosing the vectors from a fixed larger set of vectors.
Another interesting point of view, provided by random matrix theory, is that this gap corresponds to the increase in the (renormalized) MANOVA second moment $\frac{1}{p}m^{MANOVA}(\gamma,p,2)$, as $p$, $\gamma$ go to zero at the same rate ($n$ grows while $k$ and $m$ are held constant). In the limit as $p\to 0$, this becomes the second moment of the Mar\u cenko-Pastur distribution of an i.i.d matrix \cite{tulino2004random}. 
\subsection{Proof of Theorem \ref{th1}}
By standard matrix multiplication, for $1\le j,j'\le m$ 
\begin{equation} \label{XX'k}
\begin{split}
&\left((\bX\bX')^k\right)_{j,j'}= \sum_{j_2,\dots,j_k}^{m}(\bX\bX')_{j,j_2}(\bX\bX')_{j_2,j_3}\cdots
(\bX\bX')_{j_k,j'}.
\end{split}
\end{equation}
To obtain the trace, we sum over the diagonal elements, thinking of $j=j'\equiv j_1$:
\begin{equation} \label{tr XX'k}
\begin{split}
&\tr\left((\bX\bX')^d\right) = \sum_{j_1}\left((\bX\bX')^d\right)_{j_1,j_1}=
\sum_{j_1,\dots,j_d}^{m}(\bX\bX')_{j_1,j_2}(\bX\bX')_{j_2,j_3}\cdots
(\bX\bX')_{j_d,j_1}
\\&=
\sum_{j_1,\dots,j_d}^{m}\sum_{i_1,\dots,i_d}^{n}f_{j_1,i_1}f^*_{j_2,i_1}
f_{j_2,i_2}f^*_{j_3,i_2}\cdots f_{j_d,i_d}f^*_{j_1,i_d}p_{i_1,i_1}p_{i_2,i_2}\cdots p_{i_d,i_d}.
\end{split}
\end{equation}
where in the last equality we substituted $(\bF')_{i,j}=f^*_{j,i}$ and
\begin{equation} \label{XX'_ij}
(\bX\bX')_{j_t,j_{t+1}}=\sum_{i_t=1}^{n}(\bF)_{j_t,i_t}(\bP)_{i_t,i_t}(\bF)'_{i_t,j_{t+1}}.
\end{equation}
Summing over the row indices $j_1,\dots,j_d$ and using \eqref{corr}, we obtain the following sum over chains of correlations:
\begin{equation} \label{tr XX'd}
\begin{split}
&\frac{1}{n}\tr\left((\bX\bX')^d\right)
=\frac{1}{n}
\sum_{i_1,\dots,i_d}^{n}c_{i_1,i_2}c_{i_2,i_3}\cdots c_{i_d,i_1}p_{i_1,i_1}\cdots p_{i_d,i_d}.
\end{split}
\end{equation}
In order to take expectation, we break the sum into cases according to possible combinations of distinct or equal indices. When the number of distinct values in $i_1,\dots,i_d$ is $k$, the expectation is $\Ev \left[p_{i_1,i_1}\cdots p_{i_d,i_d}\right]=p^k$. The sum of $\frac{1}{n}c_{i_1,i_2}c_{i_2,i_3}\cdots c_{i_d,i_1}$ over all such combinations is denoted by $a_{d,k}(\bF)$.
Note that for $k=1$, $a_{d,1}(\bF)=\frac{1}{n}\sum_{i_1=\dots =i_d=i}^{n}c_{i,i}^d=1$. Hence, $m_d$ can be written in the following form:
\begin{equation} \label{md poly of p}
\begin{split}
m_d = p+p^2a_{d,2}(\bF)+p^3a_{d,3}(\bF)+\cdots+p^da_{d,d}(\bF)
\end{split}
\end{equation}
where $a_{d,d}(\bF)$ is of a special interest, and it corresponds to a cycle of correlations of all distinct indices:
\begin{equation} \label{a_dd(F)}
\begin{split}
a_{d,d}(\bF) = \frac{1}{n} \sum_{i_1\neq i_2\neq i_3\neq ..\neq i_d}^{n}c_{i_1,i_2}c_{i_2,i_3}\cdots c_{i_d,i_1}. 
\end{split}
\end{equation}
We now turn to consider each of the special cases $d=2,3,4$.
\\ \textbf{Second moment}:
According to \eqref{md poly of p} we have
\begin{equation} \label{m2}
\begin{split}
&m_2 =p+p^2a_{2,2}(\bF)
\end{split}
\end{equation}
where $a_{2,2}(\bF)$ corresponds to cases with $i_1\neq i_2$
\begin{equation} \label{a22}
\begin{split}
&a_{2,2}(\bF)=\frac{1}{n}\sum_{i_2\neq i_1}^{n}|c_{i_1,i_2}|^2\ge x
\end{split}
\end{equation}
where the inequality is due to the ms Welch bound \eqref{WB x}, and it is satisfied with equality iff $\bF$ is a UTF.
From \eqref{m2} and \eqref{a22}, 
\begin{equation} \label{m2 bound}
\begin{split}
&m_2 \ge  p+p^2x=m^{\rm MANOVA}(\gamma,p,2).
\end{split}
\end{equation}
\\
\textbf{Third moment}:
According to \eqref{md poly of p},
\begin{equation} \label{m3 calc}
\begin{split}
m_3 &= p + p^2a_{3,2}(\bF) + p^3a_{3,3}(\bF).
\end{split}
\end{equation}
The mid term coefficient $a_{3,2}(\bF)$ consists of all combinations of two distinct values for $i_1,i_2,i_3$:
\begin{equation} \label{a32}
\begin{split}
a_{3,2}(\bF)&=\frac{1}{n}3\sum_{i_3\neq i_1=i_2}^{n}c_{i_1,i_1}c_{i_1,i_3}c_{i_3,i_1}=\frac{1}{n}3\sum_{i_3\neq i_1}^{n}|c_{i_1,i_3}|^2=3a_{2,2}(\bF),
\end{split}
\end{equation}
where we used $c_{i,i}=1$ and \eqref{a22}.
Since \eqref{m3 calc} holds for every $p$, we can set $p=1$, and use \eqref{a32} and Lemma \ref{lemma1} for $d=3$ to obtain: 
\begin{equation} \label{m3_p=1}
\begin{split}
1 + 3a_{2,2}(\bF)+a_{3,3}(\bF)\ge \left(\frac{n}{m}\right)^2=(x+1)^2.
\end{split}
\end{equation}
Substituting \eqref{a32} and \eqref{m3_p=1} in \eqref{m3 calc} we obtain
\begin{equation} \label{m3 bound}
\begin{split}
&m_3\ge p + p^23a_{2,2}(\bF)+p^3\big[(x+1)^2-1-3a_{2,2}(\bF)\big] \\&= (p-p^3)+(p^2-p^3)3a_{2,2}(\bF)+p^3(x+1)^2.
\end{split}
\end{equation}
Since $p\le 1$, we have $p^2-p^3\ge 0$, and we use \eqref{a22} to get:
\begin{equation} \label{m3 bound2}
\begin{split}
&m_3\ge(p-p^3)+(p^2-p^3)3x+p^3(x+1)^2\\&=p + p^2 3x + p^3 (x^2-x) =m^{\rm MANOVA}(\gamma,p,3) 
\end{split}
\end{equation}
and the condition for equality in both \eqref{a22} and \eqref{m3_p=1} is the frame being a UTF.
\\ \textbf{Fourth moment}:
According to \eqref{md poly of p},
\begin{equation} \label{m4 poly of p}
\begin{split}
m_4 &= p + p^2a_{4,2}(\bF) + p^3a_{4,3}(\bF) + p^4a_{4,4}(\bF).
\end{split}
\end{equation}
Denote $h=h(\bF,\{i_l\}_{l=1}^4)=c_{i_1,i_2}c_{i_2,i_3}c_{i_3,i_4}c_{i_4,i_1} $. We first consider the second term.  
Considering all partitions of $\{i_l\}_{l=1}^4$ into two groups (two distinct values), we get:
\[
a_{4,2}=\underbrace{4\frac{1}{n}\sum_{i_2=i_3=i_4\neq i_1}^{n}h}_{a^{(1)}_{4,2}}
+\underbrace{2\frac{1}{n}\sum_{i_1=i_2\neq i_3=i_4}^{n}h}_{a^{(2)}_{4,2}}
+\underbrace{\frac{1}{n}\sum_{i_1=i_3\neq i_2=i_4}^{n}h}_{a^{(3)}_{4,2}}
\]
where $a^{(1)}_{4,2}$ corresponds to partitions consisting of three identical indices and one different - $i_1$ or $i_2$ or $i_3$ or $i_4$, $a^{(2)}_{4,2}$ corresponds to partitions consisting of two different, non-crossing, pairs of indices - $i_1=i_2,i_3=i_4$ or $i_2=i_3,i_4=i_1$, $a^{(3)}_{4,2}$ corresponds to a partition consisting of two different, crossing, pairs of indices - $i_1=i_3,i_2=i_4$.
We can rewrite these three components as:
\begin{align} \label{a42 components}	
&a^{(1)}_{4,2}=4\frac{1}{n}\sum_{i_2\neq i_1}^{n}|c_{i_1,i_2}|^2=4a_{2,2}(\bF)\\
&a^{(2)}_{4,2}=2\frac{1}{n}\sum_{i_3\neq i_1}^{n}|c_{i_1,i_3}|^2=2a_{2,2}(\bF)\\
\label{a42_3}
&a^{(3)}_{4,2}=\frac{1}{n}\sum_{i_2\neq i_1}^{n}|c_{i_1,i_2}|^4 
\end{align}
To lower bound $a^{(3)}_{4,2}$, we use Jensen's inequality:
\begin{equation} \label{a42_3_bound}
\begin{split}
\frac{1}{n(n-1)}\sum_{i_2\neq i_1}^{n}|c_{i_1,i_2}|^4\ge \left(\frac{1}{n(n-1)}\sum_{i_2\neq i_1}^{n}|c_{i_1,i_2}|^2\right)^2
\end{split}
\end{equation}
which is achieved with equality iff all absolute correlations are constant, i.e. $\bF$ is an ETF. Hence, from \eqref{a42_3} and \eqref{a42_3_bound}:
\begin{equation} \label{a42_3 bound2}
a^{(3)}_{4,2}\ge \frac{1}{n-1}\left(\frac{1}{n}\sum_{i_2\neq i_1}^{n}|c_{i_1,i_2}|^2\right)^2\ge \frac{x^2}{n-1} 
\end{equation}
where the second inequality follows from the Welch bound \eqref{WB x}. We now turn to the third term in \eqref{m4 poly of p}.
Considering all partitions of $\{i_l\}_{l=1}^4$ into three groups, i.e. three distinct values, we get:
\[
a_{4,3}=\underbrace{4\frac{1}{n}\sum_{i_1=i_2\neq i_3\neq i_4}^{n}h}_{a^{(1)}_{4,3}}
+\underbrace{2\frac{1}{n}\sum_{i_1=i_3\neq i_2\neq i_4}^{n}h}_{a^{(2)}_{4,3}}
\]
where $a^{(1)}_{4,3}$ corresponds to partitions consisting of one pair of identical indices and two different values- $i_1=i_2$ or $i_2=i_3$ or $i_3=i_4$ or $i_4=i_1$, $a^{(2)}_{4,3}$ corresponds to partitions consisting of one pair of identical indices and two different values- $i_1=i_3$ or $i_2=i_4$.
We can rewrite these two components as:
\begin{align} \label{a43 components}	
&a^{(1)}_{4,3}=4\frac{1}{n}\sum_{i_2\neq i_3\neq i_4}^{n}c_{i_2,i_3}c_{i_3,i_4}c_{i_4,i_1}=4a_{3,3}(\bF)\\
\label{a43 components2}
&a^{(2)}_{4,3}=2\frac{1}{n}\sum_{i_1\neq i_2\neq i_4}^{n}|c_{i_1,i_2}|^2|c_{i_1,i_4}|^2
\end{align}
Let $C_{i_1}$ denote the sum over all absolute correlations between $\bfv_{i_1}$ and all other frame vectors.
\begin{equation} \label{C_i}
\begin{split}
C_{i_1} = \sum_{i_2\neq i_1}^{n}|c_{i_1,i_2}|^2.
\end{split}
\end{equation}
Using this notation we can write \eqref{a43 components2} as
\begin{equation} \label{b2_bound}
\begin{split}
&\frac{1}{2}a^{(2)}_{4,3}
=\frac{1}{n}\sum_{i_1}^{n}\sum_{i_2\neq i_1}^{n}|c_{i_1,i_2}|^2\sum_{i_4\neq i_2,i_1}^{n}|c_{i_1,i_4}|^2=\frac{1}{n}\sum_{i_1}^{n}\sum_{i_2\neq i_1}^{n}|c_{i_1,i_2}|^2\left[C_{i_1}-|c_{i_1,i_2}|^2\right]\\&=\frac{1}{n}\sum_{i_1}^{n}C_{i_1}\sum_{i_2\neq i_1}^{n}|c_{i_1,i_2}|^2-\frac{1}{n}\sum_{i_1}^{n}\sum_{i_2\neq i_1}^{n}|c_{i_1,i_2}|^4=\frac{1}{n}\sum_{i_1}^{n}C_{i_1}^2-a^{(3)}_{4,2} \,\,\,\,\,\,\,\,\, 
\end{split}
\end{equation}
Thus, we can lower bound the following sum 
\begin{equation} \label{b1+b2}
\begin{split}
a^{(3)}_{4,2}+\frac{1}{2}a^{(2)}_{4,3} &= \frac{1}{n}\sum_{i_1}^{n}C_{i_1}^2 \ge \left(\frac{1}{n}\sum_{i_1}^{n}C_{i_1}\right)^2
\ge x^2
\end{split}
\end{equation}
where the first inequality is again due to Jensen and is achieved with equality if the $C_i$s are equal for all $i$, and the second inequality is the Welch bound \eqref{WB x}.
Combining all terms we have
\begin{equation} \label{m4 total}
\begin{split}
&m_4=p+p^2(6a_{2,2}+a^{(3)}_{4,2})+p^3(4a_{3,3}+a^{(2)}_{4,3})+p^4a_{4,4}.
\end{split}
\end{equation}
Now we repeat the procedure from the bound on $m_3$ with sequential substitution of all bounds and gathering of similar terms.
We set $p=1$ in \eqref{m4 total} and use Lemma \ref{lemma1}:
\begin{equation} \label{a44 bound}
\begin{split}
&a_{4,4}\ge (x+1)^3-1-6a_{2,2}-4a_{3,3}-a^{(3)}_{4,2}-a^{(2)}_{4,3}
\end{split}
\end{equation}
Assuming UTF, we could calculate directly $a_{4,4}$, as well as $a_{3,3}$ without applying the relation from Lemma \ref{lemma1} (see Appendix \ref{app:app1}).
Substituting \eqref{a44 bound} into \eqref{m4 total} we get:
\begin{equation} \label{m4 bound2}
\begin{split}
&m_4\ge p-p^4+p^4(x+1)^3+(p^2-p^4)6a_{2,2}\\&+(p^2-p^4)a^{(3)}_{4,2}+(p^3-p^4)a^{(2)}_{4,3}+(p^3-p^4)4a_{3,3}.
\nonumber
\end{split}
\end{equation}
As $(p^3-p^4)\ge 0$, we can substitute \eqref{m3_p=1} and get:
\begin{equation} \label{m4 bound3}
\begin{split}
&m_4\ge p-p^4+p^4(x+1)^3+(p^3-p^4)\left(4(x+1)^2-4\right)\\&+p^2(1-p)6a_{2,2}+(p^2-p^4)a^{(3)}_{4,2}+(p^3-p^4)a^{(2)}_{4,3}
\nonumber
\end{split}
\end{equation}
and now we use the bound on $a_{2,2}$ \eqref{a22}.
The last two terms can be reordered to become a function of $a^{(3)}_{4,2}$ and $a^{(3)}_{4,2}+\frac{1}{2}a^{(2)}_{4,3}$ for which we have bounds
\begin{equation} \label{m4 bound4}
\begin{split}
&(p^2-p^4)a^{(3)}_{4,2}+(p^3-p^4)a^{(2)}_{4,3} \\&=(p^3-p^4)2(a^{(3)}_{4,2}+\frac{1}{2}a^{(2)}_{4,3})+p^2(1-p)^2a^{(3)}_{4,2}
\nonumber
\end{split}
\end{equation}
So now we can apply \eqref{a42_3 bound2} and \eqref{b1+b2}
\begin{equation} \label{m4 bound5}
\begin{split}
&m_4\ge p+p^2(6x+\frac{1}{n-1}x^2)+p^3(6x^2-4x-2\frac{1}{n-1}x^2)+p^4(x^3-3x^2+x+\frac{1}{n-1}x^2)\\&=m^{\rm MANOVA}(\gamma,p,4)+p^2(1-p)^2\frac{x^2}{n-1}
\end{split}
\nonumber
\end{equation}
with equality iff $\bF$ is ETF. 

Note that the asymptotic lower bound $\lim\limits_{n\to \infty}m_4\ge m^{\rm MANOVA}(\gamma,p,4)$ holds with equality under the weaker condition that $\bF$ is a UTF and $C_{i_1} = \sum_{i_2\neq i_1}^{n}|c_{i_1,i_2}|^2$ is equal for all $i$ and $\frac{1}{n}\sum_{i_2\neq i_1}^{n}|c_{i_1,i_2}|^4\to 0$ as $n\to \infty$, i.e. ETF is sufficient but not necessary. 
$\Box$

\section{Asymptotic Moments of ETFs} \label{sec:asyMoments}
From the main results of Section \ref{secEWB} we know that ETFs achieve the lowest possible moments $m_d$ \eqref{moment d} for $d\le 4$. Moreover, Asymptotically these lower bounds depend only on the aspect ratios of the frame ($\gamma$) and the subsets ($\frac{p}{\gamma}$) and is equal to the moments of the MANOVA ensemble.
The asymptotic moments for all $d$, along with a concentration result, i.e, almost surely the trace of a random subset is equal to the expected value \eqref{moment d}, will provide rigorous proofs for the results of \cite{haikin2017random}.
As the first step, one would need to show that the moments $m_d$ are in fact equal to the moments of the MANOVA density for all $d$.
The empirical $d$-th moment of $\bX\bX'$, $\frac{1}{n}\tr \left((\bX'\bX)^d\right)$ is the average of the eigenvalues of $\bX'\bX$ raised to the $d$-th power. Thus, $m_d$ is the expectation of the $d$-th moment of the empirical eigenvalues distribution of $\bX'\bX$ over random erasure patterns. Asymptotically by showing that the variance vanishes, a concentration analysis of a typical subset to $m_d$ is completed. 

The Stieltjes transform connects between the typical asymptotic empirical moment and the asymptotic density of the eigenvalues.
Let $\lambda$ be a random variable, the distribution of which equals to the asymptotic spectrum of the random matrix $\bX'\bX$, $f_{\bX\bX'}$. The Stieltjes transform of the matrix (or the transform of its asymptotic spectrum) is defined by
\begin{equation} \label{Stie}
S_\lambda(z)=\Ev\left[ \frac{1}{\lambda - z}\right].
\end{equation}
Expanding the transform in a Laurent series we obtain the connection to moments:
\begin{equation} \label{Stie expend}
S_\lambda(z)=\Ev\left[ -\frac{1}{z}\left(\frac{1}{1-\frac{\lambda}{z}}\right)\right]=\Ev\left[ -\frac{1}{z} \sum_{k=0}^{\infty}\left(\frac{\lambda}{z}\right)^k\right]=-\frac{1}{z}\sum_{k=0}^{\infty} \frac{\Ev(\lambda^k)}{z^k}=-\frac{1}{z}\sum_{k=0}^{\infty} \frac{m_k}{z^k}.
\end{equation}
Given $S_\lambda(\cdot)$, the inversion formula that yields the p.d.f. of $\lambda$ is
\begin{equation} \label{Stie inv}
f_{\bX\bX'}(\lambda)=\lim\limits_{\omega\to 0^+}\frac{1}{\pi}\Im\left[S_f(\lambda+j\omega)\right].
\end{equation}

Carleman's condition \cite{akhiezer1965classical} gives a sufficient condition for the determinacy of the moment problem. That is, if the moments of subsets of ETF satisfy Carleman's condition, the spectral density must be the same as MANOVA with the same moments.

Calculating these moments assuming ETF is a simpler task than proving a lower bound.   
\subsection{Recursive Algorithm for Calculation of Erased ETF Moments}
We suggest a recursive algorithm for the calculation of higher moments in the limit $n\to \infty$, under the assumption that the frame is ETF. 
According to \eqref{md poly of p}, 
\begin{equation} \label{trd}
\begin{split}
&\lim\limits_{n\to \infty}m_d = \lim\limits_{n\to \infty}\frac{1}{n}\Ev\left[\tr \left((\bX\bX')^d\right) \right]\\&=\lim\limits_{n\to \infty}\frac{1}{n}\Ev\left[ \sum_{i_1,i_2,i_3,..,i_d}^{n}c_{i_1,i_2}c_{i_2,i_3},..,c_{i_d,i_1}p_{i_1,i_1}\cdots p_{i_d,i_d} \right] \\&= p + p^2 a_{d,2}^*(\bF) + p^3 a_{d,3}^*(\bF) + \cdots p^d a_{d,d}^*(\bF)
\end{split}
\end{equation}
where $a_{d,k}^*(\bF)$ are the asymptotic values of $a_{d,k}(\bF)$ form \eqref{md poly of p} for $n\to \infty$.
\begin{equation} \label{a_dk}
\begin{split}
a_{d,k}^*(\bF) \triangleq \lim_{n\to \infty }\frac{1}{n}\sum\limits_{k \text{ out of } d \text{ indices are distinct} }
 c_{i_1,i_2}c_{i_2,i_3},..,c_{i_d,i_1}.
\end{split}
\end{equation}
$a_{d,d}^*$ corresponds to the cycle of correlations for all distinct indices:
\begin{equation} \label{a_dd}
\begin{split}
a_{d,d}^*(\bF) \triangleq \lim_{n\to \infty } \frac{1}{n} \sum_{i_1\neq i_2\neq i_3\neq ..\neq i_d}^{n}c_{i_1,i_2}c_{i_2,i_3},..,c_{i_d,i_1}.
\end{split}
\end{equation}
We will see that $a_{d,k}^*(\bF)$, when $\bF$ is ETF, are functions of $x=\frac{n}{m}-1$ and we thus denote them by $a_{d,k}(x)$.
$a_{d,d}(x)$,  the summation over a cycle of correlations, can be computed directly. Alternatively, since ETF is also UTF we can also use Lemma \ref{lemma1}:
\begin{equation} \label{a_dd_q1}
\begin{split}
a_{d,d}(x) = \left(\frac{n}{m}\right)^{d-1} - \sum_{k=2}^{d-1} a_{d,k}(x)-1 = (x+1)^{d-1} - \sum_{k=2}^{d-1} a_{d,k}(x)-1
\end{split}
\end{equation}
As for $a_{d,k}(x), \ k<d$, it stands for all the partitions with only $k$ out of $n$ distinct indices (${i_1,...,i_d}$). Each partition induces a summation over multiplication of smaller cycles -  $a_{d,b}^{(j)}(x)$. Imagine a cyclic graph with $d$ nodes. Partitioning to $k$ groups, means uniting nodes of same group. For a given partition after uniting we get a new graph with (one or more) smaller cycles. 
We now use two facts which still require a formal proof:
\begin{itemize}
	\item Summation over multiplication of smaller cycles is equal to the multiplication of coefficients which correspond to smaller cycles:
	\begin{equation} \label{a_dk prod}
	a_{d,k}^{(j)}(x) = \prod_{l=1}^{s}a_{d_l,d_l}(x)
	\end{equation}
	where $s$ is the number of cycles and $d_l<d$ is the size of cycle $l$. 
	\item Crossing partitions result in asymptotically vanishing expressions for $a_{d,k}^{(j)}(x)$.
	
\end{itemize}
The degree of \eqref{a_dk prod} is $d_l-1$, thus the degree of $a_{d,k}(x)$ is $\sum_{l=1}^{s}(d_l-1) = \sum_{l=1}^{s}d_l - s$. If the new graph is a cycle of $k$ nodes, then $d_1 = k$. For any additional cycle, there is one more edge added to the total count, thus
\begin{equation} \label{sum d_l}
\begin{split}
\sum_{l=1}^{s}d_l = k + s -1.
\end{split}
\end{equation} 
\begin{equation} \label{a_dk_degree}
\begin{split}
\Rightarrow \text{degree}(a_{d,k}(x)) = k -1
\end{split}
\end{equation}
$\sum_{l=1}^{s}d_l\le d$ as uniting two neighboring nodes removes an edge, while uniting two non-neighboring nodes preserves the number of edges.
Thus, $k + s -1\le d$. From the other sides $d_l\ge 2 \Rightarrow k + s -1 \ge 2s$. This means that for a given $k$, the possible amount of cycles induced by the partition is $s\le \min(k-1, d-(k-1))$. For every $s$, the combination of $\{d_l\}$ which satisfy (\ref{sum d_l}) determines the expression in \eqref{a_dk prod} but the number of appearances of such combinations in the induced cycles by partition, is still a question. 
In such model, we are counting all the non-crossing partitions.

The set of all noncrossing partitions of a $d$-element set is enumerated by the Catalan numbers $C(d)$. The number of noncrossing partitions of a $d$-element set with $k$ blocks is the Narayana number 
\begin{equation} \label{narayana number}
N(d,k)=\frac{1}{d}{{d}\choose{k}}{{d}\choose{k-1}}.
\end{equation}
\noindent{\bf Examples and Visualization}
In order to visualize the method we draw two graphs for each partition. The left one contains $d$ vertices, each represents one of the indices $i_1,\dots,i_d$ from the sum in \eqref{md poly of p}. A given partition induces a coloring - same color indicates that the summation is over identical values of the relevant indices. The right graph shows the induced cycles, the vertices are the remaining distinct indices and the edges are the correlations $c_{i_l,i_t}$.
A single cycle means that the summation is over a chain of $k$ correlations ($k$ distinct indices) and the result is $a_{k,k}(x)$.

We could also use the recursive method for calculation of the asymptotic $m_4$ under the assumption of ETFness. 
\begin{equation} \label{lim m4 rec}
\begin{split}
&\lim\limits_{n\to \infty}m_4 = p + p^2 a_{4,2}(x) + p^3 a_{4,3}(x) +  p^4 a_{4,4}(x)
\end{split}
\end{equation}
For $a_{4,3}$ calculation, $d=4$, partition to $k=3$ groups gives:

\SetGraphUnit{1.8}
\SetVertexSimple[MinSize    = 10pt, LineColor = blue!60, FillColor = blue!60]

\begin{tikzpicture}[rotate=90] 
\Vertices[NoLabel]{circle}{A,B,C,D}
\AddVertexColor{red}{A,B}
\AddVertexColor{green}{C}
\AddVertexColor{yellow}{D}
\end{tikzpicture}
\hspace{2cm}
\begin{tikzpicture}[rotate=135] 
\Vertices[NoLabel]{circle}{A,B,C}
\AddVertexColor{red}{A}
\AddVertexColor{green}{B}
\AddVertexColor{yellow}{C}
\draw[-,bend right]  (A) edge (B);
\draw[-,bend right]  (B) edge (C);
\draw[-,bend right]  (C) edge (A);
\end{tikzpicture}
$ 4\cdot a_{3,3}(x)$
\begin{tikzpicture}
\node[draw,text width=3.5cm] at (0,10) {\em \textcolor{red}{Nodes: uniting identical indices (same color) \\Edges: correlations} };

\end{tikzpicture}
\\
As there are 4 ways to unite two neighboring indices in the chain, $a_{4,3}(x)$ contains $ 4a_{3,3}(x)$ in its expression.

\begin{tikzpicture}[rotate=90] 
\Vertices[NoLabel]{circle}{A,B,C,D}
\AddVertexColor{red}{A,C}
\AddVertexColor{green}{B}
\AddVertexColor{yellow}{D}
\end{tikzpicture}
\hspace{2cm}
\begin{tikzpicture}[rotate=0] 
\Vertices[NoLabel]{line}{A,B,C}
\AddVertexColor{red}{B}
\AddVertexColor{green}{A}
\AddVertexColor{yellow}{C}
\draw[-,bend right]  (A) edge (B);
\draw[-,bend right]  (B) edge (A);
\draw[-,bend right]  (B) edge (C);
\draw[-,bend right]  (C) edge (B);
\end{tikzpicture}
$ 2\cdot a_{2,2}^2(x)$
\begin{tikzpicture}
\node[draw,text width=3cm] at (0,10) {\em \textcolor{red}{summation over two smaller \\cycles} };

\end{tikzpicture}\\
And in total $a_{4,3}(x)=4a_{3,3}(x)+2a_{2,2}^2(x)=4(x^2-x)+x^2=6x^2-4x$.\\
For $a_{4,2}$ calculation, $d=4$, partition to $k=2$ groups gives:

\begin{tikzpicture}[rotate=90] 
\Vertices[NoLabel]{circle}{A,B,C,D}
\AddVertexColor{red}{A,B}
\AddVertexColor{green}{C,D}
\end{tikzpicture}
\hspace{2cm}
\begin{tikzpicture}[rotate=135] 
\Vertices[NoLabel]{circle}{A,B}
\AddVertexColor{red}{A}
\AddVertexColor{green}{B}
\draw[-,bend right]  (A) edge (B);
\draw[-,bend right]  (B) edge (A);
\end{tikzpicture}
$ 2\cdot a_{2,2}(x)$

\begin{tikzpicture}[rotate=90] 
\Vertices[NoLabel]{circle}{A,B,C,D}
\AddVertexColor{red}{A,B,C}
\AddVertexColor{green}{D}
\end{tikzpicture}
\hspace{2cm}
\begin{tikzpicture}[rotate=0] 
\Vertices[NoLabel]{line}{A,B}
\AddVertexColor{red}{A}
\AddVertexColor{green}{B}
\draw[-,bend right]  (A) edge (B);
\draw[-,bend right]  (B) edge (A);
\end{tikzpicture}
$ 4\cdot a_{2,2}(x)$

\begin{tikzpicture}[rotate=90] 
\Vertices[NoLabel]{circle}{A,B,C,D}
\AddVertexColor{red}{A,C}
\AddVertexColor{green}{B,D}
\end{tikzpicture}
\hspace{2cm}
Crossing partition! $\to 0$
\begin{tikzpicture}
\node[draw,text width=2.7cm] at (0,10) {\em \textcolor{red}{$i_1=i_3,i_2=i_4$} };

\end{tikzpicture}\\

And in total $a_{4,2}(x)=6a_{2,2}(x)=6x$.

Using Lemma \ref{lemma1} with equality we have:
\begin{equation}
a_{4,4} = (x+1)^3-1-6x-(6x^2-4x) =x^3-3x^2+x
\end{equation}
Substituting $a_{4,2}(x)$,$a_{4,3}(x)$ and $a_{4,4}(x)$ we get:
\begin{equation}
\lim\limits_{n\to \infty}m_4 = p + p^2 6x + p^3 (6x^2-4x) +  p^4 (x^3-3x^2+x).
\end{equation}

Visualization for $d=6$, partitions to $k=3$ groups and the induced cycles:
\SetGraphUnit{1.8}
\SetVertexSimple[MinSize    = 10pt, LineColor = blue!60, FillColor = blue!60]

\begin{tikzpicture}[rotate=90] 
\Vertices[NoLabel]{circle}{A,B,C,D,E,F}
\AddVertexColor{red}{F,A,B,C}
\AddVertexColor{green}{D}
\end{tikzpicture}
\hspace{2cm}
\begin{tikzpicture}[rotate=10] 
\Vertices[NoLabel]{circle}{A,B,C}
\AddVertexColor{red}{B}
\AddVertexColor{green}{C}
\draw[-,bend right]  (A) edge (B);
\draw[-,bend right]  (B) edge (C);
\draw[-,bend right]  (C) edge (A);
\end{tikzpicture}
$6\cdot a_{3,3}(x)$

\vspace{1cm}
\begin{tikzpicture}[rotate=90] 
\Vertices[NoLabel]{circle}{A,B,C,D,E,F}
\AddVertexColor{red}{A,B,C}
\AddVertexColor{green}{D}
\end{tikzpicture}
\hspace{2cm}
\begin{tikzpicture}[rotate=30] 
\Vertices[NoLabel]{circle}{A,B,C}
\AddVertexColor{red}{B}
\AddVertexColor{green}{C}
\draw[-,bend right]  (A) edge (B);
\draw[-,bend right]  (B) edge (C);
\draw[-,bend right]  (C) edge (A);
\end{tikzpicture}
$2\cdot 6\cdot a_{3,3}(x)$

\vspace{1cm} 
\begin{tikzpicture}[rotate=90] 
\Vertices[NoLabel]{circle}{A,B,C,D,E,F}
\AddVertexColor{red}{A,B}
\AddVertexColor{green}{C,D}
\end{tikzpicture}
\hspace{2cm}
\begin{tikzpicture}[rotate=10] 
\Vertices[NoLabel]{circle}{A,B,C}
\AddVertexColor{red}{B}
\AddVertexColor{green}{C}
\draw[-,bend right]  (A) edge (B);
\draw[-,bend right]  (B) edge (C);
\draw[-,bend right]  (C) edge (A);
\end{tikzpicture}
$2\cdot a_{3,3}(x)$

\vspace{1cm}
\begin{tikzpicture}[rotate=90] 
\Vertices[NoLabel]{circle}{A,B,C,D,E,F}
\AddVertexColor{red}{A,B,C}
\AddVertexColor{green}{F,D}
\end{tikzpicture}
\hspace{2cm}
\begin{tikzpicture}[rotate=-45] 
\Vertices[NoLabel]{line}{A,B,C}
\AddVertexColor{red}{A}
\AddVertexColor{green}{B}
\draw[-,bend right]  (A) edge (B);
\draw[-,bend right]  (B) edge (C);
\draw[-,bend right]  (B) edge (A);
\draw[-,bend right]  (C) edge (B);
\end{tikzpicture}
$6\cdot a_{2,2}^2(x)$

\vspace{1cm}
\begin{tikzpicture}[rotate=90] 
\Vertices[NoLabel]{circle}{A,B,C,D,E,F}
\AddVertexColor{red}{A,B,D}
\AddVertexColor{green}{F,E}
\end{tikzpicture}
\hspace{2cm}
\begin{tikzpicture}[rotate=45] 
\Vertices[NoLabel]{line}{A,B,C}
\AddVertexColor{red}{B}
\AddVertexColor{green}{C}
\draw[-,bend right]  (A) edge (B);
\draw[-,bend right]  (B) edge (C);
\draw[-,bend right]  (B) edge (A);
\draw[-,bend right]  (C) edge (B);
\end{tikzpicture}
$12\cdot a_{2,2}^2(x)$

\vspace{1cm}
\begin{tikzpicture}[rotate=90] 
\Vertices[NoLabel]{circle}{A,B,C,D,E,F}
\AddVertexColor{red}{A,B,D,E}
\AddVertexColor{green}{F}
\end{tikzpicture}
\hspace{2cm}
\begin{tikzpicture}[rotate=45] 
\Vertices[NoLabel]{line}{A,B,C}
\AddVertexColor{red}{B}
\AddVertexColor{green}{C}
\draw[-,bend right]  (A) edge (B);
\draw[-,bend right]  (B) edge (C);
\draw[-,bend right]  (B) edge (A);
\draw[-,bend right]  (C) edge (B);
\end{tikzpicture}
$3\cdot a_{2,2}^2(x)$

\vspace{1cm}
\begin{tikzpicture}[rotate=90] 
\Vertices[NoLabel]{circle}{A,B,C,D,E,F}
\AddVertexColor{red}{A,B,C,E}
\AddVertexColor{green}{F}
\end{tikzpicture}
\hspace{2cm}
\begin{tikzpicture}[rotate=45] 
\Vertices[NoLabel]{line}{A,B,C}
\AddVertexColor{red}{B}
\AddVertexColor{green}{C}
\draw[-,bend right]  (A) edge (B);
\draw[-,bend right]  (B) edge (C);
\draw[-,bend right]  (B) edge (A);
\draw[-,bend right]  (C) edge (B);
\end{tikzpicture}
$6\cdot a_{2,2}^2(x)$

\vspace{1cm}
\begin{tikzpicture}[rotate=90] 
\Vertices[NoLabel]{circle}{A,B,C,D,E,F}
\AddVertexColor{red}{A,B}
\AddVertexColor{green}{C,F}
\end{tikzpicture}
\hspace{2cm}
\begin{tikzpicture}[rotate=-45] 
\Vertices[NoLabel]{line}{A,B,C}
\AddVertexColor{red}{A}
\AddVertexColor{green}{B}
\draw[-,bend right]  (A) edge (B);
\draw[-,bend right]  (B) edge (C);
\draw[-,bend right]  (B) edge (A);
\draw[-,bend right]  (C) edge (B);
\end{tikzpicture}
$3\cdot a_{2,2}^2(x)$

And in total $20a_{3,3}(x) +30a_{2,2}^2(x)$

Visualization for $d=6$, partitions to $k=4$ groups and the induced cycles:

\begin{tikzpicture}[rotate=90] 
\Vertices[NoLabel]{circle}{A,B,C,D,E,F}
\AddVertexColor{red}{A,B,F}
\AddVertexColor{green}{D}
\AddVertexColor{yellow}{E}
\end{tikzpicture}
\hspace{2cm}
\begin{tikzpicture}[rotate=90] 
\Vertices[NoLabel]{circle}{A,B,C,D}
\AddVertexColor{red}{A}
\AddVertexColor{green}{C}
\AddVertexColor{yellow}{D}
\draw[-,bend right]  (A) edge (B);
\draw[-,bend right]  (B) edge (C);
\draw[-,bend right]  (C) edge (D);
\draw[-,bend right]  (D) edge (A);
\end{tikzpicture}
$ 6\cdot a_{4,4}(x)$

\vspace{1cm}
\begin{tikzpicture}[rotate=90] 
\Vertices[NoLabel]{circle}{A,B,C,D,E,F}
\AddVertexColor{red}{A,B}
\AddVertexColor{green}{D}
\AddVertexColor{yellow}{E,F}
\end{tikzpicture}
\hspace{2cm}
\begin{tikzpicture}[rotate=90] 
\Vertices[NoLabel]{circle}{A,B,C,D}
\AddVertexColor{red}{A}
\AddVertexColor{green}{C}
\AddVertexColor{yellow}{D}
\draw[-,bend right]  (A) edge (B);
\draw[-,bend right]  (B) edge (C);
\draw[-,bend right]  (C) edge (D);
\draw[-,bend right]  (D) edge (A);
\end{tikzpicture}
$ 6\cdot a_{4,4}(x)$

\vspace{1cm}
\begin{tikzpicture}[rotate=90] 
\Vertices[NoLabel]{circle}{A,B,C,D,E,F}
\AddVertexColor{red}{A,B}
\AddVertexColor{green}{D,E}
\AddVertexColor{yellow}{F}
\end{tikzpicture}
\hspace{2cm}
\begin{tikzpicture}[rotate=90] 
\Vertices[NoLabel]{circle}{A,B,C,D}
\AddVertexColor{red}{A}
\AddVertexColor{green}{C}
\AddVertexColor{yellow}{D}
\draw[-,bend right]  (A) edge (B);
\draw[-,bend right]  (B) edge (C);
\draw[-,bend right]  (C) edge (D);
\draw[-,bend right]  (D) edge (A);
\end{tikzpicture}
$ 3\cdot a_{4,4}(x)$

\vspace{1cm}
\begin{tikzpicture}[rotate=90] 
\Vertices[NoLabel]{circle}{A,B,C,D,E,F}
\AddVertexColor{red}{A,B,E}
\AddVertexColor{green}{D}
\AddVertexColor{yellow}{F}
\end{tikzpicture}
\hspace{2cm}
\begin{tikzpicture}[rotate=90] 
\Vertices[NoLabel]{circle}{A,B,C,D}
\AddVertexColor{red}{A}
\AddVertexColor{green}{C}
\AddVertexColor{yellow}{D}
\draw[-,bend right]  (A) edge (B);
\draw[-,bend right]  (B) edge (C);
\draw[-,bend right]  (C) edge (A);
\draw[-,bend right]  (D) edge (A);
\draw[-,bend right]  (A) edge (D);
\end{tikzpicture}
$ 12\cdot a_{3,3}(x)\cdot a_{2,2}(x)$

\vspace{1cm}
\begin{tikzpicture}[rotate=90] 
\Vertices[NoLabel]{circle}{A,B,C,D,E,F}
\AddVertexColor{red}{A,B}
\AddVertexColor{green}{D}
\AddVertexColor{yellow}{E}
\end{tikzpicture}
\hspace{2cm}
\begin{tikzpicture}[rotate=90] 
\Vertices[NoLabel]{circle}{A,B,C,D}
\AddVertexColor{red}{A}
\AddVertexColor{green}{C}
\AddVertexColor{yellow}{D}
\draw[-,bend right]  (A) edge (B);
\draw[-,bend right]  (B) edge (A);
\draw[-,bend right]  (B) edge (C);
\draw[-,bend right]  (C) edge (D);
\draw[-,bend right]  (D) edge (B);
\end{tikzpicture}
$ 6\cdot a_{3,3}(x)\cdot a_{2,2}(x)$

\vspace{1cm}
\begin{tikzpicture}[rotate=90] 
\Vertices[NoLabel]{circle}{A,B,C,D,E,F}
\AddVertexColor{red}{A,B}
\AddVertexColor{green}{D,F}
\AddVertexColor{yellow}{E}
\end{tikzpicture}
\hspace{2cm}
\begin{tikzpicture}[rotate=90] 
\Vertices[NoLabel]{circle}{A,B,C,D}
\AddVertexColor{red}{A}
\AddVertexColor{green}{C}
\AddVertexColor{yellow}{D}
\draw[-,bend right]  (A) edge (B);
\draw[-,bend right]  (C) edge (A);
\draw[-,bend right]  (B) edge (C);
\draw[-,bend right]  (C) edge (D);
\draw[-,bend right]  (D) edge (C);
\end{tikzpicture}
$ 12\cdot a_{3,3}(x)\cdot a_{2,2}(x)$

\vspace{1cm}
\begin{tikzpicture}[rotate=90] 
\Vertices[NoLabel]{circle}{A,B,C,D,E,F}
\AddVertexColor{red}{B,D,F}
\AddVertexColor{green}{C}
\AddVertexColor{yellow}{E}
\end{tikzpicture}
\hspace{2cm}
\begin{tikzpicture}[rotate=90] 
\Vertices[NoLabel]{circle}{A,B,C}
\node[draw, inner sep=0pt] (D) {};

\AddVertexColor{red}{D}
\AddVertexColor{green}{B}
\AddVertexColor{yellow}{C}
\draw[-,bend right]  (A) edge (D);
\draw[-,bend right]  (D) edge (A);
\draw[-,bend right]  (C) edge (D);
\draw[-,bend right]  (D) edge (C);
\draw[-,bend right]  (B) edge (D);
\draw[-,bend right]  (D) edge (B);

\end{tikzpicture}
$ 2\cdot a_{2,2}^3(x)$

\vspace{1cm}
\begin{tikzpicture}[rotate=90] 
\Vertices[NoLabel]{circle}{A,B,C,D,E,F}
\AddVertexColor{red}{A,C}
\AddVertexColor{green}{D,F}
\AddVertexColor{yellow}{E}
\end{tikzpicture}
\hspace{2cm}
\begin{tikzpicture}[rotate=-45] 
\Vertices[NoLabel]{line}{A,B,C,D}

\AddVertexColor{red}{C}
\AddVertexColor{green}{B}
\AddVertexColor{yellow}{A}
\draw[-,bend right]  (D) edge (C);
\draw[-,bend right]  (C) edge (D);
\draw[-,bend right]  (C) edge (B);
\draw[-,bend right]  (B) edge (C);
\draw[-,bend right]  (B) edge (A);
\draw[-,bend right]  (A) edge (B);

\end{tikzpicture}
$ 3\cdot a_{2,2}^3(x)$

And in total $15a_{4,4}(x) +30a_{3,3}(x)a_{2,2}(x)+5a_{2,2}^3(x)$\\

\subsection{Results and Obsrevations}
Moments $m_2$, $m_3$ and $m_4$ which are defined in \eqref{Manova moments}, are confirmed with this method. Moments $m_5$ and $m_6$ are also calculated in this manner and are indeed in agreement with the moments of
Wachter's MANOVA ensemble, \cite{dubbs2015infinite}.
We bring here the resulting expressions for the moments, below each expression note the the sum of coefficients comprising $a_{d,k}(x)$ is the Narayana number $N(d,k)$:
\begin{equation} \label{m2 recursion}
\begin{split}
m_2 &= p + p^2\cdot \underbrace{x}_{a_{2,2}}
\end{split}
\end{equation}
\begin{equation} \label{m3 recursion}
\begin{split}
m_3 &= p + p^2\cdot \textbf{3}a_{2,2}(x) + p^3((x+1)^2-a_{3,2}(x)-1) \\&= p + p^2 \textcolor{blue}{3}x + p^3 \underbrace{(x^2-x) }_{a_{3,3}}
\end{split}
\end{equation}
\begin{equation} \label{m3 narayana}
\begin{split}
\textbf{3} = N(3,2) 
\end{split}
\end{equation}
\begin{equation} \label{m4 recursion}
\begin{split}
\lim\limits_{n\to \infty}m_4 &= p + p^2\cdot \textbf{6}a_{2,2}(x) + p^3(\textbf{4}a_{3,3}(x)+\textbf{2}a_{2,2}^2(x)) + p^4((x+1)^2-a_{4,3}(x)-a_{4,2}(x)-1) \\&= p + p^2 \textcolor{blue}{6}x + p^3 (\textcolor{blue}{6}x^2-\textcolor{blue}{4}x)  + p^4 \underbrace{(x^3 - \textcolor{blue}{\bf 3}x^2 + x)  }_{a_{4,4}} 
\end{split}
\end{equation}
\begin{equation} \label{m4 narayana}
\begin{split}
\textbf{6} = N(4,2), \ \  \textbf{4}+\textbf{2} = 6 = N(4,3)
\end{split}
\end{equation}
\begin{equation} \label{m5 recursion}
\begin{split}
\lim\limits_{n\to \infty}m_5 &= p + p^2\cdot \textbf{10}a_{2,2}(x) + p^3(\textbf{10}a_{3,3}(x)+\textbf{10}a_{2,2}^2(x)) +p^4(\textbf{5}a_{4,4}(x)+\textbf{5}a_{3,3}(x)\cdot a_{2,2}(x)) \\&+ p^5((x+1)^2-a_{5,4}(x)-a_{5,3}(x)-a_{5,2}(x)-1) \\&= p + p^2 \textcolor{blue}{10}x + p^3 (\textcolor{blue}{20}x^2-\textcolor{blue}{10}x)  + p^4 (\textcolor{blue}{10}x^3 - \textcolor{blue}{20}x^2 + \textcolor{blue}{5}x) + p^5 \underbrace{(x^4-\textcolor{ blue}{\bf 6}x^3 + \textcolor{blue}{\bf 6}x^2 -x) }_{a_{5,5}}   
\end{split}
\end{equation}
\begin{equation} \label{m5 narayana}
\begin{split}
\textbf{10} = N(5,2), \ \  \textbf{10}+\textbf{10} = 20 = N(5,3), \ \  \textbf{5}+\textbf{5} = 10 = N(5,4)
\end{split}
\end{equation}
\begin{equation} \label{m6 recursion}
\begin{split}
\lim\limits_{n\to \infty}m_6 &= p + p^2\cdot \textbf{15}a_{2,2}(x) + p^3(\textbf{20}a_{3,3}(x)+\textbf{30}a_{2,2}^2(x)) \\&+p^4(\textbf{15}a_{4,4}(x)+\textbf{30}a_{3,3}(x)\cdot a_{2,2}(x)+\textbf{5}a_{2,2}^3(x)) \\&+p^5(\textbf{6}a_{5,5}(x)+\textbf{6}a_{4,4}(x)\cdot a_{2,2}(x)+\textbf{3}a_{3,3}^2(x)) \\&+ p^6((x+1)^2-a_{6,5}(x)-a_{6,4}(x)-a_{6,3}(x)-a_{6,2}(x)-1) \\&= p + p^2 \textcolor{blue}{15}x + p^3 (\textcolor{blue}{50}x^2-\textcolor{blue}{20}x)  + p^4 (\textcolor{blue}{50}x^3 - \textcolor{blue}{75}x^2 + \textcolor{blue}{15}x) + p^5 (\textcolor{blue}{15}x^4-\textcolor{blue}{60}x^3 + \textcolor{blue}{45}x^2 -\textcolor{blue}{6}x)  \\&+ p^6 \underbrace{(x^5-\textcolor{blue}{\bf 10}x^4+\textcolor{blue}{\bf 20}x^3 -\textcolor{blue}{\bf 10}x^2 +x)   }_{a_{6,6}} 
\end{split}
\end{equation}
\begin{equation} \label{m6 narayana}
\begin{split}
\textbf{15} = N(6,2), \ \  \textbf{20}+\textbf{30} = 50 = N(6,3), \ \  \textbf{15}+\textbf{30}+\textbf{5} = 50 = N(6,4), \ \  \textbf{6}+\textbf{6}+\textbf{3} = 15 = N(6,5)
\end{split}
\end{equation}
In \cite{dubbs2015infinite}, the moments of the MANOVA ensemble are derived through direct computation of the integral on the density. Dubbs and Edelman extract pyramids of numbers from the coefficients of the moments. We can identify the coefficients of $\{a_{i,d}\}_{i=2}^{d}$ for $2\le d\le6$ (blue coefficients in \eqref{m2 recursion}-\eqref{m6 recursion}) in the first 5 pyramids.

Surprisingly, all $a_{d,d}(x)$ are the $N_{d-1}(x)$ Narayana polynomials (Narayana numbers $N(d-1,j)$ as coefficients), which means that $a_{d,k}(x)$ can be expressed as a function of $N_{i}(x)$s for $1 \le i\le k-2$ and $m_d(x)$ as a function of $N_{i}(x)$s for $1 \le i\le k-1$.

\subsection{Future Work}
A rigorous formulation and proof, as well as generalized closed expression for the recursion coefficients, is still part of a current work.
A quite similar method for moments analysis was part of Wachter's work for the MANOVA spectral density calculation \cite{wachter1978strong}.
Another possible approach is to understand whether the assumptions of randomness of the MANOVA ensemble are valid for random subsets of ETFs, and then the same proof can be applied.  
Alternatively, maybe asymptotic equiangular properties hold for random MANOVA matrices.  




	\chapter{Conclusions}
	\label{chapter:Conclusions}
		\section{Summary}
This thesis was motivated by a source coding problem with erasure side information at the encoder. We suggested and studied an analog coding approach based on frames, along with an independent research of frames. In the latter study we explored the properties of structured frames, which lead us far beyond the application of analog coding. 
Since a previously suggested scheme, based on bandlimited interpolation, suffered from severe signal amplification, we turned to examination of possible extensions. 
The first scheme we proposed was designed with a random i.i.d frame with redundancy. 
Next, we considered different representation of signals by frames, from irregular spectrum (difference set or random spectrum) through various ETFs.
We studied the performance of the schemes based on these different approaches, and significant improvement was achieved using different types of ETFs.

The surprising observation that many ETFs and near-ETFs share similar performance, as well as MANOVA limiting spectral distribution, led to a comprehensive empirical study of universality properties for the submatrix ensemble corresponding to various well-known deterministic frames.
A common challenge for many applications, such as sparse signals representation and erasure channels, is a design of frames with favorable properties of subframes. Thus, such observation enables evaluation of numerous measures on a whole family of frames.

Finally, we have proved two exciting results on the moments of frames. One is a tight lower bound on unit norm frames, achieved with equality for ETFs. Second is the already expected fact (from our empiric study), that the asymptotic moments of ETFs match the moments of MANOVA density. 

If we could prove that ETFs satisfy a lower bound on all moments with equality, including the inverse moment, our assumption on the superiority of ETFs over all possible frames will be approved.
This will also serve as a converse for the achievable performance using analog coding for erasure channels or source with erasures.
Specifically, it will prove that the optimal information-theoretic solution cannot be achieved by means of analog coding.
\section{Further Research}
We bring here few possible directions which are natural extensions of our work. 
\begin{enumerate}
	\item {\bf Analysis of optimal rates using ETFs in finite SNR/SDR regime}. Performance evaluation at different erasure fractions and optimal $\beta$ (redundancy factor).
	
	\item {\bf Are ETFs the best we can do with analog coding approach?} Extension of the Erasure Welch Bound to higher moments, $d\ge 5$ may bring us closer to the answer. Another open questions: Bernoulli erasure model versus fixed $n-k$ out of $n$ (the analysis in \ref{secEWB} assumed Bernoulli), robustness to worst case versus average erasure patterns.

	\item {\bf Rigorous mathematical study of the 
		fascinating phenomena described in Chapter \ref{chapter:PNAS}}. We have started with the analysis of expectation of moments over random subsets. Derivation of a closed-form expression for higher asymptotic moments of ETF must be completed.
	In order to prove that the limiting spectral density of subsets of ETF is MANOVA, a concentration analysis is required, i.e, showing that the variance of moments vanishes asymptotically.
	
	\item {\bf Understanding of the origin of the close relationship between MANOVA distribution, an object from random matrix theory, and deterministic construction of ETFs}. This may enable an easy application of Wachter's derivation on ETFs.
	
\end{enumerate}

	\appendix

	\chapter{A Direct Calculation of $a_{3,3}$ and $a_{4,4}$}
	\label{app:app1}
	In this appendix we will show that
assuming UTF we could also compute $a_{3,3}(F)$ and $a_{4,4}(F)$  directly without using \eqref{md bound_p=1}. The idea is to start with summation over column indices: 
\begin{equation} \label{cycle3 calc}
\begin{split}
a_{3,3}(F)&=\frac{1}{n}\sum_{j_1,j_2,j_3}^{m}\sum_{i_1\neq i_2\neq i_3}^{n}F_{j_1,i_1}F_{j_2,i_1}^*F_{j_2,i_2}F_{j_3,i_2}^*F_{j_3,i_3}F_{j_1,i_3}^* \\&= \frac{1}{n}\sum_{j_1,j_2,j_3}^{m}\sum_{i_1\neq i_2}^{n}F_{j_1,i_1}F_{j_2,i_1}^*F_{j_2,i_2}F_{j_3,i_2}^*\left[ \frac{n}{m}\delta_{j_3,j_1}-F_{j_3,i_1}F_{j_1,i_1}^*- F_{j_3,i_2}F_{j_1,i_2}^* \right] \\&=\frac{1}{n}\frac{n}{m}\sum_{j_1,j_2}^{m}\sum_{i_1\neq i_2}^{n}F_{j_1,i_2}^*F_{j_1,i_1}F_{j_2,i_1}^*F_{j_2,i_2}\\&-\frac{1}{n}\sum_{j_1,j_2,j_3}^{m}\sum_{i_1\neq i_2}^{n}|F_{j_1,i_1}|^2F_{j_2,i_1}^*F_{j_2,i_2}F_{j_3,i_2}^*F_{j_3,i_1}\\&-\frac{1}{n}\sum_{j_1,j_2,j_3}^{m}\sum_{i_1\neq i_2}^{n}F_{j_1,i_2}^*F_{j_1,i_1}F_{j_2,i_1}^*F_{j_2,i_2}|F_{j_3,i_2}|^2\\&=\frac{1}{n}\frac{n}{m}\sum_{i_1=1}^{n}\sum_{i_3\neq i_1}^{n}|c_{i_1,i_3}|^2-2\frac{1}{n}\sum_{i_1=1}^{n}\sum_{i_2\neq i_1}^{n}|c_{i_1,i_2}|^2\\&=\frac{n}{m}a_{2,2}(F)-2a_{2,2}(F) =x^2-x
\end{split}
\end{equation}
where the second equality is due to UTF property (\ref{UTF}) and the fact that 
$i_3\ne i_2 $ and $i_3\ne i_2$, and the last equality is due to \eqref{a22} and \eqref{WB x}.

\begin{equation} \label{cycle4 calc}
\begin{split}
&a_{4,4}(F)=\frac{1}{n}\sum_{j_1,j_2,j_3,j_4}^{m}\sum_{i_1\neq i_2\neq i_3\neq i_4}F_{j_1,i_1}F_{j_2,i_1}^*F_{j_2,i_2}F_{j_3,i_2}^*F_{j_3,i_3}F_{j_4,i_3}^*F_{j_4,i_4}F_{j_1,i_4}^*\\&=\frac{1}{n}\sum_{j_1,j_2,j_3,j_4}^{m}\sum_{i_1\neq i_2\neq i_3}F_{j_1,i_1}F_{j_2,i_1}^*F_{j_2,i_2}F_{j_3,i_2}^*F_{j_3,i_3}F_{j_4,i_3}^*\left[\frac{n}{m}\delta_{j_4,j_1}-F_{j_4,i_1}F_{j_1,i_1}^*-F_{j_4,i_2}F_{j_1,i_2}^*- F_{j_4,i_3}F_{j_1,i_3}^* \right]\\&=\frac{1}{n}\sum_{j_1,j_2,j_3}^{m}\sum_{i_1\neq i_2\neq i_3}\frac{n}{m}F_{j_1,i_1}F_{j_2,i_1}^*F_{j_2,i_2}F_{j_3,i_2}^*F_{j_3,i_3}F_{j_1,i_3}^*\\&-\frac{1}{n}\sum_{\bcancel{j_1},j_2,j_3,j_4}^{m}\sum_{i_1\neq i_2\neq i_3}\bcancel{|F_{j_1,i_1}|^2}F_{j_2,i_1}^*F_{j_2,i_2}F_{j_3,i_2}^*F_{j_3,i_3}F_{j_4,i_3}^*F_{j_4,i_1}
\\&-\frac{1}{n}\sum_{j_1,j_2,j_3,j_4}^{m}\sum_{i_1\neq i_2\neq i_3}F_{j_1,i_1}F_{j_2,i_1}^*F_{j_2,i_2}F_{j_3,i_2}^*F_{j_3,i_3}F_{j_4,i_3}^*F_{j_4,i_2}F_{j_1,i_2}^*
\\&-\frac{1}{n}\sum_{j_1,j_2,j_3,\bcancel{j_4}}^{m}\sum_{i_1\neq i_2\neq i_3}F_{j_1,i_1}F_{j_2,i_1}^*F_{j_2,i_2}F_{j_3,i_2}^*F_{j_3,i_3}\bcancel{|F_{j_4,i_3}|^2}F_{j_1,i_3}^*\\&=\frac{n}{m}a_{3,3}(F)-2a_{3,3}(F)-\frac{1}{n}\sum_{i_1\neq i_2\neq i_4}^{n}|c_{i_1,i_2}|^2|c_{i_3,i_2}|^2=(x+1)(x^2-x)-2(x^2-x)-\frac{n-2}{n-1}x^2\\&=x^3-3x^2+x+\frac{1}{n-1}x^2
\end{split}
\end{equation}
where the evaluation of $\frac{1}{2}a^{(2)}_{4,3}=\frac{1}{n}\sum_{i_1\neq i_2\neq i_4}^{n}|c_{i_1,i_2}|^2|c_{i_3,i_2}|^2$ required ETF. For $a_{3,3}(F)$ we used \eqref{cycle3 calc}.
The final expression can be identified as the term multiplied by $p^4$ in \eqref{m4 bound5}.

	\chapter{Matrices - Short Background}
	\label{app:app2}
	In this appendix we present a brief background on matrices - decompositions, eigenvalues etc. 
Let $\bfA$, $\bfB$ be an $n\times m$, $m\times n$ matrices respectively and  $r=\min(m,n)$. A Gram matrix is defined as $\bfG = \bfA \bfA'$. Let $\bfC$, $\bfD$ be $n\times n$ square matrices.

\section{Matrix Decompositions}
The Singular Value Decomposition (SVD) of $\bfA$ is 
\begin{equation}
\bfA = \bfU \bf\Sigma \bfV'
\end{equation}
where $\bfU$ is an $n\times n$ unitary matrix ($\bfU \bfU'= \bfU' \bfU=I_n$),  $\bfV$ is an $m\times m$ unitary matrix ($\bfV \bfV'= \bfV' \bfV=I_m$), and $\bf\Sigma$ is an $n\times m$ diagonal matrix with the singular values of $\bfA$, $\sigma_1\ge \dots \ge \sigma_r\ge 0$, in its diagonal.

A diagonalizable matrix $\bfC$ is similar to a diagonal matrix. Its diagonalization is 
\begin{equation}
\bfC = \bfP \bf\Lambda \bfP^{-1}
\end{equation}
where $\bf\Lambda$ is an $n\times n$ diagonal matrix  with the eigenvalues of $\bfC$, $\lambda_1, \dots ,\lambda_n$, in its diagonal.
For Hermitian (or real symmetric) matrix $\bfC$, $\lambda_i$ are real values and $\bfP$ is a unitary matrix, i.e, $\bfC = \bfP \bf\Lambda \bfP'$.
For a positive semi-definite matrix $\lambda_i\ge 0$, thus its diagonalization is equivalent to its SVD (with $\bfU=\bfV$).
 
The square $n\times n$ Gram matrix is diagonalizable and is similar to a diagonal matrix $\bf\Sigma\bf\Sigma'$
\begin{equation}
\bfA \bfA' = \bfU \bf\Sigma \bfV' \bfV \bf\Sigma' \bfU' = \bfU \bf\Sigma\bf\Sigma'\bfU'.
\end{equation}
The diagonal of $\Lambda=\bf\Sigma\bf\Sigma'$ consists of the eigenvalues of the Gram matrix $\lambda_1, \dots \lambda_r$ and $n-r$ zeros.
Since diagonalization is unique, $\lambda_i = |\sigma_i|^2$.
Similarly,
\begin{equation}
\bfA' \bfA = \bfV \bf\Sigma'\bf\Sigma\bfV'.
\end{equation}
The diagonal of $\bf\Sigma'\bf\Sigma$ consists of the eigenvalues of the Gram matrix $\lambda_1, \dots \lambda_r$ and $m-r$ zeros.
Thus $\bfA' \bfA$ shares the same non-zero eigenvalues as $\bfA \bfA'$.
Note that if $\bfA$ is full-rank, $\lambda_i$ (and $\sigma_i$) are non-zero for $i=1,\dots r$.

The inverse of a diagonalizable matrix is 
\begin{equation} \label{inverseEig}
\bfC^{-1} =  \bfP \bf\Lambda^{-1}\bfP^{-1} \Rightarrow \lambda(\bfC^{-1})=\frac{1}{\lambda(\bfC)}.
\end{equation}

\section{"Energy" and  Other Means}
Throughout this thesis we consider properties of matrices which involve functions of their eigenvalues. Here we demonstrate the connection between the eigenvalues and the trace and determinant of a matrix:
\begin{equation}\label{traceC}
\tr(\bfA\bfB) = \tr(\bfB\bfA)\Rightarrow\tr(\bfC) = \tr(\bfP \bf\Lambda \bfP^{-1})=\tr( \bf\Lambda \cancel{\bfP^{-1}\bfP})=\tr( \bf\Lambda)=\sum \lambda_i.
\end{equation}
\begin{equation}
\text{det}(\bfC\bfD) = \text{det}(\bfC)\text{det}(\bfD)\Rightarrow\text{det}(\bfC) = \cancel{\text{det}(\bfP)}\text{det}(\bf{\Lambda})\cancel{\text{det}(\bfP^{-1})}=\text{det}( \bf\Lambda)=\prod \lambda_i.
\end{equation} 
Using \eqref{inverseEig} and \eqref{traceC} we get the expression for the trace of the inverse of a Gram matrix (or more generally of an invertible matrix):
\begin{equation}
\tr(\bfG^{-1}) = \sum \lambda_i(\bfG^{-1})=\sum \frac{1}{\lambda(\bfG)}.
\end{equation}

The following measures are all different means of the eigenvalues of a Gram matrix
\begin{enumerate}
	\item "Energy": $\frac{1}{n}\tr(\bfG) = \text{ arithmetic mean of eigenvalues}$.
	\item "Volume": $\sqrt[n]{\text{det}(\bfG)} = \text{ geometric mean of eigenvalues}$.
	\item 1 / "Inverse Energy": $\bigg[\frac{1}{n}\tr(\bfG^{-1})\bigg]^{-1} = \text{ harmonic mean of eigenvalues}$.
\end{enumerate}

\bibliographystyle{plain}
\bibliography{Refs}



\section*{Supplementary material}

## תקציר

קידוד אנלוגי הינו שיטה יעילה להתמודד עם מחיקות אשר מבוססת על יתירות לינארית במרחב הסיגנל. עבודה קודמת בחנה קודים המבוססים על התמרת פורייה (בדידה) חסומת סרט עבור ערוצים גאוסיים עם מחיקות. אנחנו מרחיבים את הגישה הזו ומראים שניתן לשפר משמעותית את הביצועים של התמרה חסומת סרט ע״י שימוש בספקטרום לא רגולרי, או באופן כללי באמצעות מסגרות מיוחדות – equiangular tight frames (ETF).

מסגרות הינן בסיסים מורחבים אשר מהווים מוקד עניין במתמטיקה, מדעי המחשב, הנדסה וסטטיסטיקה, בשל תרומתם להשגת ייצוג יציב ורובסטי. המוטיבציה לתיכנון מסגרות בעלות תכונות רצויות עבור תתי המסגרות מגיעה מתחומים שונים, כולל סדרות מריחה לריבוי משתמשים, חישה דחוסה וקידוד אנלוגי.

אנחנו מציגים קשר חדשני בין מסגרות דטרמיניסטיות ותורת המטריצות האקראיות. אנחנו מראים באופן אמפירי כי אנסמבל MANOVA מציע תיאור אוניברסלי של תתי קבוצות אקראיות מתוך מסגרות דטרמיניסטיות מסוג ETF (או כמעט ETF). כמו כן, אנחנו מפתחים מסגרת עבודה אנליטית ומביאים הוכחה פורמלית לחלק מהתוצאות האמפיריות, בפרט שהצורה האסימפטוטית עבור מומנטים גבוהים של תתי קבוצות של ETFים תואמות לזו של MANOVA.

לבסוף, כשעוסקים בבסיסים מורחבים, חסם Welch הוא חסם תחתון על שורש ממוצע הריבועים של קרוס קורלציות בין הוקטורים. אנחנו מרחיבים את החסם למתאר של מחיקות, בו הבסיס נוצר מתת קבוצה אקראית של וקטורי המסגרת שנבחרו באופן Bernoulli. החסם התחתון הוא הדוק, מושג בשוויון עבור ETF ואסימטוטית מתלכד עם המומנטים של MANOVA. התוצאות האלו מציעות נקודת מבט חדשה על העליונות של ETFים ביחס למסגרות אחרות.

הפקולטה להנדסה ע״ש איבי ואלדר פליישמן
המחלקה להנדסת חשמל מערכות

# מסגרת לקידוד אנלוגי

חיבור זה הוגש כעבודת גמר לקראת התואר ״מוסמך האוניברסיטה״
בהנדסת חשמל ואלקטרוניקה באוניברסיטת תל אביב

מאת

## מרינה חייקין

העבודה נעשתה במחלקה להנדסת חשמל – מערכות
בהנחייתם של

### פרופ׳ רם זמיר
דר׳ מתן גביש

סיון תשע״ח

# מסגרת לקידוד אנלוגי

מאת

**מרינה חייקין**



\end{document}